\documentclass[aps,prb,showpacs,twocolumn,superscriptaddress]{revtex4-2}
\usepackage{siunitx}
\usepackage{amsmath}    
\usepackage{diagbox}
\usepackage{amsfonts}
\usepackage{chemformula}
\usepackage{amssymb}
\usepackage[colorlinks=true,citecolor=blue,linkcolor=red]{hyperref}
\usepackage{graphicx}
\usepackage{bbold}					% for \mathbb{1} as a nice unit matrix symbol

\usepackage{float}

\usepackage{bm} 

\usepackage{dcolumn}
\usepackage{mathtools}

\usepackage{array,booktabs,ragged2e}	
\usepackage{soul}						% strike out text using \st{}
\usepackage{cancel}						% detto, but more clearly visible when reading
\usepackage{tabularx}% for table in the CHS-model discussion
\usepackage{multirow}
\usepackage{hhline}
\usepackage{adjustbox}
\usepackage{sidecap}
\usepackage{threeparttable}
\usepackage{colortbl} %added by JX, for cellcolor in tables.
\usepackage{makecell}
\usepackage[utf8]{inputenc}
\usepackage[T1]{fontenc}
\usepackage{titlesec}
\def\bra#1{\left<{#1}\right|}					% "bra"-state
\def\ket#1{\left|{#1}\right>}					% "ket"-state
\def\braket#1#2{\left<{#1}|{#2}\right>}			% "bra-ket"-product
\newcommand{\angstrom}{\mbox{\normalfont\AA}}	% Angstrom, 10^{-10} meters.

\def\degree {^{\circ}}

\newcommand{\br}{\pmb{r}}

\newcolumntype{P}[1]{>{\raggedleft\arraybackslash}p{#1}}
\newcolumntype{R}[1]{>{\centering\arraybackslash}p{#1}}

\usepackage{bm,amsmath,amssymb,mathrsfs,latexsym,graphicx,psfrag,mathtools}

\usepackage{xcolor}

\usepackage{empheq}%\begin{empheq}[box=\fbox]{align} ... \end{empheq}

\newcommand{\bsl}[1]{\boldsymbol{#1}}

\newcommand{\ii}{\mathrm{i}}

\let\oldAA\AA
\renewcommand{\AA}{\text{\normalfont\oldAA}}

\newcommand{\ie}{{\emph{i.e.}}}

\newcommand{\G}{\mathcal{G}}

\newcommand{\Q}{\mathcal{Q}}

\newcommand{\K}{\text{K}}

\newcommand{\tmt}{\text{$t$MoTe$_2$}}

\usepackage{comment}

\usepackage{cleveref}
\crefname{appendix}{App.}{Apps.}
\crefname{equation}{Eq.}{Eqs.}
\crefname{figure}{Fig.}{Figs.}
\crefname{table}{Tab.}{Tabs.}
\crefname{section}{Sec.}{Secs.}
\creflabelformat{appendix}{[#2#1#3]}
\usepackage[caption=false]{subfig}
\crefmultiformat{figure}{Figs.~#2#1\xdef\mycreffirstarg{#1}#3}%
 { and~#2#1#3}{, #2#1#3}{ and~#2#1#3}
 \crefname{figure}{Fig.}{Figs.}
\newcommand{\bpm}{\begin{pmatrix}}
\newcommand{\epm}{\end{pmatrix}}

\newcommand{\bea}{\begin{equation} \begin{aligned}}
\newcommand{\eea}{\end{aligned} \end{equation} }

\titlespacing{\section}{0pt}{3.5ex}{1.5ex}
\begin{document}

%%%%%%%%%%%%%%%%%%%%%%%%%
%%% TITLE INFORMATION %%%
%%%%%%%%%%%%%%%%%%%%%%%%%

\title{Chern-Selective multi-valley Flat Bands in Twisted Mono-Bilayer and Mono-Trilayer MoTe$_2$}

 \author{Ziyue Qi}
\thanks{These authors contributed equally.}
\affiliation{Beijing National Laboratory for Condensed Matter Physics and Institute of physics,
Chinese academy of sciences, Beijing 100190, China}
\affiliation{University of Chinese academy of sciences, Beijing 100049, China}

\author{Hanqi Pi}
\thanks{These authors contributed equally.}
\affiliation{Beijing National Laboratory for Condensed Matter Physics and Institute of physics,
Chinese academy of sciences, Beijing 100190, China}
\affiliation{University of Chinese academy of sciences, Beijing 100049, China}

\author{Yan Zhang}
\thanks{These authors contributed equally.}
\affiliation{Beijing National Laboratory for Condensed Matter Physics and Institute of physics,
Chinese academy of sciences, Beijing 100190, China}
\affiliation{University of Chinese academy of sciences, Beijing 100049, China}

\author{Jiaxuan Liu}
\affiliation{Beijing National Laboratory for Condensed Matter Physics and Institute of physics,
Chinese academy of sciences, Beijing 100190, China}
\affiliation{University of Chinese academy of sciences, Beijing 100049, China}

\author{Nicolas Regnault}
\affiliation{Center for Computational Quantum Physics, Flatiron Institute, 162 5th Avenue, New York, NY 10010, USA}
\affiliation{Department of Physics, Princeton University, Princeton, New Jersey 08544, USA}
\affiliation{Laboratoire de Physique de l’Ecole normale sup\'erieure,
ENS, Universit\'e PSL, CNRS, Sorbonne Universit\'e,
Universit\'e Paris-Diderot, Sorbonne Paris Cit\'e, 75005 Paris, France}

\author{Hongming Weng}
\affiliation{Beijing National Laboratory for Condensed Matter Physics and Institute of physics,
Chinese academy of sciences, Beijing 100190, China}
\affiliation{University of Chinese academy of sciences, Beijing 100049, China}
\affiliation{Songshan Lake Materials Laboratory, Dongguan, Guangdong 523808, China}

\author{B. Andrei Bernevig}
\email{bernevig@princeton.edu}
\affiliation{Department of Physics, Princeton University, Princeton, New Jersey 08544, USA}
\affiliation{Donostia International Physics Center, P. Manuel de Lardizabal 4, 20018 Donostia-San Sebastian, Spain}
\affiliation{IKERBASQUE, Basque Foundation for Science, Bilbao, Spain}

\author{Jiabin Yu}
\affiliation{Department of Physics, Princeton University, Princeton, New Jersey 08544, USA}
\affiliation{Department of Physics, University of Florida, Gainesville, FL, USA}
\affiliation{Quantum Theory Project, University of Florida, Gainesville, FL, USA}

\author{Quansheng Wu}
\email{quansheng.wu@iphy.ac.cn}
\affiliation{Beijing National Laboratory for Condensed Matter Physics and Institute of physics,
Chinese academy of sciences, Beijing 100190, China}
\affiliation{University of Chinese academy of sciences, Beijing 100049, China}
\date{\today}

\begin{abstract}

The interplay between moir\'e flat bands originating from different valleys can give rise to a variety of exotic  quantum phases. In this work, we investigate the electronic properties of twisted mono-bilayer (A-AB) and mono-trilayer (A-ABA) MoTe$_2$ using first-principles calculations and continuum models.  
Unlike previous studies on twisted bilayer systems, in which low-energy flat bands originate solely from the $K/K'$ valleys, in 
A-AB and A-ABA twisted MoTe$_2$ (\tmt) the moiré bands at low energies arise from both the $\Gamma$ and $K/K'$ valleys, with spin Chern numbers $C_s=0$ (for $\Gamma$) and $C_{\uparrow/\downarrow}=\pm1$ (for $K/K'$), respectively. 
We show that the multi-valley moiré flat bands are governed by interlayer-hybridization effects, and that different stacking configurations and thicknesses tune the relative energy alignment between the $\Gamma$  and $K$ valley moiré flat bands. 
By constructing valley-resolved continuum models and performing Wannierization for the low-energy moir\'e bands, we further uncover that the Berry curvature and quantum metric distributions can be effectively tuned by the layer number and stacking configuration. 
Unlike other moir\'e systems, where only one kind of valley influenced the low energy physics, the simultaneous appearance of two distinct types of valleys, with different symmetries, establish A-AB and A-ABA \tmt\ as ideal platforms for studying layer-controlled multi-valley physics.

\end{abstract} 

\maketitle

\section{Introduction} 
Moiré superlattices in twisted transition metal dichalcogenides (TMDs)~\cite{Wu_Hub_PhysRevLett.121.026402,wu_topological_2019,PhysRevB.103.155142,devakul_magic_2021} %has offered a versatile platform to explore isolated nearly flat bands independent 
offer an exemplary platform to explore a variety of correlated electron states, including Mott~\cite{Wu_Hub_PhysRevLett.121.026402,Regan2020,Tang2020,Wang2020,Li2021} and charge-transfer~\cite{Zhang_PhysRevB.102.201115,Fu_PhysRevX.12.021031,Ryee2023,Campbell2024,Wei2025} insulators, generalized Wigner crystals~\cite{Regan2020,Zhou2021,Li2021_wigner} and correlated topological phases~\cite{Li2021_QAHE,Fu_PhysRevX.12.021031}. Remarkably, fractional Chern insulator (FCI) states~\cite{neupert,sheng,regnault,tang11,Sun2011,PhysRevResearch.3.L032070,PhysRevB.107.L201109} 
%\textbf{BAB: you're forgetting MacDonald Papers while citing a lot of Liang Fu. We should also cite macdonald for MoTe2, and Jiabin can mention other people to cite} 
have recently been observed~\cite{cai2023signatures,zeng2023integer,park2023,PhysRevX.13.031037,xu2023maximally,Ji2024,Redekop2024,Park2025,Xu2025,park2025observationhightemperaturedissipationlessfractional,chang2025evidencecompetinggroundstates,xu2025signaturesunconventionalsuperconductivitynear} in twisted bilayer MoTe$_2$ without an external magnetic field, sparking a surge of research on this system~\cite{reddy_fractional_2023, wang_fractional_2024,MacDonald_PhysRevLett.132.096602,MacDonald_PhysRevB.110.035130,qiu_interaction-driven_2023, crepel_Chiral_2024, dong_composite_2023, yu_fractional_2024, abouelkomsan_band_2024, jia2023moir, wang_topology_2023, reddy_toward_2023, xu_Multiple_2024,highlandau_PhysRevLett.134.076503,composite_PhysRevLett.133.186602,Xu_second_2025,Thompson2025} and on other relevant platforms, such as rhombohedral graphene superlattices~\cite{Lu2024,Lu2025,Choi_2025,xie2025tunablefractionalcherninsulators,PhysRevLett.133.206502,PhysRevLett.133.206503,PhysRevLett.133.206504,yu_PhysRevB.109.205122,kwan2023moirefractionalcherninsulators,yu2024moirefractionalcherninsulators,Dong_PhysRevLett.133.206503,Liu_PhysRevB.110.075109,PhysRevB.110.115146,tan_2024_wavefunctionapproachfractionalanomalous,Xie_2024_PhysRevB.109.L241115,Das_2024_PhysRevB.110.155148,li2025multibandexactdiagonalizationiteration}.

Going beyond bilayer TMDs, previous studies on untwisted multilayer TMDs~\cite{Kuc_2011_confinement,Jin_2013_Measurement,Zhang_2014_observation,Neil_2017_Neil,Movva_2018_population} have revealed that the interlayer coupling of $d_{z^2}$ orbitals at $\Gamma$ valley is stronger than that of $d_{x^2-y^2}\pm id_{xy}$ orbitals at $K/K'$ valleys, and that adding layers progressively shifts the remote $\Gamma$ valley bands towards the low-energy $K/K'$ valley bands, as illustrated in \cref{schematic:a} to \cref{schematic:c}. 
Based on this mechanism, twisted multilayer TMDs may serve as promising platform for realizing valley-controlled correlated states, such as valley-charge-transfer insulator and tunable metal-insulator transition, which have been observed in other twisted multilayer TMDs~\cite{Foutty2023,Campbell2024,Wei2025,Ma2025}. 

In this work, we perform first-principles calculations on A-AB and A-ABA \tmt\ and 
show that we can indeed tune the energy difference between low-energy moir\'e flat bands from different valleys by changing the number of stacked layers.
Unlike bilayer twisted TMDs~\cite{wang2024fractional, reddy_fractional_2023, jia2023moir, devakul_magic_2021, zhang2024universal}---where only $K/K'$ valleys lie at low energies and $\Gamma$ valley remains far from Fermi level, both $\Gamma$ and $K/K'$ valleys in our multilayer structures appear at low energies.
We construct continuum models of A–AB and A–ABA \tmt\ using both the faithful nonfitted method~\cite{zhang2024universal} and a fitting approach, and further derive tight-binding models by Wannierizing their low-energy bands. These moiré models incorporate layer-hybridized low-energy orbitals, naturally capturing the distinct interlayer-hybridization patterns at different valleys.
In A–AB \tmt\ at $3.89^\circ$, the isolated top valence bands primarily originate from the $K/K'$ valleys but exhibit a strong redistribution of Berry curvature and quantum metric, whose maxima shift from the $\Gamma_M$ to the $K_M$ point compared to AA bilayer \tmt.
Moreover, in A–ABA \tmt, the isolated top valence bands arise simultaneously from both $\Gamma$ and $K/K'$ valleys, enabling the study of multi-valley physics under electronic correlations.
The layer-controlled quantum geometry and multi-valley flat bands provide a new degree of freedom for manipulating correlated electronic states in moir\'e materials.

\section{Untwisted multilayer MoTe$_2$}
\label{sec:untwisted_multilayer}
\begin{figure}[t]
    \centering 
    \subfloat{\label{schematic:a}}%
  \subfloat{\label{schematic:b}}%
   \subfloat{\label{schematic:c}}%
    \includegraphics[width=0.48\textwidth]{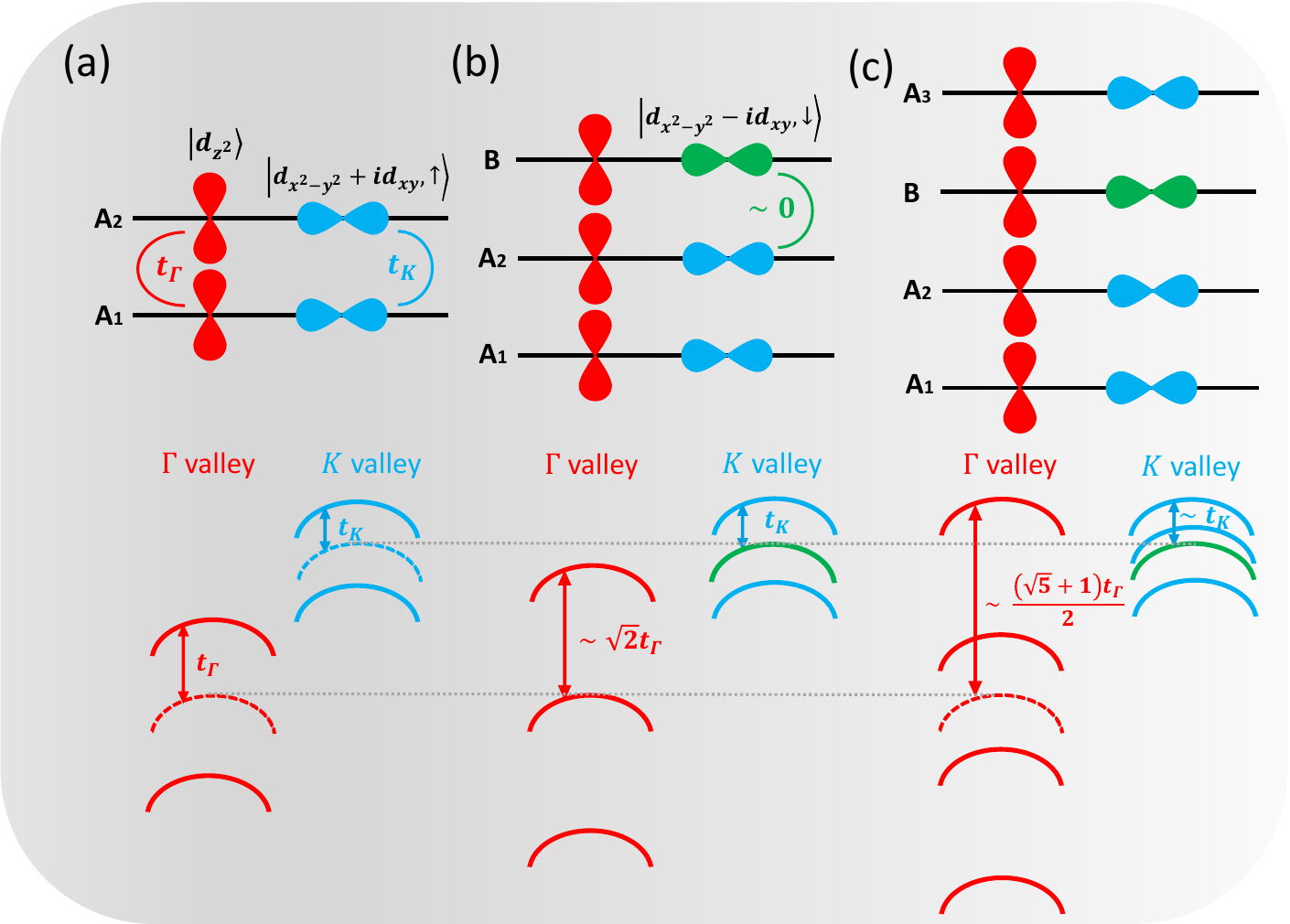}
    \caption{
    Schematic diagrams of the interlayer-coupling mechanism at the $\Gamma$ valley ($d_{z^2}$,colored red)  and the $K$ valley ($d_{x^2-y^2}\pm id_{xy}$ , colored blue/green) in untwisted multilayer  MoTe$_2$: (a) AA bilayer, (b) AAB trilayer, (c) AABA tetralayer. To illustrate the trend, we label the energy shift of the maximum energy (upper solid lines) at $\Gamma$ and $K$ in each multilayer system relative to the monolayer case (dashed) without interlayer coupling. $t_\Gamma$ and $t_K$ denote the interlayer coupling strength of $d_{z^2}$ and $d_{x^2-y^2}+id_{xy}$ between adjacent layers, respectively. It satisfies $t_\Gamma > t_K$.
    \label{fig:schematic}
    }
\end{figure}

Before delving into the moir\'e systems, we first investigate the electronic structures of untwisted multilayer MoTe$_2$ and provide a detailed discussion of the layer-hybridization mechanism that was only briefly mentioned in Refs.~\cite{Jin_2013_Measurement,Zhang_2014_observation,Neil_2017_Neil}. 
For convenience, we label the layers from bottom to top as A$_1$, A$_2$, B and A$_3$ (see \cref{fig:schematic}), where the B layer is rotated by 180$^\circ$ relative to the A layers.

\begin{figure*}[htbp]
    \centering 
    \includegraphics[width=0.95\textwidth]{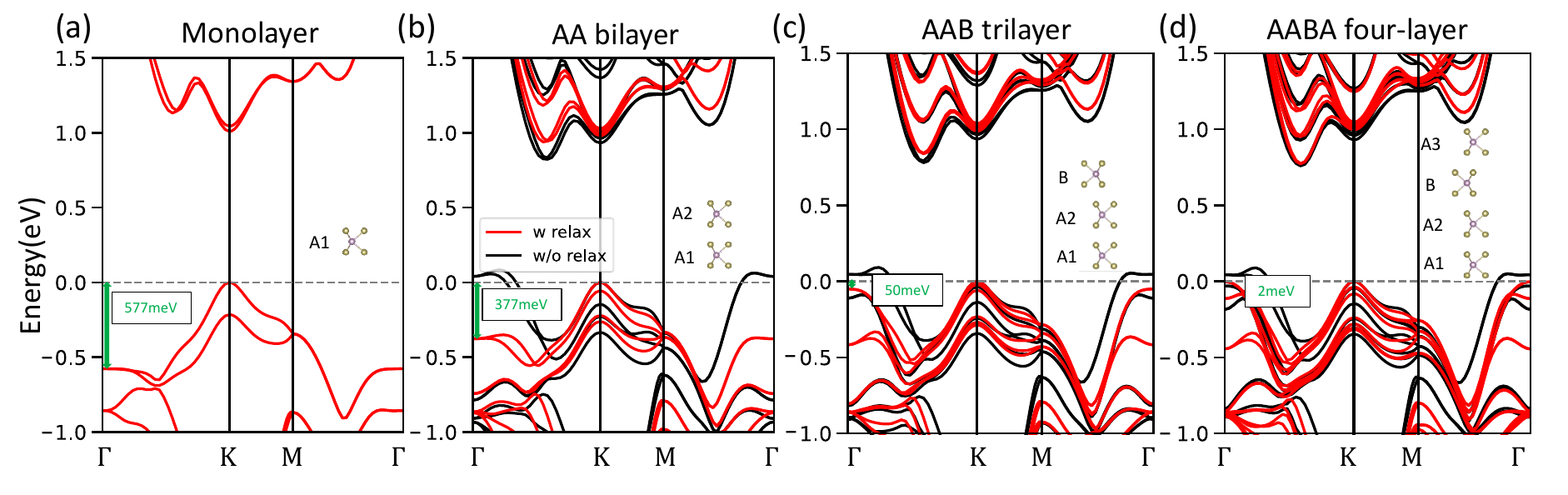}
    \caption{ Atomic and electronic structures of untwisted multilayer MoTe$_2$ with (red lines) and without (black lines) lattice relaxation. The calculations include SOC effects. (a)-(d) show the band structures of monolayer, AA bilayer, AAB trilayer and AABA tetralayer MoTe$_2$. The green numbers denote the energy difference between the maximum energies at $\Gamma$ and $K$ valley in relaxed band structures (red). The insets in (a)-(d) exhibit the atomic structures viewed along the $x$-axis. Lattice relaxation is performed using the Grimme DFT-D2 van der Waals corrections (IVDW=10) implemented in VASP.  \label{untwisted_band}}
\end{figure*}

The band structures of untwisted MoTe$_2$, ranging from monolayer to tetralayer stacks, with and without lattice relaxation, are depicted in \cref{untwisted_band}.  Lattice relaxation 
increases the interlayer distance between adjacent A-layers in multilayer structures from 7.0 \AA \ to 7.6 \AA\ after relaxation (\cref{Fig-relax-process} in \cref{appendix:atomic_structure}). This increase reduces the interlayer coupling and suppresses the maximum energy at the $\Gamma$ valley by 422 meV (AA bilayer), 95 meV (ABA trilayer), and 45 meV (AABA tetralayer) compared to unrelaxed band structures. Nonetheless, due to the different layer hybridization strengths at $\Gamma$ and $\K/\K'$,
the maximum energy at $\Gamma$ in the relaxed band structures is progressively raised to approach that at $K$ with the increasing number of stacked layers, as shown in  \cref{untwisted_band}.  
The close energies of the low-energy bands around $\Gamma$ and $K/K'$ points in the untwisted AAB and AABA structures are clearly different from the large energy difference in the AA-stacked MoTe$_2$.

To understand this layer hybridization mechanism in detail, let us first look at the orbital components (see \cref{appendix:untwist_orbital}).
In multilayer MoTe$_2$, the $d_{z^2}$ and $d_{x^2-y^2}\pm id_{xy}$ orbitals of Mo atoms in different layers dominate the highest valence bands (HVBs). 
Because the $d_{z^2}$ orbital extends in the z direction while the $d_{x^2-y^2}\pm id_{xy}$  orbitals extend in the $x$–$y$ direction, it is natural to expect stronger interlayer coupling at the $\Gamma$ valley than that at the $K/K'$ valleys, as illustrated in \cref{fig:schematic}.

Moreover, for both AAB and AABA MoTe$_2$, the low-energy valence states at the $\Gamma$ valley consist of $d_{z^2}$ orbitals across all the layers. At the $K$ valley, spin-orbit coupling (SOC) induces spin-valley locking in each layer, and the low-energy states are formed only by $d_{x^2-y^2}+ id_{xy}$ orbitals from A-layers.
This occurs because the A and B layers carry opposite spins at the $K$-valley, which suppresses their interlayer coupling, whereas the A layers are spin-aligned and therefore hybridize to form the low-energy states.
The absence of the B-layer contribution due to spin-valley locking at the $K/K'$ valley further weakens the interlayer coupling compared to the $\Gamma$ valley.
As a result, the energy splitting between different layer-hybridized states at the $\Gamma$ valley is much larger than that at the $K/K'$ valley, and this contrast is enhanced as the number of layers increases.
Combined with the fact that the HVB at $\Gamma$ lies much lower in energy than that at $K/K'$ in monolayer MoTe$_2$, it is reasonable that the HVB at $\Gamma$ will energetically approach that at $K/K'$ as the number of layers increases.
The close energies of HVB at $\Gamma$ and $K/K'$ in the untwisted AAB and AABA structures lead to the close energies of the low-energy bands in the $\Gamma$ and $K/K'$ valleys in both A-AB and A-ABA \tmt, as discussed in the following.

\section{A-AB and A-ABA \tmt}
\begin{figure*}[htbp]
    \centering 
    \subfloat{\label{twist_band:a}}
    \subfloat{\label{twist_band:b}}
    \subfloat{\label{twist_band:c}}
    \subfloat{\label{twist_band:d}}
    \subfloat{\label{twist_band:e}}
    \subfloat{\label{twist_band:f}}
    \includegraphics[width=0.9\textwidth]{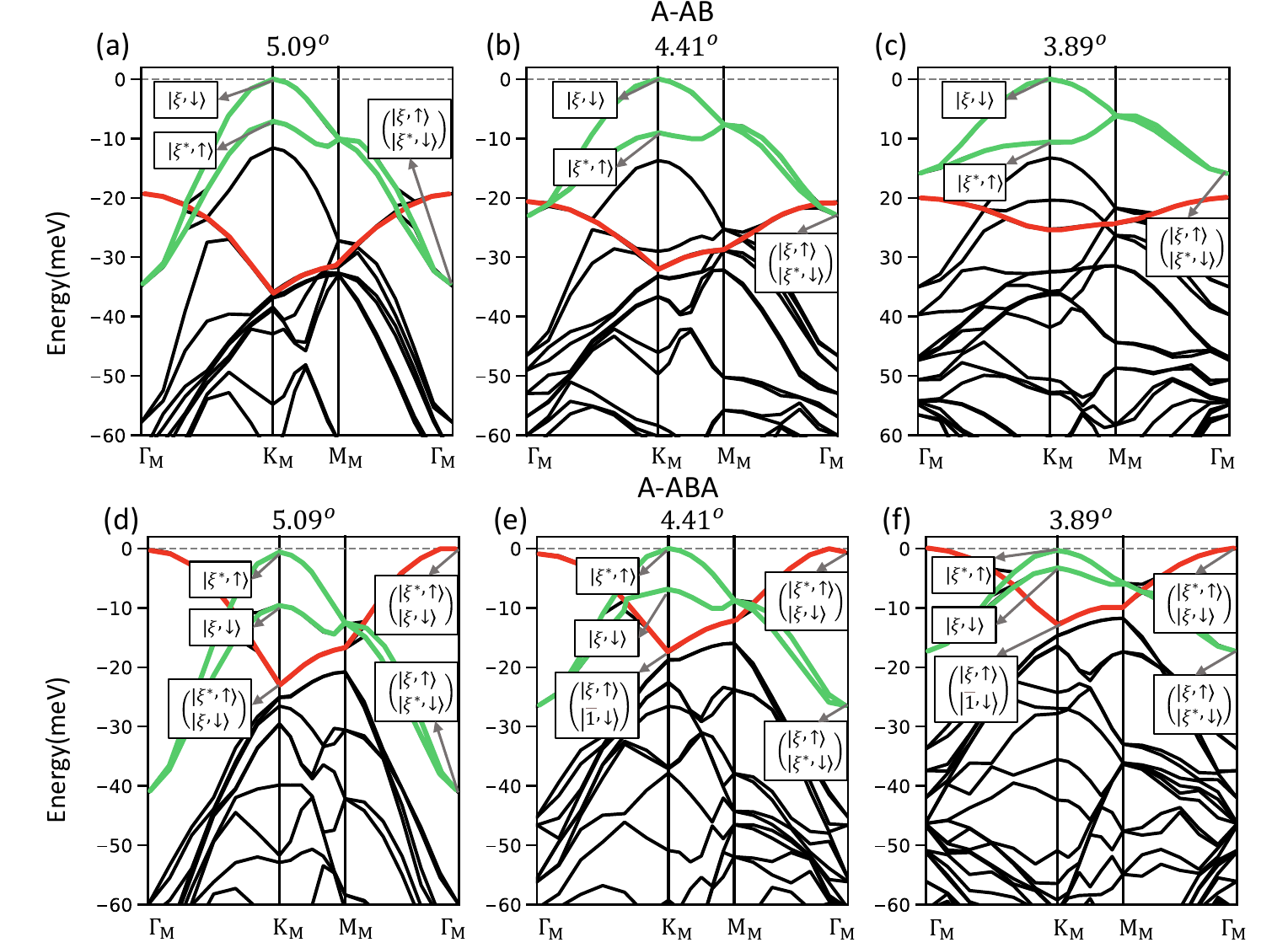}
    \caption{ Moir\'e bands and irreducible representations at high-symmetry points $\Gamma_M$ and $K_M$. (a)-(c) are A-AB $t$MoTe$_2$, and (d)-(e) are A-ABA $t$MoTe$_2$ with twist angles ranging from 5.09$^{\circ}$ to 3.89$^{\circ}$. We use $\xi=e^{i\pi/3}$, $\xi^*=e^{-i\pi/3}$, and $\overline{1}=-1$ to represent the spinful $C_3$ eigenvalues. $C_3$ eigenvalue at $K'_M$ is related to that at $K_M$ by $\mathcal{T}$ symmetry as $\xi_{(K_M',\uparrow)}^{(n)}=\xi_{(K_M,\downarrow)}^{*(n)}$. Green and red lines represent the top two valence bands from the $K/K'$ and $\Gamma$ valley, respectively. The low-energy $\Gamma$-valley bands are nearly spin-degenerate due to Kramers' degeneracy at the $\Gamma$ valley. 
    \label{twist_band} 
    }
\end{figure*}

With the knowledge of untwisted multilayer systems, we now turn to the twisted systems.  The atomic relaxation of A-AB and A-ABA \tmt\ is described in detail in \cref{appendix:atomic_structure}. In the following discussion, we refer to the bottom A-layer in A-AB/A-ABA \tmt\ as the monolayer sector, and the top untwisted AB/ABA portion as 2H-stacked multilayer sector.
Based on the relaxed lattice structures, we perform large-scale DFT calculations on these systems using Vienna Ab initio Simulation Package (VASP)~\cite{KRESSE199615,PhysRevB.47.558,PhysRevB.48.13115,PhysRevB.49.14251,PhysRevB.54.11169}. 
\cref{twist_band} presents the electronic structures of A–AB and A–ABA \tmt\ at different twist angles, including contributions from all valleys ($K$, $K'$ and $\Gamma$).

Let us first focus on the A-AB \tmt. As the twist angle decreases, the top two valence bands from the $K/K'$ valley become isolated at $3.89^\circ$, as shown in \cref{twist_band:a} to \cref{twist_band:c}.
Layer-resolved orbital analysis (See \cref{appendix:moire_band_orbital}) reveals that the top isolated bands originate from the $d_{x^2-y^2}\pm id_{xy}$ orbitals of the two adjacent A-layers \ie, A-A in A-AB, with negligible contribution from the B-layer. 
%Similar to the untwisted case, this is because of spin-valley locking in each layer, which requires the $K$ valley top valence band of the B-layer to have opposite spin to that of the top valence band of the A-layers.
This occurs because spin–valley locking enforces opposite spin orientations at the $K$ valley between the A and B layers, thereby suppressing interlayer coupling between them at the $K/K'$ valleys.
However, the presence of the B layer makes the relaxed structure break the $C_{2y}$ symmetry (and the associated effective inversion symmetry~\cite{jia2023moir}) compared with the AA bilayer \tmt, which leads to energy splittings between the low-energy valence bands within each valley ($K$ and $K'$) along the $\Gamma_M$–$K_M$–$M_M$ path (see the symmetry analysis in \cref{appendix:symmetry_analysis}).
For the $\Gamma$ valley, the valence band maximum is 20 meV below that at $K/K'$ valleys for different twist angles in A-AB \tmt, and a direct gap ($\sim$ 5 meV) between the top $\Gamma$ and $K$ valley bands appears at $3.89^\circ$. 
The low-energy $\Gamma$-valley bands mainly originate from layer-hybridized $d_{z^2}$ orbitals of the A$_2$ and B layers  (See \cref{appendix:moire_band_orbital}).
And notably, $d_{z^2}$ orbital from A$_1$ layer lies at energies far from ($\sim$350 meV) Fermi level and has negligible contribution to the low-energy states at the $\Gamma$ valley.  
%\textbf{BAB: whats A1? Also why are we talking about dz2 orbital? I thought it was d+id that makes the K valley}

More interestingly, in A–ABA \tmt, at twist angles below $5.09^\circ$, the low-energy region contains four valence bands in total -- two nearly spin-degenerate bands from the $\Gamma$ valley and one from each of the $K$ and $K'$ valleys. These bands coexist within a narrow energy window (17 meV for $3.89^\circ$) and are well separated from the other bands, by approximately 20 meV for the $K/K'$ valleys and 3 meV for the $\Gamma$ valley, as shown in \cref{twist_band:d} to \cref{twist_band:f}. This behavior is distinct from that in the bilayer or A–AB \tmt.
As for the layer component, the top $K/K'$ valley bands originate from $d_{x^2-y^2}\pm id_{xy}$ orbital from the A$_1$ layer and the layer-hybridized $d_{x^2-y^2}\pm id_{xy}$ state between the A-layers ($A_2,A_3$) in 2H-stacked multilayer sector, and the top $\Gamma$ valley bands mainly originate from the layer-hybridized $d_{z^2}$ states from A$_2$, B and A$_3$ layers, as shown in \cref{appendix:moire_band_orbital}. The layer component in A-ABA \tmt\ is governed by a similar interlayer hybridization mechanism as in A–AB \tmt, retaining identical spin-valley locking at $K/K'$ valley and comparable strength of interlayer coupling at each valley---the only change being the extra A$_3$ layer in the 2H-stacked multilayer sector.
The moir\'e bands of A-AB and A-ABA \tmt $\;$ at other twist angles are shown in \cref{appendix:moire_band_orbital}.

We also investigate the topological properties and charge density distributions of A-AB and A-ABA \tmt. 
Because the monolayer MoTe$_2$ has the in-plane mirror symmetry $M_z$ and the low-energy orbitals share the same spinless mirror eigenvalue ($+1$), the spin $U(1)$ symmetry is preserved in the monolayer, and thereby approximately maintained in the low-energy moir\'e models.
Consequently, we can classify the moir\'e bands by  the spin Chern number ($C_s$), which can be indicated by the spinful $C_3$ eigenvalues $\xi_{(\boldsymbol{k},s)}^{(n)}$ at high-symmetry points via~\cite{PhysRevB.86.115112}
\bea
\label{eq:c3_eigenvalue}
C_s=\sum_{n} \operatorname{ln}(-\xi_{(\Gamma_M,s)}^{(n)}\xi_{(K_M,s)}^{(n)}\xi_{(K'_M,s)}^{(n)})/(i\dfrac{2\pi}{3})\operatorname{mod}3
\eea
Here, $s=\uparrow/\downarrow$ denotes spin-up/down. 
The spinful $C_3$ eigenvalues of the top bands from each valley at high symmetry points are labeled in \cref{twist_band}. More detailed analysis on the topological properties, including spin Chern numbers computed using the Berry curvature method~\cite{doi:10.1143/JPSJ.74.1674} in continuum models, is presented in the following discussion. 
We further calculate the charge density distribution of isolated top bands at $\Gamma_M$ and $K_M$ of  $4.41^\circ$ and $5.09^\circ$ A-ABA \tmt $\;$ and $3.89^\circ$ A-AB \tmt, which form a hexagonal lattice at the $K/K'$ valleys and a triangular lattice at the $\Gamma$ valley. The detailed analysis of charge density distribution can be found  in \cref{appendix:moire_band_orbital}. 
%\textbf{BAB: do we have the wilson loop for the Gamma point bands please?}

\begin{figure*}[htbp]
    \centering 
    \subfloat{\label{full_band_3.89:a}}
    \subfloat{\label{full_band_3.89:b}}
    \subfloat{\label{full_band_3.89:c}}
    \subfloat{\label{full_band_3.89:d}}
    \subfloat{\label{full_band_3.89:e}}
    \subfloat{\label{full_band_3.89:f}}
    \subfloat{\label{full_band_3.89:g}}
    \subfloat{\label{full_band_3.89:h}}
    \includegraphics[width=1\textwidth]{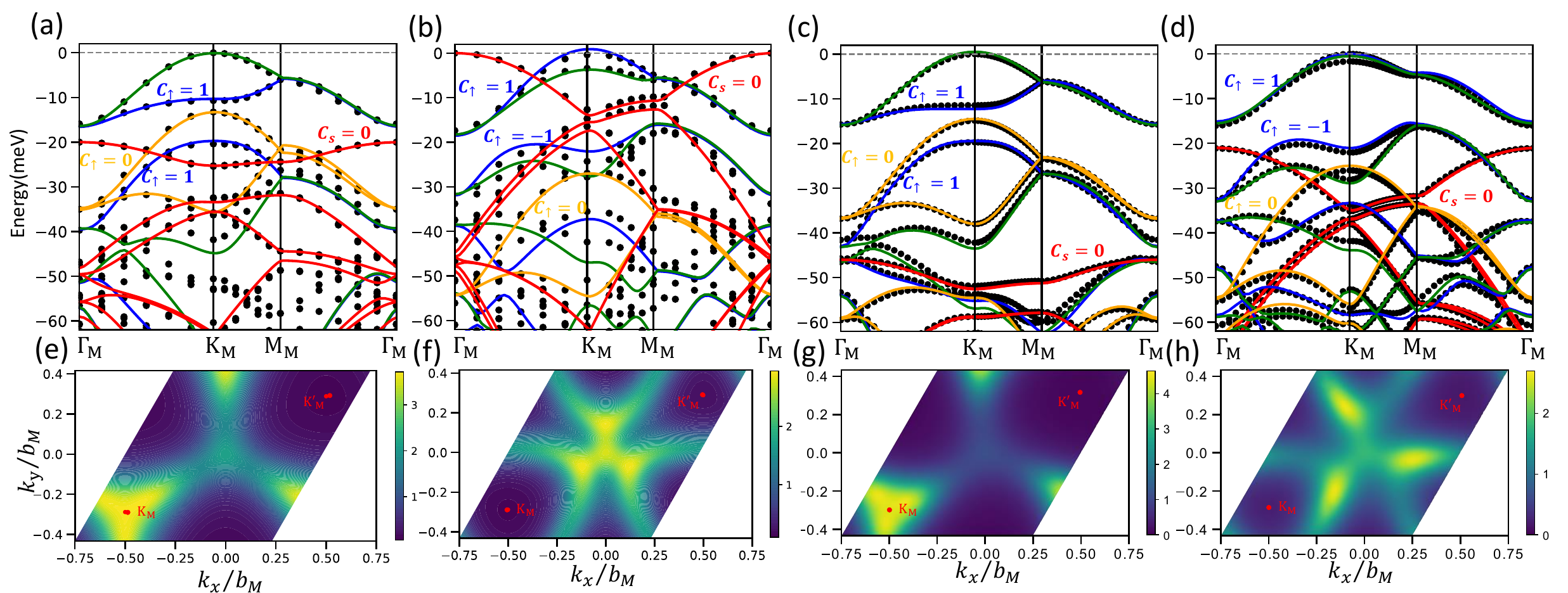}
    \caption{ (a) and (b) show the continuum-model fitting to DFT bands calculated from VASP for A-AB and A-ABA tMoTe$_2$, (c) and (d) present the moir\'e bands calculated from accurate continuum models at different valleys and DFT bands calculated from OpenMX for A-AB and A-ABA \tmt. The twist angle is $3.89^\circ$. Black dot lines are DFT bands, while the other solid lines are the moir\'e bands calculated by continuum models. Red lines are $\Gamma$ valley bands , blue/green lines are $K/K'$ valley bands of A-layers, and orange lines are $K/K'$ valley bands of B-layer. Spin Chern numbers of the several low-energy bands are labeled in (a)-(d). (e)-(h) show  the Berry curvature distribution of the top moir\'e bands from the $K$ valley calculated by corresponding continuum models in (a)-(d).  \label{full_band_3.89} 
    }
\end{figure*}

\section{Continuum models}

Based on DFT calculations, we construct continuum models for both the $K/K'$ and $\Gamma$ valleys in A–AB and A–ABA \tmt\ using two complementary approaches: (i) fitted continuum models parameterized from VASP results, and (ii) accurate continuum models~\cite{zhang2024universal} directly extracted from ab initio wavefunctions obtained by OpenMX. As will be shown below, both approaches yield consistent results, while the accurate continuum models provide a more faithful description of the electronic structure.

We first construct the fitted continuum models. For the $K/K'$ valley, due to spin-valley locking, the A-layers and the B-layer are effectively decoupled. Therefore, continuum models for the A-layers and the B-layer can be constructed separately for the $K/K'$ valleys.
A two-orbital continuum model for the A-layers in both A-AB and A-ABA \tmt\ is constructed as follow
\begin{equation}
\label{eq:continum_2orb}
H_{\eta}=\int d^2r\left ( \begin{matrix}
  \psi^{\dagger}_{\eta,\textbf{r},b}& \psi^{\dagger}_{\eta,\textbf{r},t}
\end{matrix} \right ) \begin{pmatrix}
 h_{\eta,b}(\textbf{r}) & t_{\eta}(\textbf{r})\\
 t^{*}_{\eta}(\textbf{r}) &  h_{\eta,t}(\textbf{r})
\end{pmatrix}\begin{pmatrix}
 \psi_{\eta,\textbf{r},b}\\
 \psi_{\eta,\textbf{r},t}
\end{pmatrix}
\end{equation}
where $\eta=\pm 1 $ denotes the $K/K'$ valley, $\psi_{\eta,\textbf{r},b}$ represents the $d_{x^2-y^2} \pm id_{xy}$ orbital from the A$_1$ layer, $\psi_{\eta,\textbf{r},t}$ represents the $d_{x^2-y^2} \pm id_{xy}$ orbital from the A$_2$ layer in A-AB \tmt\ and the highest-energy layer-hybridized
$d_{x^2-y^2} \pm id_{xy}$ state from A$_2$ and A$_3$ layers in A-ABA 
\tmt, respectively.
The diagonal terms in Eq.\ref{eq:continum_2orb} are given by
\begin{equation}
\label{k_intralyer_potential}
h_{\eta,l}(\textbf{r}) = (-\frac{\hbar^2 \nabla^2}{2 m^*_{l}}+(-)^l \frac{\epsilon_K}{2}) + V_{\eta,l}(\textbf{r})
\end{equation}
The absence of $C_{2y}$ symmetry allows $V_{\eta,b}(\textbf{r})\neq V_{\eta,t}(\textbf{r})$, and moir\'e bands from $K$ and $K'$ valley split along $\Gamma_M-K_M-M_M$ path (See \cref{twist_band} and \cref{full_band_3.89}). 
We find that we can fit the the top two valence bands at the $K/K'$ valley by using this model with only the first harmonic terms of $V_{\eta,l}(\textbf{r})$ and $t_{\eta}(\textbf{r})$, as shown in \cref{full_band_3.89:a} for A-AB \tmt\ and \cref{3orbital_band-A-ABA} for A-ABA \tmt. For the B-layer, we also construct a single-orbital continuum model to describe the moir\'e bands from the $K/K'$ valley. Details on the above models can be found in \cref{appendix:continuum_k}.

Beyond the two-orbital model, we also construct a three-orbital continuum model at the $K/K'$ valley for A-ABA \tmt.
Unlike the two-orbital model, which includes only the $d_{x^2-y^2}\pm i d_{xy}$ orbital from the A$_1$ layer and the highest-energy layer-hybridized $d_{x^2-y^2}\pm i d_{xy}$ state formed by the A$_2$ and A$_3$ layers, and thus can describe at most two low-energy bands at each valley, the three-orbital model explicitly includes the $d_{x^2-y^2}\pm i d_{xy}$ orbitals from all three A-layers (A$_1$, A$_2$, and A$_3$), allowing it to capture up to three low-energy bands at each valley. As shown by the blue and green lines in \cref{full_band_3.89:b}, the three-orbital model provides an excellent fit to the DFT bands of A–ABA \tmt\ at $3.89^\circ$, successfully reproducing the three topmost valence bands originating from the A-layers at each valley.
Additionally, this model includes layer-dependent on-site potential energies, enabling the study of the influence of an external displacement field on the electronic structure of the A–ABA \tmt\ system. Details of the three-orbital model are provided in \cref{appendix:continuum_three_orbital}. 

For the continuum model at the $\Gamma$ valley, it is sufficient to consider the highest-energy layer-hybridized $d_{z^2}$ state from A$_2$, B layers in A-AB, and A$_2$, B, A$_3$ layers in A-ABA \tmt\ to describe the low-energy bands.
This single-orbital model works well because the $d_{z^2}$ orbital from the A$_1$ layer lies significantly lower in energy and therefore couples only weakly to the higher-energy layer-hybridized states.
As shown in \cref{moire-orbital} in \cref{appendix:moire_band_orbital}, this layer-hybridized state dominates the low-energy bands at the $\Gamma$ valley, justifying the use of a single-orbital continuum model. 
The single-orbital continuum model at the $\Gamma$ valley takes the general form

\bea
\label{gamma_ham}
H_{\Gamma}=\int d^2r \psi^{\dagger}_{\br,t}
 (\frac{\hbar^2 \nabla^2}{2 m^*_{\Gamma,t}} s_0 + V_{\Gamma,t}(\br) ) \psi_{\br,t}
\eea
Here, $\psi^{\dagger}_{\br,t}=(\psi^{\dagger}_{\br,t,\uparrow},\ \psi^{\dagger}_{\br,t,\downarrow})$ denotes the highest-energy layer-hybridized $d_{z^2}$ state with spin degree of freedom. We find that we can fit the low-energy bands from the $\Gamma$ valley well using this single-orbital continuum model for both A-AB and A-ABA \tmt, as shown by the red lines \cref{full_band_3.89:a} and \cref{full_band_3.89:b}.
The detailed discussion of this model can be found in \cref{appendix:continuum_gamma}.

More detailed information on the fitting results and parameters of the continuum models at different valleys, including the 
 comparison between the results from the two-orbital and three-orbital model at $K/K'$ valleys for A-ABA \tmt, are provided in \cref{appendix:result_parameter}.

Furthermore, we use our universal Moiré-Model-
Building method without fitting~\cite{zhang2024universal} to obtain the accurate continuum models for A-AB and A-ABA \tmt. The electronic structures of these systems used in our method are calculated using the Truncated Atomic Plane Wave (TAPW)~\cite{PhysRevB.107.125112,Chen_2024,shi2024moireopticalphononsdancing,zhang2024universal}  Hamiltonian
combined with the OpenMX~\cite{ozaki_Variationally_2003,ozaki_Numerical_2004,ozaki_Efficient_2005} software package, as discussed in \cref{appendix:tapw}. The procedure for constructing these accurate continuum models and the detailed results are discussed in \cref{appendix:universal_model}. 
The results calculated from OpenMX and accurate continuum models at different valleys exhibit good agreement with those from VASP and fitted continuum models, as shown in \cref{full_band_3.89:c} and \cref{full_band_3.89:d}, except for a deviation ($\sim$20meV) in the relative energy positions between the $\Gamma$ (red) and $K/K'$ (blue/green) valleys arising from the different DFT bands calculated from VASP and OpenMX. 
The accurate continuum models, which are directly extracted from first-principles wavefunctions, can precisely capture the electronic structure and wavefunction characteristics in the DFT results, providing reliable and unbiased description.
Moreover, we can see from \cref{fig:spin_polar_reduce_gamma} in \cref{appendix:universal_model} that good spin-polarization and spin $U(1)$ symmetry of the low-energy $\Gamma$ valley bands are also inherited in the accurate continuum model.

As for the topological properties, we computed the spin Chern numbers of the low-energy moir\'e bands from different valleys using both the Berry curvature method~\cite{doi:10.1143/JPSJ.74.1674} in continuum models and the $C_3$ eigenvalues obtained from the DFT calculations based on Eq.~\ref{eq:c3_eigenvalue}, which give consistent results, as labeled in \cref{twist_band} and \cref{full_band_3.89}.
The spin Chern number of the top band from the $K$ valley is $C_\uparrow = 1$, with the Berry curvature concentrated near $K_M$ in A–AB (\cref{full_band_3.89:e}) rather than $\Gamma_M$ in A–A \cite{jia2023moir} and A–ABA \tmt\ (\cref{full_band_3.89:f}). This redistribution likely arises from distinct lattice relaxation and the resulting modulation of the moiré potential and interlayer couplings in A-AB \tmt, which significantly alters the valley-resolved band topology.
The corresponding Berry curvature distributions calculated from accurate continuum models for the top bands at $K$ valley  are also shown in \cref{full_band_3.89:g} and \cref{full_band_3.89:h}, which show good agreement with those from the fitted models (\cref{full_band_3.89:e} and \cref{full_band_3.89:f}) but break the mirror symmetry $M_y$ due to the higher harmonic terms in the accurate continuum models. 
The quantum metric exhibits a similar spatial distribution, which is discussed in \cref{appendix:curvature and metric}.

Moreover, the spin Chern number of the second top band from $A$-layers at the $K$ valley is $C_\uparrow=1$ for A-AB \tmt\ but $C_\uparrow=-1$ for A-ABA \tmt. 
The top bands from $K/K'$ valleys in B layer and $\Gamma$ valley all have $C_s=0$, despite a band inversion occurring at the $\Gamma$ valley in A–ABA \tmt\ between $5.09^\circ$ and $4.41^\circ$ (see the change of $C_3$ eigenvalues in 
\cref{twist_band:d} and \cref{twist_band:e}). 
We further calculate the Wilson loops \cite{Wilsonloop_PhysRevB.84.075119} of the nearly spin-degenerate top $\Gamma$-valley bands using both the accurate continuum model and the fitted continuum model, as shown in \cref{fig:WilsonLoop} in \cref{appendix:universal_model}. The Wilson loops exhibit a pronounced change before and after the band inversion, despite their $Z_2$ indices remaining zero.
More detailed results on the Berry curvature and the quantum metric distribution of the low-energy bands at both $\Gamma$ and $K/K'$ valleys are provided in \cref{appendix:curvature and metric}.

\begin{figure}[htbp]
    \centering 
    \subfloat{\label{TB_3.89:a}}
    \subfloat{\label{TB_3.89:b}}
    \subfloat{\label{TB_3.89:c}}
    \subfloat{\label{TB_3.89:d}}
    \includegraphics[width=0.5\textwidth]{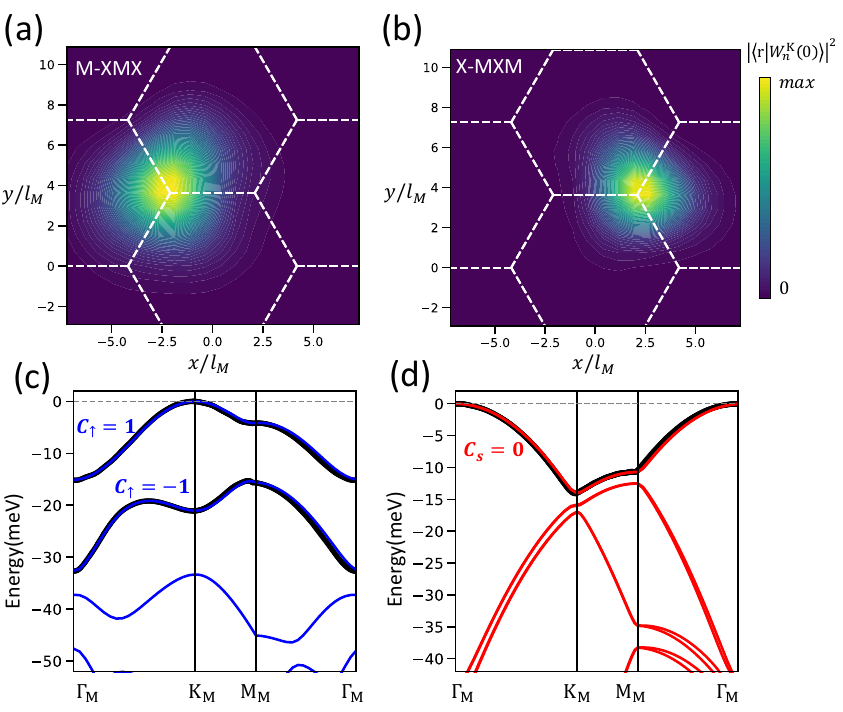}
    \caption{ (a) and (b) show the Wannier function distributions $|\braket{\textbf{r}}{W^K_n(\textbf{0})}|^2$ of the two orbitals corresponding to the top two bands at the $K$ valley in A-ABA \tmt, which are extracted from the accurate continuum model with a twist angle of $3.89^\circ$ and localized at the M-XMX and X-MXM regions, respectively. (c)and (d) show the comparison of energy bands calculated from the tight-binding models (black) and accurate continuum models at $K$ (blue) and $\Gamma$ (red) valley with twist angle $3.89^\circ$ respectively, which exhibit excellent consistency.\label{fig:TB_3.89}}
\end{figure}

\section{Localized Wannier function and tight-binding model}

For the convenience of the future correlated study on the interplay between multiple valleys, we extract the maximally localized Wannier functions (MLWFs) and tight-binding (TB) models for the top two bands at the $K$ valley ($C_\uparrow=\pm1$) and the top nearly spin-degenerate bands at the $\Gamma$ valley ($C_s=0$) in A-ABA \tmt. The TB models at the $K'$ valley can be obtained via the $\mathcal{T}$ symmetry. We use a general method to obtain accurate TB models from continuum models based on WANNIER90 package~\cite{MOSTOFI2008685}, which has also been implemented in magic-angle twisted graphene to construct topological heavy fermion models~\cite{song_1,song_2,Yu_2023_PhysRevB,herzogarbeitman2024}. This Wannierization method based on continuum models is introduced in \cref{appendix:wann_method}. For the $K$ valley, the two MLWFs in real space  $|\braket{\textbf{r}}{W^K_n(\textbf{0})}|^2$  extracted from the accurate continuum model are localized at M-XMX and X-MXM regions (Wyckoff position $1b$ and $1c$) respectively and form the hexagonal lattice, as shown in \cref{TB_3.89:a} and \cref{TB_3.89:b}, which is consistent with the case in AA bilayer \tmt~\cite{devakul_magic_2021}. However, the spread of the two MLWFs are not identical owing to the $C_{2y}$ symmetry breaking in A-ABA \tmt. 
For the $\Gamma$ valley, the Wannier center of the top valence bands shifts from the M–XMX region (Wyckoff position $1b$) at $3.89^\circ$ and $4.41^\circ$ to the A–ABA region (Wyckoff position $1a$) at $5.09^\circ$, as shown in \cref{fig:wann_G} in \cref{appendix:wann_result}. 
This is caused by the band inversion between $4.41^\circ$ and $5.09^\circ$ and complies with the variation trend observed in the charge density distributions (\cref{A_ABA_Gamma_fit}) and the Wilson loops (\cref{fig:WilsonLoop}). 
The definitions of the M-XMX, X-MXM and A-ABA regions can be found in \cref{appendix:atomic_structure}. 
Energy bands calculated from TB models match precisely with those from the accurate continuum models at both $K/K'$ and $\Gamma$ valley for a twist angle of $3.89^\circ$, as shown in \cref{TB_3.89:c} and \cref{TB_3.89:d}.
More detailed results on MLWFs and TB models at different valleys are listed in \cref{appendix:wann_result}.

\section{Summary and Outlook}

In summary, we systematically studied the electronic structures of A-AB and A-ABA \tmt\ by using first principle calculations and constructing continuum models. Due to the different layer hybridization strengths between $d_{z^2}$ and $d_{x^2-y^2}\pm i d_{xy}$ orbitals, top moir\'e bands from both the $\Gamma$ ($C_s=0$) and $K/K'$ valleys ($C_{\uparrow/\downarrow}=\pm1$) appear at low energies in these systems, and become isolated from other bands with twist angles below $5.09^\circ$ in A-ABA \tmt. The relative energy difference between top moir\'e bands from different valleys can be further regulated by the displacement fields, which is applicable in experiments.
By constructing a series of continuum models and tight-binding models for different valleys, we further reveal distinct distributions of Berry curvature and quantum geometry across various stacking configurations. Our study provides an powerful pathway for engineering multi-valley physics and quantum geometry in twisted multilayer TMDs.

More intriguingly, as electron-electron interaction is considered, these systems will serve as a tunable playground for simulating the multi-orbital Hubbard model and investigating various valley-controlled correlated phases, such as the valley-charge-transfer insulator~\cite{Foutty2023,Campbell2024,Wei2025}, valley-controlled quantum anomalous Hall effect and valley-selective fractional Chern insulator. The constructed continuum models and tight-binding models provide versatile theoretical tools for further correlated studies. Our work also calls for extended investigations of other multilayer twisted heterostructures to explore rich multi-valley physics. 

Note that so far, the literature has studied $K, \Gamma$ and $M$ valley physics separately, but never in conjunction. It is usually not the case that two different types of valleys appear at the same energy. The fortunate appearance of the $K$ and $\Gamma$ valleys at the same energy in the A-ABA system allows for completely new physics.

\textbf{Note added:} A confidential draft of this work was sent to Prof. Shengwei Jiang (SJTU) on June 3, 2025. 

\section{Acknowledgments}
This work was supported by the Ministry of Science and Technology of China (Grant No. 2023YFA1607400, 2022YFA1403800), the Science Center of the National Natural Science Foundation of China (Grant No. 12188101) and the National Natural Science Foundation of China (Grant No.12274436). H.W. acoknowledge support from the New Cornerstone Science Foundation through the XPLORER PRIZE.
B.A.B.’s work was primarily supported by the Gordon and Betty Moore Foundation’s EPiQS Initiative (Grant No. GBMF11070), the Office of Naval Research (ONR Grant No. N00014-20-1-2303), the Global Collaborative Network Grant at Princeton University, the Simons Investigator Grant No. 404513, the Princeton Global Network, the NSF-MERSEC (Grant No. MERSEC DMR 2011750), the Simons Collaboration on New Frontiers in Superconductivity (Grant No. SFI-MPS-NFS-00006741-01 and SFI-MPS-NFS-00006741-06), the Schmidt Foundation at the Princeton University, and the Princeton Catalysis Initiative. N.R. also acknowledges support from the QuantERA II Programme that has received funding from the European Union’s Horizon 2020 research and innovation programme under Grant Agreement No 101017733 and from the European Research Council (ERC) under the European Union’s Horizon 2020 Research and Innovation Programme (Grant Agreement No. 101020833). 
The Flatiron Institute is a division of the Simons Foundation.
J. Y.'s work at Princeton University is supported by the Gordon and Betty Moore Foundation through Grant No. GBMF8685 towards the Princeton theory program.
J. Y.'s work at University of Florida is supported by startup funds at University of Florida.

\bibliography{refs}
\bibliographystyle{apsrev4-2}

\onecolumngrid

\newpage
\clearpage

\tableofcontents

\appendix

\section{DFT calculations}
\label{appendix:DFT_details}

\subsection{Electronic structures and orbital analysis of untwisted multilayer MoTe$_2$ }
\label{appendix:untwist_orbital}
\begin{figure*}[htbp]
    \centering 
    \includegraphics[width=1\textwidth]{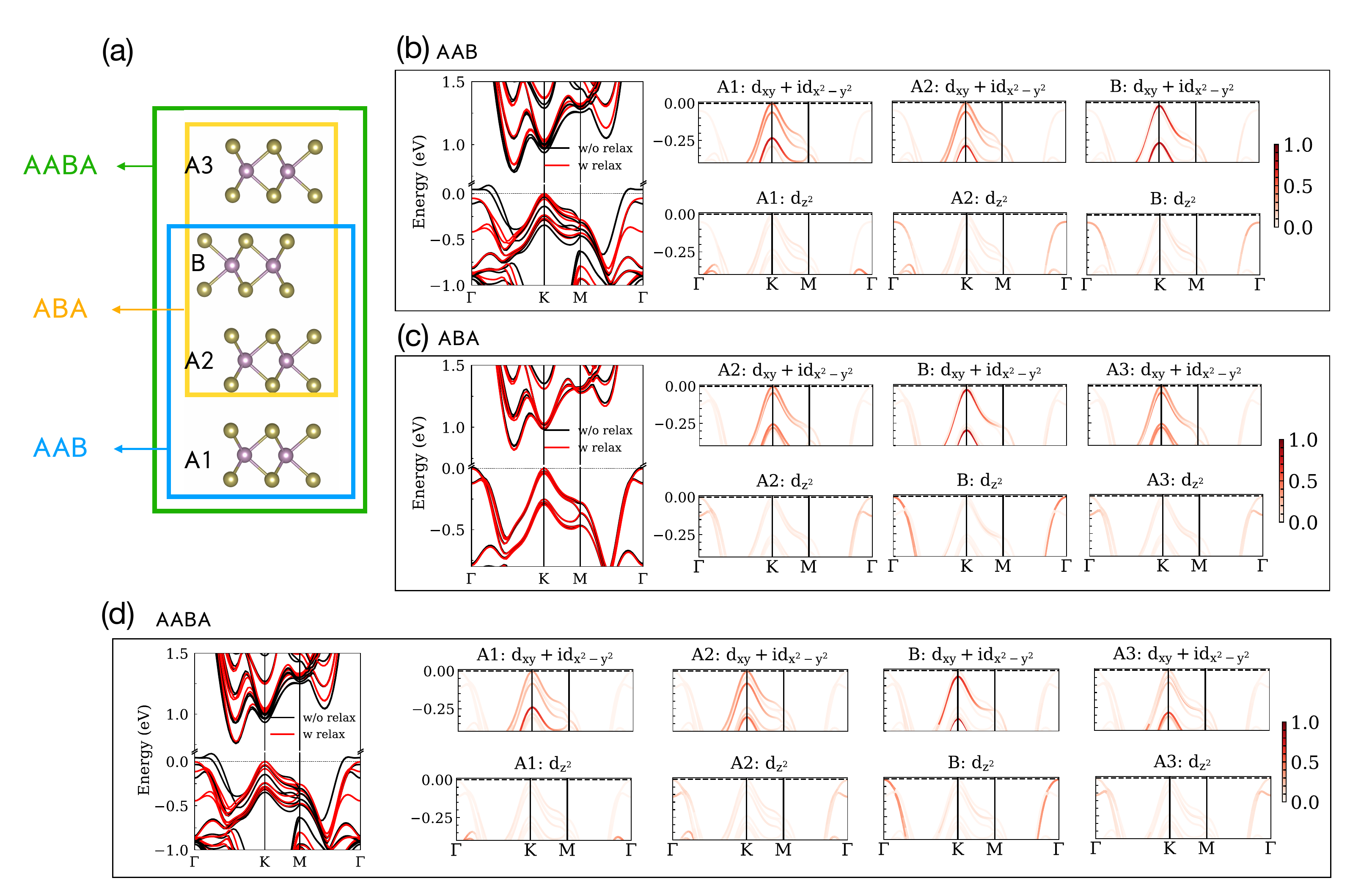}
    \subfloat{\label{untwist_orbital:a}}%
     \subfloat{\label{untwist_orbital:b}}%
     \subfloat{\label{untwist_orbital:c}}%
     \subfloat{\label{untwist_orbital:d}}%
    \caption{ Crystal and band structures of untwist multibilayer \ch{MoTe2}. (a) The crystal structure of AABA-stacking MoTe$_2$, where the top and bottom three layers correspond to the ABA and AAB trilayer structures, respectively. The band structures with/without lattice relaxation and the orbital analysis for each layer in (b) AAB, (c) ABA, and (d) AABA structures. \label{untwist_orbital}}
\end{figure*}
In this section, we study untwisted \ch{MoTe2} multilayers with different stacking configurations and discuss the influence of the interlayer coupling on the low-energy valence bands. We use ``A'' to denote a monolayer in a reference orientation and ``B'' for a monolayer rotated by $C_{2z}$ with respect to the A-monolayer. We focus on two trilayer (AAB and ABA) and one tetralayer (AABA) structures of MoTe$_2$. \cref{untwist_orbital:a} depicts the crystal structure of AABA-stacked MoTe$_2$, where the ABA and AAB trilayers form the bottom and top three layers of the AABA configuration, respectively. All three structures belong to space group $P3m1$ (No.156), generated by $C_{3z}$ and the in-plane mirror $\mathcal M_{100}$. In the rigid structure, the interlayer distance, defined as the vertical separation between Mo atomic planes, is set to \SI{7.0}{\angstrom}. After DFT relaxation, the A–B spacing remains \SI{7.0}{\angstrom}, while the A–A spacing increases to \SI{7.6}{\angstrom}. For the sake of clarity in subsequent discussions, we label the layers in the AABA structure as A$_1$, A$_2$, B, and A$_3$ from bottom to top as shown in \cref{untwist_orbital:a}. 

The left panels of \cref{untwist_orbital:b,untwist_orbital:c,untwist_orbital:d} are the band structures of different multilayer configurations. Because each structure preserves $\mathcal M_{100}$, the combined symmetry $\mathcal M\mathcal T$ leaves the $\Gamma\!-\!M$ line invariant and enforces twofold degeneracy along $\Gamma\!-\!M$. Compared to the monolayer band structure of \ch{MoTe2}~\cite{jia2023moir}, interlayer coupling in multilayer MoTe$_2$ significantly reshapes the valence bands in both the $\Gamma$ and $K$ valleys. In the ABA trilayer, relaxation has little effect, but the low-energy bands in the AAB trilayer and AABA tetralayer are strongly modified. This contrast arises from reducing A–A interlayer coupling strength due to the increased A–A spacing after relaxation.

To elucidate the role of interlayer coupling, we analyze the orbital composition of the untwisted band structures. The top and bottom panels of \cref{untwist_orbital:b,untwist_orbital:c,untwist_orbital:d}  show Mo-$d_{x^2-y^2}\pm id_{xy}$ and Mo-$d_{z^2}$ orbital components, which dominate the $\Gamma$ and $K$-valley, respectively. At $\Gamma$, $d_{z^2}$ orbitals from all layers hybridize. The strongest interlayer coupling is between adjacent A-B layers due to the smaller interlayer distance. Consequently, the topmost state at $\Gamma$ is mainly from both the B and A layers in AAB [bottom panel of \cref{untwist_orbital:b}], and predominantly from the B layer in ABA and AABA structures [bottom panels of \cref{untwist_orbital:c,untwist_orbital:d}]. Because the B layer is rotated by \SI{180}{\degree} relative to the A layer, the B-layer state at $K$ corresponds to the A-layer state at $-K$. With spin-valley locking, the B-layer state has the opposite spin to the A-layer state at $K$. Interlayer coupling between A and B at $K$ is therefore forbidden. In AAB (ABA), the B-layer decouples from the A-layers at $K$ so that its state sits between the antibonding and bonding states formed by A$_1$–A$_2$ (A$_2$–A$_3$), as seen in the top panels of \cref{untwist_orbital:b,untwist_orbital:c}. In AABA, A$_3$ is sufficiently distant from the other A-layers, so the A$_3$-state behaves analogously to the B layer and lies between the A$_1$–A$_2$ antibonding and bonding states [top panel of \cref{untwist_orbital:d}].

\subsection{Crystal structure and relaxation of A-AB and A-ABA $t$MoTe$_2$  }
\label{appendix:atomic_structure}

\begin{figure*}[htbp]%
    \centering %
    \includegraphics[width=0.95\textwidth]{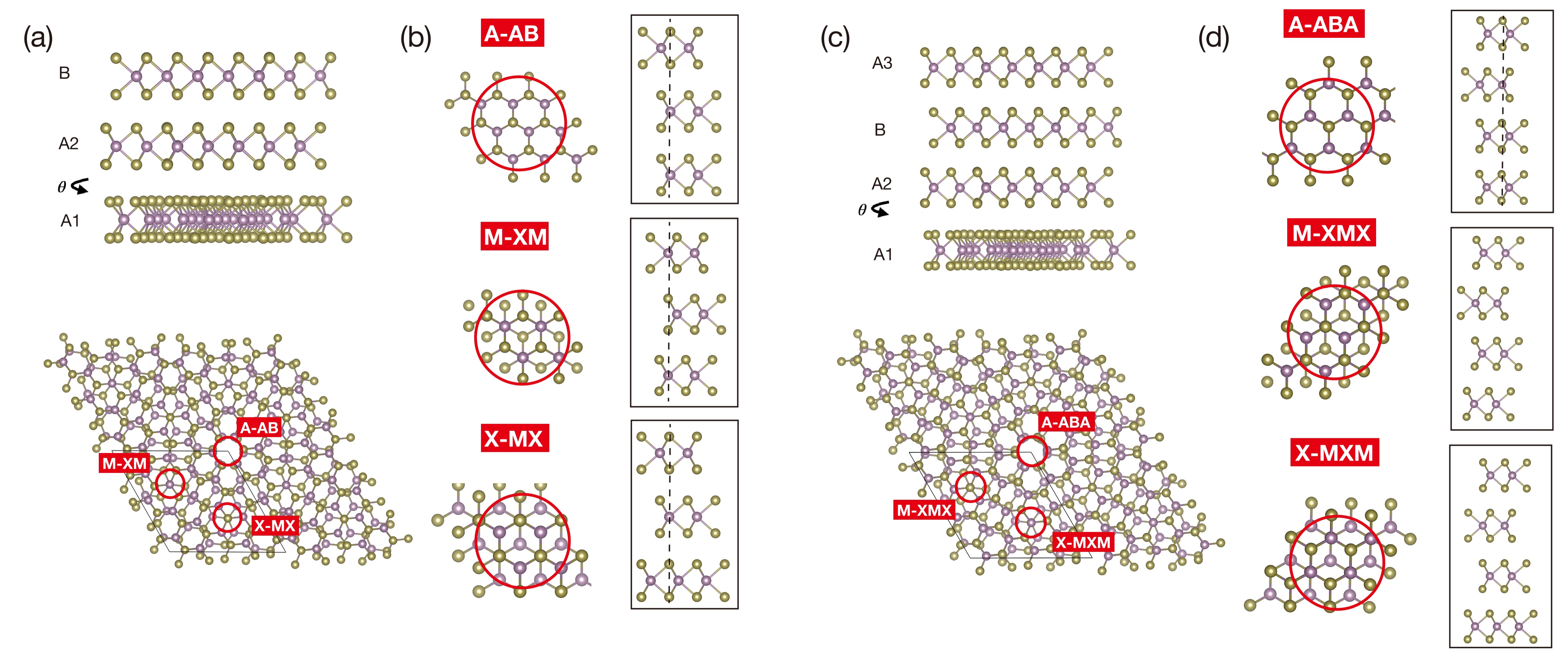}
    \subfloat{\label{fig1:a}}%
     \subfloat{\label{fig1:b}}%
     \subfloat{\label{fig1:c}}%
     \subfloat{\label{fig1:d}}%
    \caption{ Crystal structure of trilayer and bilayer MoTe$_2$ twisted on monolayer MoTe$_2$. (a) 13.2$^\circ$ $t$MoTe$_2$ with A-AB configuration. (b) Top and side views of the atomic stacking arrangement in the A-AB, M-XM, and X-MX region of the moir\'e cell. Molybdenum and chalcogen atoms are labeled as M and X, respectively. (c) 13.2$^\circ$ $t$MoTe$_2$ with A-ABA configuration. (d) Top and side views of atomic stacking arrangement in A-ABA, M-XMX, and X-MXM regions of the moir\'e cell. Both twisted structures have $C_{3z}$ symmetry perpendicular to the plane. X-MXM and X-MX denote the region where the chalcogen atom in the bottom layer is directly below the molybdenum atom in the second-lowest layer, and similarly for M-XMX and M-XM regions.   \label{fig1}}
\end{figure*}

We consider twisted mono–multilayer structures and focus on two moir\'e systems: a monolayer twisted relative to an AB bilayer and a monolayer twisted relative to an ABA trilayer, as shown in \cref{fig1:a,fig1:c}. In the absence of twisting, the monolayer is AA stacked with the adjacent layer of the bilayer/trilayer. In the following discussion, we refer to the twisted mono-bilayer and mono-trilayer structures as A-AB and A-ABA $t$MoTe$_2$, respectively. Both twisted crystal structures belong to SG $P3$ (No. 143). Similar to twisted bilayer MoTe$_2$, A-AB(A-ABA) $t$MoTe$_2$ features three $C_{3z}$-symmetric local regions, \ie, A-AB (A-ABA), M-XM (M-XMX), and X-MX (X-MXM) regions, which are denoted with red circles in \cref{fig1:b,fig1:d}.

\begin{figure*}[htbp]
    \centering 
    \includegraphics[width=1.00\textwidth]{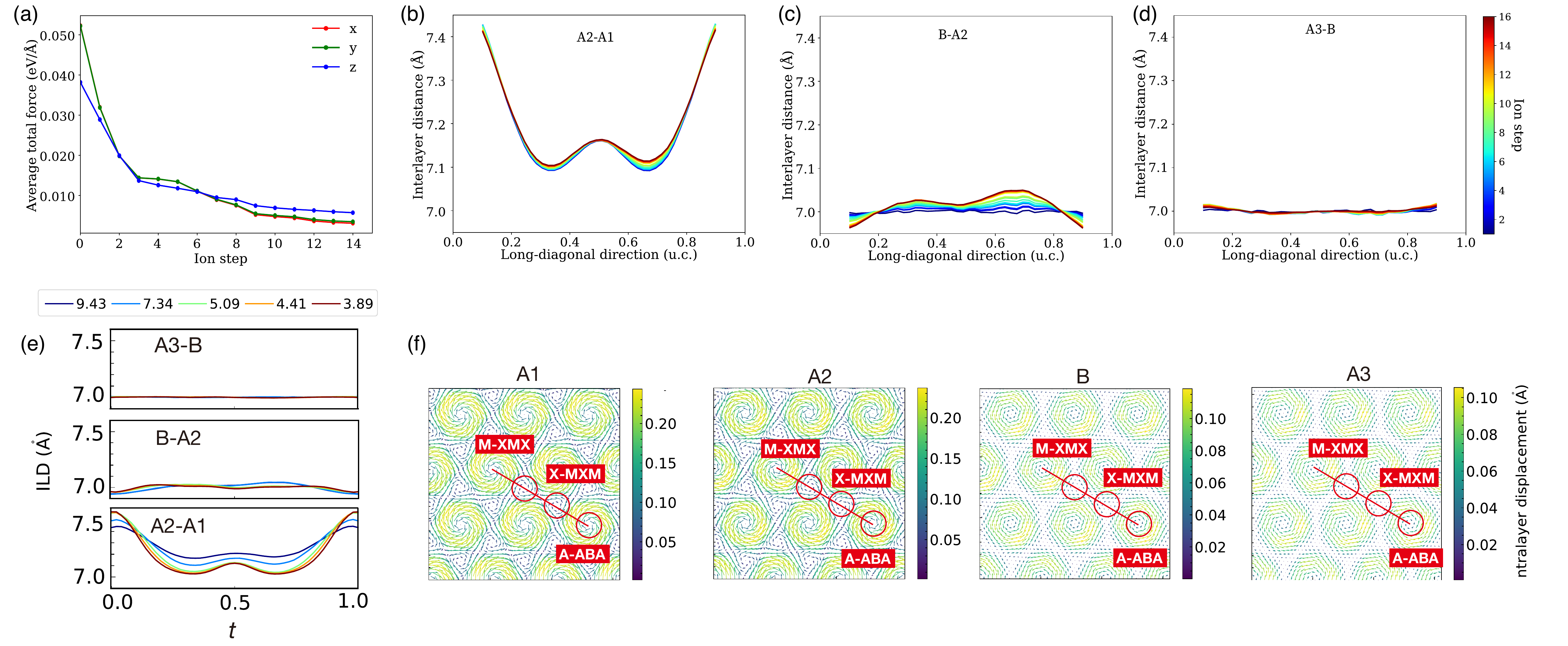}
    \subfloat{\label{Fig-relax-process:a}}%
    \subfloat{\label{Fig-relax-process:b}}%
    \subfloat{\label{Fig-relax-process:c}}%
    \subfloat{\label{Fig-relax-process:d}}%
    \subfloat{\label{Fig-relax-process:e}}%
    \subfloat{\label{Fig-relax-process:f}}%
    \caption{ Properties of relaxed A–ABA \tmt\ structures. (a) Average force per atom in different directions as a function of ionic step during relaxation. Interlayer distances along the long diagonal of the moir\'e unit cell defined in \cref{fig1} between (b) A$_2$–A$_1$, (c) B–A$_2$, and (d) A$_3$–B. Colors indicate ionic steps, with the colorbar at right. (e) Interlayer distances versus twist angle from 9.43$^\circ$ to 3.89$^\circ$. (f) Intralayer displacement fields of all layers at 3.89$^\circ$. \label{Fig-relax-process}}
\end{figure*}

To accelerate the relaxation calculation, we initialized the A–AB and A–ABA \tmt\ geometries from a fully relaxed AA bilayer by adding extra B layer and (B, A$_3$) layers, respectively. The AA bilayer was first relaxed with a machine-learned force field (MLFF) and then refined by full DFT relaxation~\cite{jia2023moir}. The initial A–AB and A–ABA \tmt\ structures were further optimized with DFT until the forces on each atom were below 0.01 eV/\AA. As shown in \cref{Fig-relax-process:a}, the average force per atom fell below this threshold within seven ionic steps for A–ABA \tmt\ at \SI{7.34}{\degree}, indicating that the starting geometries were close to equilibrium. 
We further track the relaxed interlayer distance along the long diagonal of the moir\'e cell as a function of ionic step in \cref{Fig-relax-process:b,Fig-relax-process:c,Fig-relax-process:d}. The interlayer distances change little during relaxation, corroborating the near-equilibrium nature of the initial structures. The A$_2$–A$_1$ interlayer distance pattern (\cref{Fig-relax-process:b}) closely matches that of twisted AA–\ch{MoTe2}, while the B–A$_2$ interlayer distance pattern resembles that of twisted AB–\ch{MoTe2} but with smaller variance~\cite{jia2023moir}. From A$_2$–A$_1$ to A$_3$–B, the interlayer distance variance decreases monotonically, consistent with a moir\'e potential that weakens with distance from the twisted bilayer. 
We also examine the twist angle dependence of the interlayer distance and intralayer displacements after relaxation of A–ABA \tmt. As shown in \cref{Fig-relax-process:e}, the variance for the interlayer distance of A$_1$-A$_2$ increases as the twist angle decreases, with the maximum and minimum approaching the interlayer distance of the corresponding untwisted multilayer. \cref{Fig-relax-process:f} shows the intralayer displacement of every layer with a twist angle of 3.89$^{\circ}$. The vortex pattern in A$_1$ rotates opposite to that in the other three layers, and the displacement magnitudes in B and A$_3$ are much smaller than in A$_1$ and A$_2$.

\subsection{Moir\'e band structures of A-AB and A-ABA \tmt}
\label{appendix:moire_band_orbital}

\begin{figure*}[htbp]
    \centering 
    \includegraphics[width=0.95\textwidth]{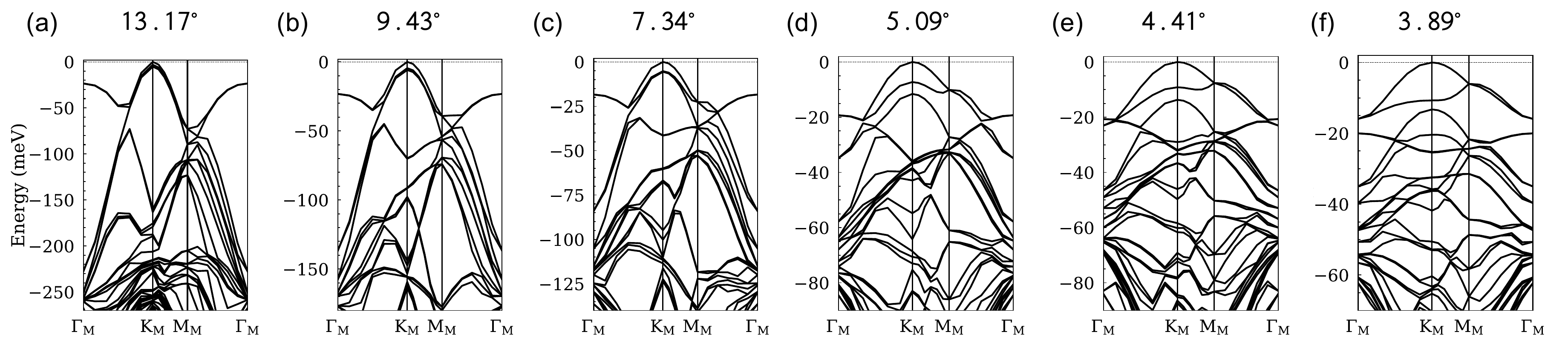}
    \caption{Band structures of A-AB $t$MoTe$_2$ with the twist angles ranging from 13.17$^{\circ}$ to 3.89$^{\circ}$. The top two valence bands in A-AB $t$MoTe$_2$ become narrow with decreasing twist angle and are separated from the other bands when the twist angle is smaller than 3.89$^{\circ}$. }\label{fig4}
\end{figure*}

\begin{figure*}[htbp]
    \centering 
    \includegraphics[width=0.95\textwidth]{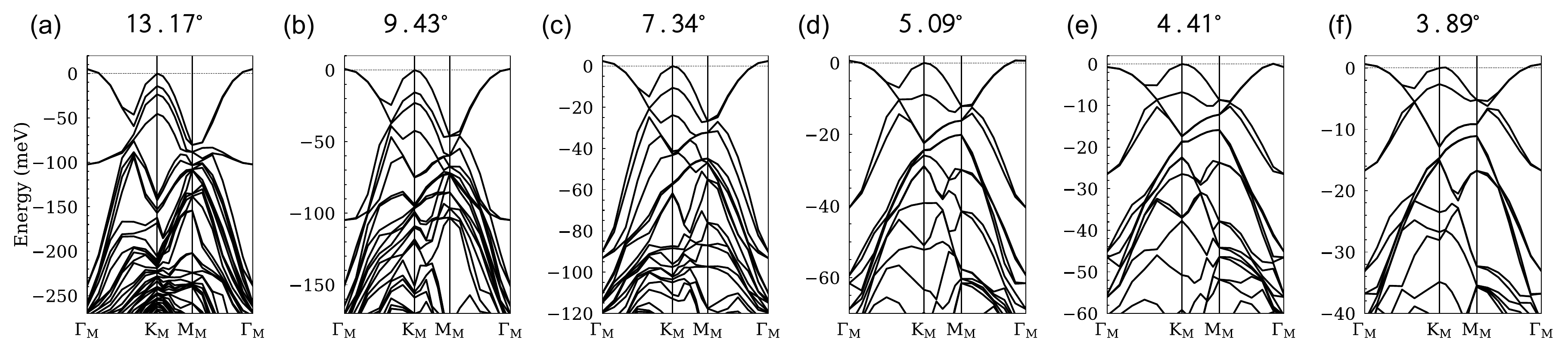}
    \caption{Band structures of A-ABA $t$MoTe$_2$ with the twist angle ranging from 13.17$^{\circ}$ to 3.89$^{\circ}$. The top four valence bands in A-ABA $t$MoTe$_2$ become narrow in energy with decreasing twist angle and are separated from the other bands when the twist angle is smaller than 5.09$^{\circ}$. } \label{fig3}
\end{figure*}

In this section, we discuss the valence band structures of relaxed A-AB and A-ABA $t$MoTe$_2$ with the twist angle ranging from 13.17$^{\circ}$ to 3.89$^{\circ}$. As shown in \cref{fig4} and \cref{fig3}, the band structure of A-AB and A-ABA $t$MoTe$_2$ differs significantly from that of twisted bilayer MoTe$_2$ (see Ref.~\cite{jia2023moir}). In twisted bilayer \ch{MoTe2} the bands are nearly doubly degenerate along $\Gamma_M\!-\!K_M\!-\!M_M$, whereas in A–AB and A-ABA \tmt\ they split substantially due to broken $C_{2y}$ symmetry (See \cref{appendix:symmetry_analysis}). Moreover, in A-AB $t$MoTe$_2$ shown in \cref{fig4}, the top two valence bands become narrower as the twist angle decreases and form an isolated pair at 3.89$^\circ$. In A-ABA $t$MoTe$_2$, the top four valence bands from both the $\Gamma$ and $ K/K'$ valley coexist in the same low-energy region and become separated from other bands when the twist angle is smaller than \SI{5.09}{\degree}.

\begin{figure}[H]
    \centering 
    \subfloat{\label{moire-orbital:a}}
    \subfloat{\label{moire-orbital:b}}
    \includegraphics[width=0.85\textwidth]{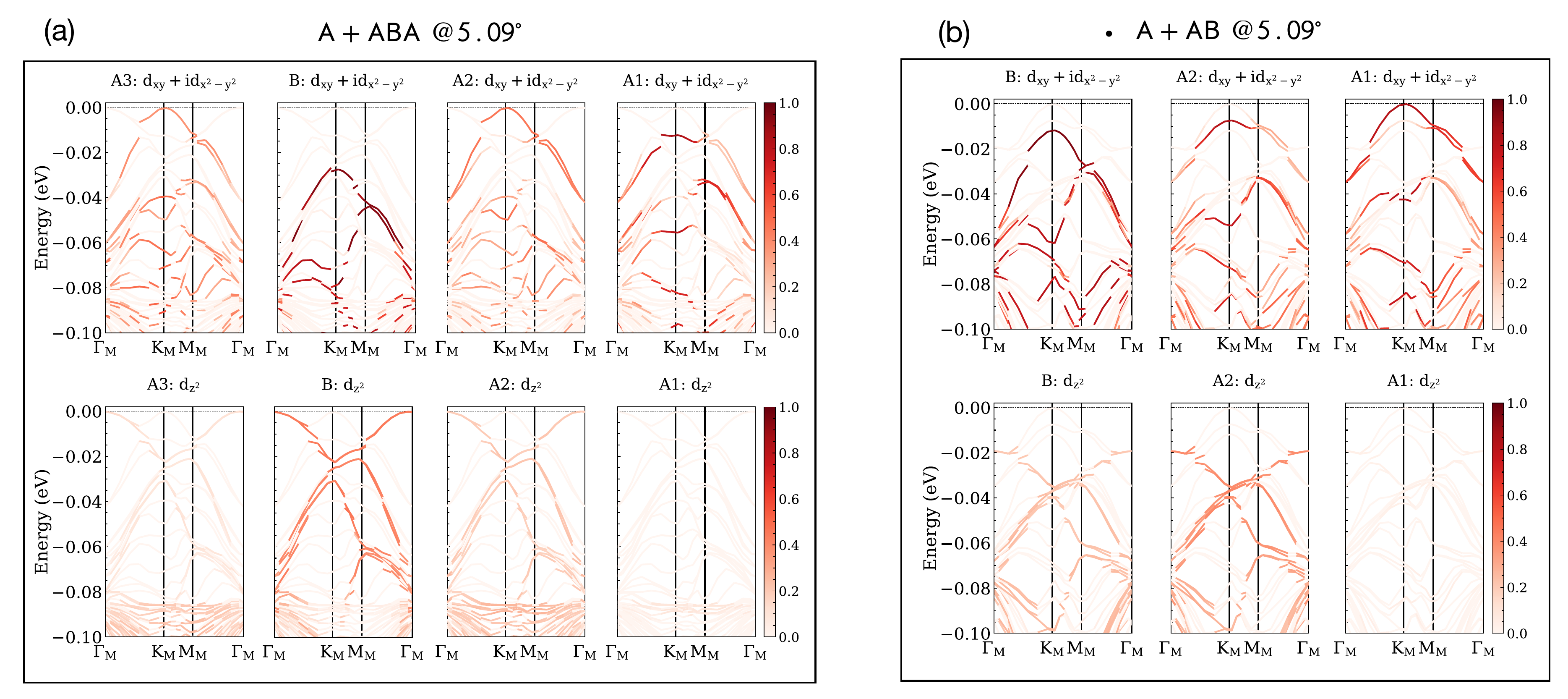}
    \caption{ Layer-resolved orbital analysis for the moir\'e band structure in (a) A-ABA and (b) A-AB-$t$MoTe$_2$ with a twist angle of 5.09$^{\circ}$. In both twisted structures, A$_1$ layer does not contribute to the $\Gamma$ valley bands. The $\Gamma$ valley bands and $K$ valley bands contributed by B layer share a large resemblance. \label{moire-orbital}}
\end{figure}

To further elucidate the isolated moi\'e bands, we perform the layer-resolved orbital analysis for the band structures of A-AB and A-ABA \tmt$\;$ at \SI{5.09}{\degree}, as shown in {\cref{moire-orbital}}. For brevity, we refer to the moir\'e energy bands dominated by $d_{z^2}$ and $d_{x^2-y^2}\pm id_{xy}$ characteristics as the $\Gamma$ valley bands and the $K$ valley bands, respectively. Different from the twisted bilayer \ch{MoTe2} where the $\Gamma$-valley bands lies about \SI{100}{meV} below $K$-valley~\cite{jia2023moir}, the $\Gamma$-valley bands in A–AB \tmt\ is only about \SI{20}{meV} below $K$-valley bands, and in A–ABA \tmt\ the top valance bands from both the $\Gamma$ and $K/K'$ valleys appear in the same low-energy region. In A–ABA \tmt, two of the four isolated bands with nearly spin degeneravy originate from the $\Gamma$ valley and two from the $K/K'$ valley. The $K/K'$-valley pair is mainly from the three A layers, reflecting weak A–B hybridization due to spin-valley locking. By contrast, the $\Gamma-$valley pair is dominated by the B layer through hybridization with its adjacent A$_2$ and A$_3$ layers. The A–AB case is similar. The top two K-valley bands are dominated by the two A layers because hybridization between A layers is stronger. For the $\Gamma$ valley, the stronger hybridization between the untwisted layers makes the B and A$_2$ layers dominate the low-energy $\Gamma$ bands.

\begin{figure}[H]
    \centering 
    \includegraphics[width=0.7\textwidth]{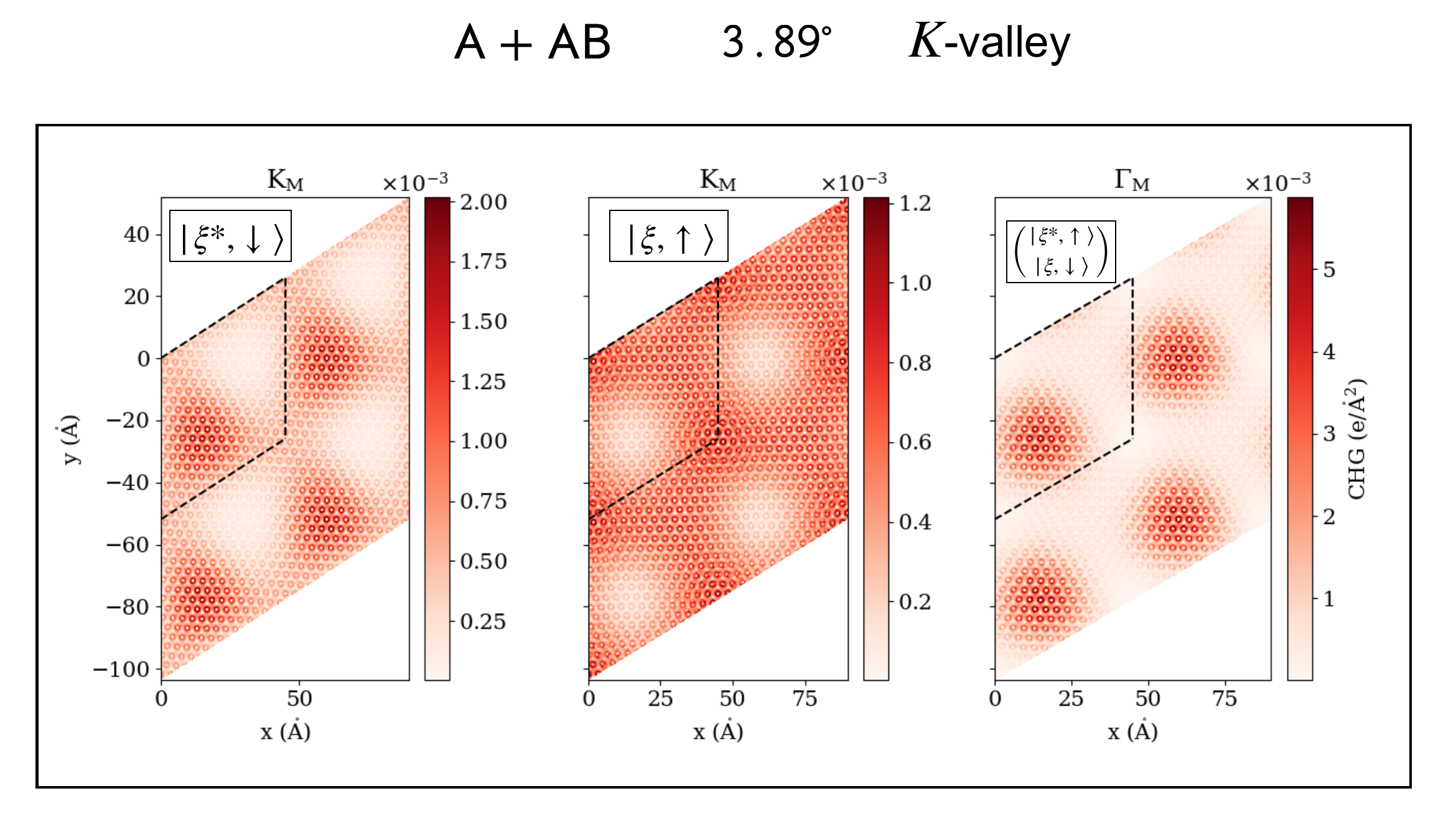}
    \caption{The distribution of charge density of the top $K/K'$-valley band in A-AB-$t$MoTe$_2$ at 3.89$^{\circ}$.  \label{fig8}}
\end{figure}

\begin{figure}[H]
    \centering 
    \includegraphics[width=0.9\textwidth]{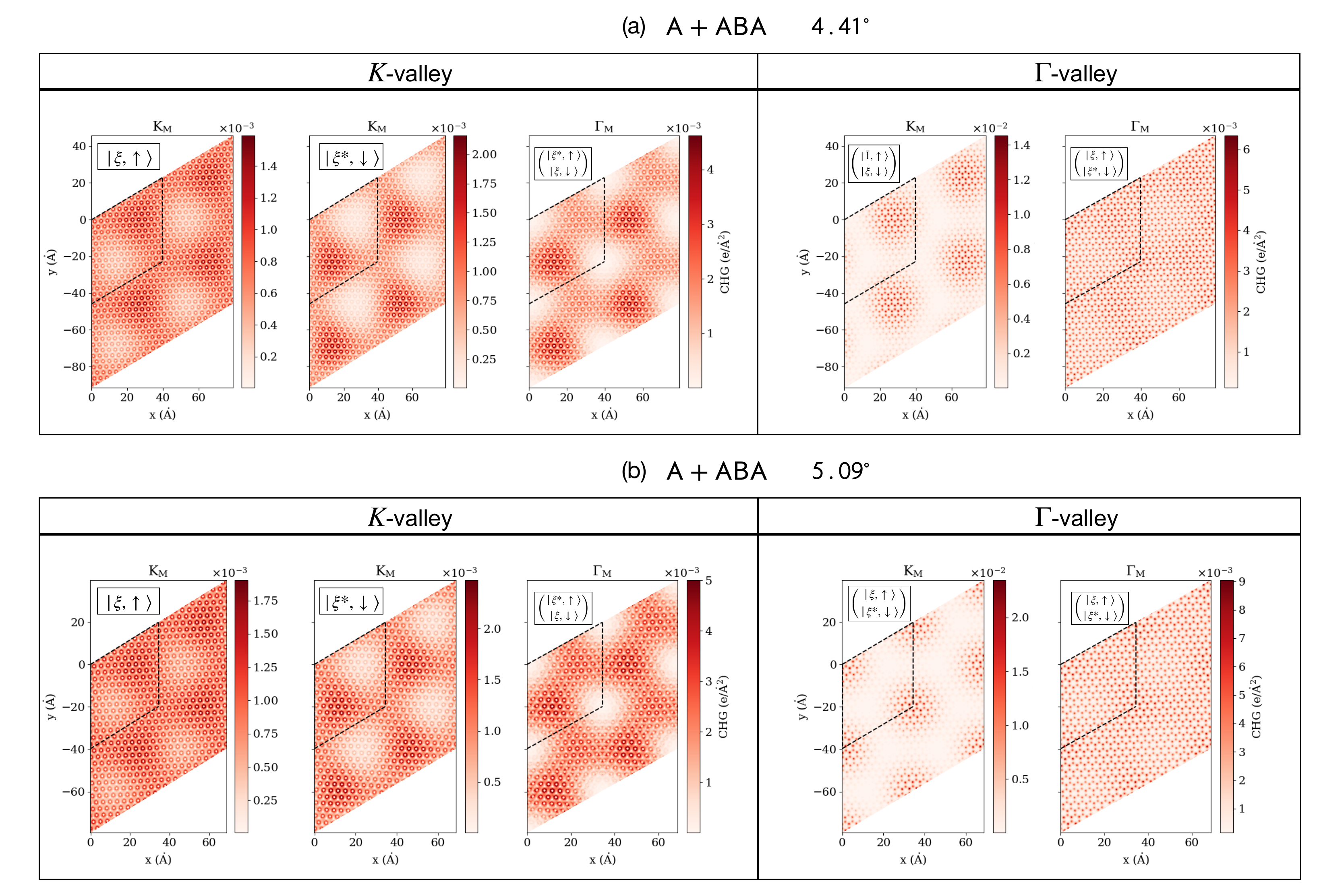}
    \caption{The distribution of charge density of the top $\Gamma$-valley and $K/K'$-valley bands in A-ABA $t$MoTe$_2$ at 4.41$^{\circ}$ (a) and 5.09$^{\circ}$ (b).  \label{fig7}}
\end{figure}

We plot the valley-resolved charge density of the top isolated moir\'e valence band at the $C_3$-invariant momenta ($\Gamma_M$, $K_M$) for A–AB \tmt\ at \SI{3.89}{\degree} (\cref{fig8}) and for A–ABA \tmt\ at \SI{5.09}{\degree} and \SI{4.41}{\degree} (\cref{fig7}).
For the $K$-valley, the patterns at $K_M$ are similar in A–ABA and A–AB. At $\Gamma_M$, A–ABA shows weight in both the M–XMX and X–MXM regions, forming a hexagonal lattice, whereas A–AB concentrates only in the X–MX region. For the $\Gamma$-valley, in A–ABA \tmt\ the charge density at $K_M$ is localized in the M–XMX region at \SI{4.41}{\degree} but in the A–ABA domain at \SI{5.09}{\degree}, form triangular lattices. As shown later, this shift arises from a band inversion of $\Gamma$-valley at $K_M$ between \SI{5.09}{\degree} and 
\SI{4.41}{\degree}.


\section{Continuum model}
\label{appendix:continuum_model}

\subsection{Symmetry analysis in moir\'e Brillouin zone (MBZ)}
\label{appendix:symmetry_analysis}

\begin{figure}[H]
    \centering 
    \includegraphics[width=0.65\textwidth]{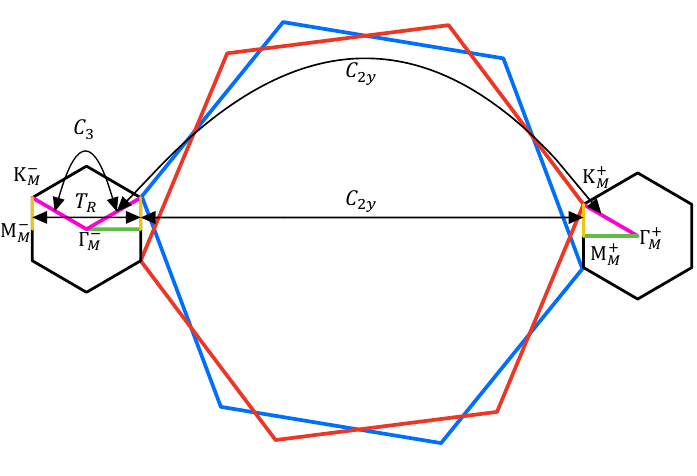}
    \caption{    Schematic illustration of the moir\'e Brillouin zone and the symmetry-protection mechanisms along the high-symmetry-point (HSP) paths. The points $\Gamma_M^\eta$, $K_M^\eta$, and $M_M^\eta$ are the high-symmetry points in the moir\'e Brillouin zone at valley index $\eta$, where $\eta=\pm 1$ corresponds to the $K$ and $K'$ valleys. Pink lines indicate the HSP path $\Gamma_M^+-K_M^+$ protected by $C_{2y}$ and $C_3$ symmetries; orange lines indicate the path $K_M^+-M_M^+$ protected by $C_{2y}$ and translation symmetry $T_R$; and green lines indicate the path $\Gamma_M^+-M_M^+$ protected solely by $C_{2y}$ symmetry.
\label{symmetry_moire}}
\end{figure}

Before constructing the continuum model for A-AB and A-ABA \tmt, we compare the symmetries of the three moir\'e stackings (AA, A-AB, and A-ABA) and discuss the influence on moir\'e band structures. 
The AA-stacked\ \tmt\ has SG $P321$ (No. 150), generated by $C_3$, $C_{2y}$, time-reversal symmetry $\mathcal{T}$, and translation symmetry $T_R$. 
A-AB and A-ABA stacked \ \tmt\ belong to SG $P3$ (No. 143), retaining $C_{3}$, $\mathcal T$, and $T_R$ but lacking $C_{2y}$.
In AA-stacked \tmt, the bands from the $K$ and $K'$ valleys are doubly degenerate along the $\Gamma_M-K_M-M_M$ path.
This valley degeneracy is protected by $C_{2y}C_3$ along $\Gamma_M^+-K_M^+$ (pink lines in \cref{symmetry_moire}) and by $C_{2y}$ along $K_M^+-M_M^+$ (orange lines in \cref{symmetry_moire}). On the contrary,
in the A-AB and A-ABA \tmt\ where $C_{2y}$ is broken, the valley degeneracy along $\Gamma_M-K_M-M_M$ is lifted, as illustrated in \cref{twist_band}. 
The absence of $C_{2y}$ is the key difference between the AA and A-AB/A-ABA \tmt\, which determines the basis for constructing the continuum model.

\subsection{Continuum models at $\Gamma$ and $K$ of untwisted multilayer MoTe$_2$}
\label{appendix:untwisted_kp}

In this subsection, we construct continuum models for the low-energy valence bands at both $\Gamma$ and $K$ valleys of untwisted multilayer MoTe$_2$, capturing the interlayer hybridization effect and approximately extracting several parameters for the moir\'e continuum models. As discussed in the main text, the moir\'e continuum models are obtained by dividing the A-AB/A-ABA \tmt\ structure into two sectors: the monolayer sector and the 2H-stacked multilayer sector, which are rotated relative to each other. Here, we focus on the 2H-stacked multilayer sector, which corresponds to an AB-stacked bilayer in the case of A-AB \tmt\ and an ABA-stacked trilayer in the case of A-ABA \tmt.

First, we construct the continuum model for the $\Gamma$ valley. 
The HVBs of the AB-stacked bilayer MoTe$_2$ originate mainly from the $d_{z^2}$ orbital on the A$_2$ and B layers. 
Therefore, the general form of the continuum model at the $\Gamma$ valley for the AB-stacked bilayer MoTe$_2$, in the basis $\{c^{\dagger}_{A_2\uparrow}(\mathbf{k}), c^{\dagger}_{A_2\downarrow}(\mathbf{k}), c^{\dagger}_{B\uparrow}(\mathbf{k}), c^{\dagger}_{B\downarrow}(\mathbf{k})\}$, can be written as follows:
\bea
\label{eq_AB_gamma}
H^{AB}_{\Gamma}(\textbf{k})=\begin{bmatrix}
 h_{\Gamma,A_2}(\textbf{k}) & t_{\Gamma}(\textbf{k}) \\
 t_{\Gamma}^\dagger(\textbf{k}) & h_{\Gamma,B}(\textbf{k})
\end{bmatrix}
\eea
where $h_{\Gamma,l}(\textbf{k})$ (with $l=A_2,B$) and $t_{\Gamma}(\textbf{k})$ are both $2\times 2$ matrices acting on the spin degree of freedom. 
The symmetry group of the AB-stacked bilayer MoTe$_2$ is generated by the threefold rotation symmetry $C_3$, the twofold rotation symmetry $C_{2x}$, time-reversal symmetry $\mathcal{T}$, and translation symmetry $T_\textbf{R}$. 
The representations of these point-symmetry operations in reciprocal space are given by

\bea 
\label{eq_symm_untwist}
&C_3 c^{\dagger}_{l,\sigma}(\textbf{k}) C_3^{-1} =c^{\dagger}_{l,\sigma'}(C_3\textbf{k}) (e^{-i\frac{\pi}{3}s_z})_{\sigma'\sigma}  \\
&C_{2x} c^{\dagger}_{l,\sigma}(\textbf{k}) C_{2x}^{-1} =c^{\dagger}_{\overline{l},\sigma'}(C_{2x}\textbf{k}) (e^{-i\frac{\pi}{2}s_x})_{\sigma'\sigma} \\
&\mathcal{T} c^{\dagger}_{l,\sigma}(\textbf{k}) \mathcal{T}^{-1} =c^{\dagger}_{l,\sigma'}(-\textbf{k}) (is_y)_{\sigma'\sigma} 
\eea
Under the same basis as Eq.~\ref{eq_AB_gamma}, the representations in Eq.~\ref{eq_symm_untwist} can be written as the following matrices

\bea 
\label{eq_symm_op_AB}
&C_3 = e^{-i\frac{\pi}{3}s_z} \sigma_0 \\
&C_{2x} = -is_x\sigma_x \\
&\mathcal{T} = is_y\sigma_0 \mathcal{K} 
\eea
Here, $s_i$ and $\sigma_i$ ($i=0,x,y,z$) are the Pauli matrices for the spin degree of freedom $\{\uparrow,\downarrow\}$ and the layer degree of freedom $\{A_2,B\}$, respectively. 
The symmetry constraints on $h_{\Gamma,l}(\textbf{k})$ are given by

\bea 
\label{eq_con_AB_h}
&e^{-i\frac{\pi}{3}s_z} h_{\Gamma,l}(\textbf{k}) e^{i\frac{\pi}{3}s_z}= h_{\Gamma,l}(C_3\textbf{k}) \\
&s_x h_{\Gamma,l}(\textbf{k}) s_x = h_{\Gamma,\overline{l}}(C_{2x}\textbf{k})\\
&s_y h_{\Gamma,l}^*(\textbf{k}) s_y = h_{\Gamma,l}(-\textbf{k})
\eea
Which forbid the nonzero $s_x$, $s_y$, and $s_z$ terms and constrain the quadratic dispersion to be the lowest-order harmonic at $\Gamma$. 
The symmetry constraints on $t_{\Gamma}(\textbf{k})$ are given by

\bea 
\label{eq_con_AB_t}
&e^{-i\frac{\pi}{3}s_z} t_{\Gamma}(\textbf{k}) e^{i\frac{\pi}{3}s_z}= t_{\Gamma}(C_3\textbf{k}) \\
&s_x t_{\Gamma}^{\dagger}(\textbf{k}) s_x = t_{\Gamma}(C_{2x}\textbf{k})\\
&s_y t_{\Gamma}^*(\textbf{k}) s_y = t_{\Gamma}(-\textbf{k})
\eea
Taking all the symmetry constraints into consideration, and retaining only the zeroth harmonic in $t_\Gamma(k)$ and the quadratic term in $h_{\Gamma,l}(k)$, the continuum model of the AB-stacked bilayer MoTe$_2$ at $\Gamma$ can be written as follows
 
\bea
\label{eq_AB_gamma_final}
H^{AB}_{\Gamma}(\textbf{k})=\begin{bmatrix}
 -\frac{\hbar^2 k^2}{2m_{\Gamma,A_2}}s_0 & t^0_{\Gamma}s_0 \\
 t^0_{\Gamma}s_0 & -\frac{\hbar^2 k^2}{2m_{\Gamma,B}}s_0
\end{bmatrix}
\eea
Where $C_{2x}$ symmetry constrain $m_{\Gamma,A_2}=m_{\Gamma,B}\equiv m_{\Gamma}$.

The continuum model at the $\Gamma$ valley of the ABA-stacked trilayer MoTe$_2$ is similar to that of the AB-stacked bilayer MoTe$_2$. 
The symmetry group of this system is generated by the threefold rotation symmetry $C_3$, the twofold rotation symmetry $C_{2y}$, time-reversal symmetry $\mathcal{T}$, and translation symmetry $T_R$. 
Under these symmetry constraints, and retaining the same harmonic terms as in Eq.~\ref{eq_AB_gamma_final}, the continuum model of the ABA-stacked trilayer MoTe$_2$ at $\Gamma$ can be written as follows

\bea
\label{eq_ABA_gamma_final}
H^{ABA}_{\Gamma}(\textbf{k})=\begin{bmatrix}
 -\frac{\hbar^2 k^2}{2m_{\Gamma}}s_0 & t^0_{\Gamma,1}s_0& t^0_{\Gamma,2}s_0 \\
 t^0_{\Gamma,1}s_0 & -\frac{\hbar^2 k^2}{2m_{\Gamma}}s_0 & t^0_{\Gamma,1}s_0\\
 t^0_{\Gamma,2}s_0&t^0_{\Gamma,1}s_0 & -\frac{\hbar^2 k^2}{2m_{\Gamma}}s_0
\end{bmatrix}
\eea

\begin{figure}[H]
    \centering 
    \includegraphics[width=1\textwidth]{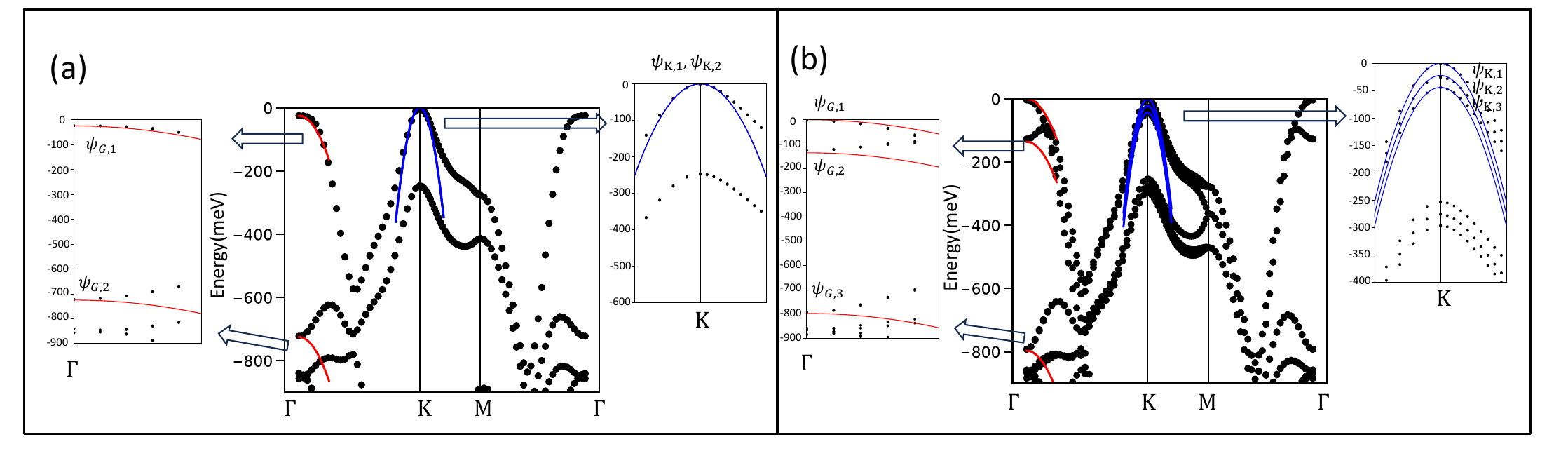}
    \caption{Continuum-model fitting to the HVB calculated by DFT for AB (a) and ABA (b) stacked MoTe$_2$. 
Black dots represent the DFT bands, and the red/blue lines show the energy dispersions obtained from the continuum model at the $K$ and $\Gamma$ valleys. 
\label{untwisted_fit}}
\end{figure}

Using the models in Eq.~\ref{eq_AB_gamma_final} and Eq.~\ref{eq_ABA_gamma_final}, we qualitatively fit the HVB and analyze its wavefunction components for the AB- and ABA-stacked MoTe$_2$ at the $\Gamma$ valley. 
The fitting parameters are listed in Table~\ref{table:untwist_gamma_parameter}. The bands of continuum models fitted to DFT at the $\Gamma$ valley are shown as blue curves in Fig.~\ref{untwisted_fit}. 
The wavefunction components and eigenenergies of the HVB obtained from the continuum model are summarized in Table~\ref{table:untwist_gamma_wave}, which explicitly reveal the hybridization of the HVB. 
These results are consistent with the orbital analysis and energy splitting of the AB and ABA-stacked MoTe$_2$, as shown in \cref{untwist_orbital} in \cref{appendix:untwist_orbital}.

Moreover, the HVB at the $\Gamma$ valley of the A-AB/A-ABA \tmt\ originates mainly from $\psi_{G,1}$ of the AB- or ABA-stacked MoTe$_2$, which has negligible interlayer coupling with the $d_{z^2}$ orbital on the A$_1$ layer. 
Several parameters of the moir\'e continuum model at $\Gamma$ valley can be approximately extracted from the DFT data. 
First, the displacement energy between the top and bottom sectors in Eq.~\ref{eq2} is $\epsilon_\Gamma=350$\,meV for the A-AB \tmt\ and $\epsilon_\Gamma=311$\,meV for the A-ABA \tmt, independent of the twist angle. 
This value corresponds to the eigenenergy of $\psi_{G,1}$ (see Table~\ref{table:untwist_gamma_wave}) and represents the energy difference between the $d_{z^2}$ orbital of the A$_1$ layer and the hybridized $d_{z^2}$ states of the 2H-stacked multilayer sector. 
Second, the effective mass $m_{\Gamma,l}^*$ can also be extracted for the untwisted systems, with $m_{\Gamma,l}^*\!\sim\!2.4\,m_e$ for the A-AB \tmt\ and $m_{\Gamma,l}^*\!\sim\!1.6\,m_e$ for the A-ABA \tmt\ (see Table~\ref{table:untwist_gamma_parameter}).

\begin{table}[H]
\centering
\begin{tabular}{|c|c|c|c|c|}
\hline \multicolumn{2}{|c|}{ $\Gamma$ valley AB MoTe$_2$ } &\multicolumn{3}{c|}{ $\Gamma$ valley ABA MoTe$_2$}\\
\hline $m_\Gamma(m_e)$ & $t_\Gamma^0(meV)$& $m_\Gamma(m_e)$ & $t_{\Gamma,1}^0(meV)$ & $t_{\Gamma,2}^0(meV)$ \\
\hline \textcolor{red}{2.4} & -350 & \textcolor{red}{1.6} & -275 & -175 \\
\hline
\end{tabular}
\caption{The continuum model parameters of untwisted AB (Eq.\ref{eq_AB_gamma_final})/ABA (Eq.\ref{eq_ABA_gamma_final}) MoTe$_2$ at $\Gamma$ valley fitting to the DFT bands.}
\label{table:untwist_gamma_parameter}
\end{table}

\begin{table}[H]
\centering
\begin{tabular}{|c|c|c|c|}
\hline \multicolumn{4}{|c|}{Hybridized states of $\Gamma$ valley AB MoTe$_2$ } \\
\hline
\diagbox{$\textbf{Wavefunction component}$\\
}{$\textbf{layer}$} &A$_2$&B & eigenenergy (meV) \\
\hline
$\psi_{G,1}$ &$0.500$ & $0.500$ & \textcolor{red}{350}  \\
\hline
$\psi_{G,2}$ &$0.500$ & $0.500$ & {-350} \\
\hline
\end{tabular}

\begin{tabular}{|c|c|c|c|c|}
\hline \multicolumn{5}{|c|}{Hybridized states of $\Gamma$ valley ABA MoTe$_2$ } \\
\hline
\diagbox{$\textbf{Wavefunction component}$}{$\textbf{layer}$} &A$_2$&B & A$_3$ & eigenenergy (meV) \\
\hline
$\psi_{G,1}$ &$0.195$ & $0.610$ & $0.195$& \textcolor{red}{311}  \\
\hline
$\psi_{G,2}$ &$0.500$ & $0.000$ & $0.500$& {175}  \\
\hline
$\psi_{G,3}$ &$0.305$ & $0.390$ & $0.305$ & {-486} \\
\hline
\end{tabular}
\caption{Wavefunction components of HVB fitting by continuum model of AB/ABA MoTe$_2$ at $\Gamma$ valley. $\{\psi_{G,1},\psi_{G,2},\psi_{G,3}\}$ are denoted in \cref{untwisted_fit}.}
\label{table:untwist_gamma_wave}
\end{table}

Next, we construct the continuum model at the $K$ valley of untwisted AB- and ABA-stacked MoTe$_2$. 
As explained in the main text (see Sec.\ref{sec:untwisted_multilayer}), spin-valley locking constrains the interlayer coupling to exist only between the A layers at $K/K'$, while the coupling between the A layers and the B layer is negligible. 
Thus, the continuum model of the AB-stacked bilayer MoTe$_2$ at the $K/K'$ valleys can be written as follows
\bea
\label{eq_AB_K_final}
H^{AB}_{\eta}(\textbf{k})=\begin{bmatrix}
 -\frac{\hbar^2 (k-\eta K)^2}{2m_{\eta,A_2}} & 0 \\
0 & -\frac{\hbar^2 (k-\eta K)^2}{2m_{\eta,B}}
\end{bmatrix}
\eea
The continuum model of the ABA-stacked trilayer MoTe$_2$ at the $K/K'$ valleys, considering only the zeroth harmonic of the interlayer coupling between $A_2$ and $A_3$, can be written as follows
\bea
\label{eq_ABA_K_final}
H^{ABA}_{\eta}(\textbf{k})=\begin{bmatrix}
 -\frac{\hbar^2 (k-\eta K)^2}{2m_{\eta,A_2}} & 0 & t  \\
0 & -\frac{\hbar^2 (k-\eta K)^2}{2m_{\eta,B}} & 0 \\
t^* & 0 & -\frac{\hbar^2 (k-\eta K)^2}{2m_{\eta,A_3}}
\end{bmatrix}
\eea
Here, we use $\eta=\pm 1$ to denote $\{\ket{\uparrow,K},\ket{\downarrow,K'}\}$. 
For the AB-stacked MoTe$_2$, the $C_{2x}$ and $\mathcal{T}$ symmetries constrain $m_{\eta,A_2}=m_{\eta,B}\equiv m$ in Eq.~\ref{eq_AB_K_final}. 
For the ABA-stacked MoTe$_2$, the $\mathcal{M}_z$, $C_{2y}$, and $\mathcal{T}$ symmetries impose the constraints $m_{\eta,A_2}=m_{\eta,A_3}\equiv m_A$, $m_{\eta,B}\equiv m_B$, and $t=t^*\in \mathbb{R}$ in Eq.~\ref{eq_ABA_K_final}.

\begin{table}[H]
\centering
\begin{tabular}{|c|c|c|c|}
\hline \multicolumn{1}{|c|}{ $K$ valley AB MoTe$_2$ } &\multicolumn{3}{c|}{ $K$ valley ABA MoTe$_2$}\\
\hline $m(m_e)$ & $m_A(m_e)$ & $m_B(m_e)$ & $t(meV)$ \\
\hline \textcolor{red}{0.6} & 0.6 & \textcolor{red}{0.6} & 21 \\
\hline
\end{tabular}
\caption{Continuum-model parameters of untwisted AB (Eq.~\ref{eq_AB_K_final}) and ABA (Eq.~\ref{eq_ABA_K_final}) MoTe$_2$ at the $K$ valley, fitted to the DFT bands.}
\label{table:untwist_k_parameter}
\end{table}

\begin{table}[H]
\centering
\begin{tabular}{|c|c|c|c|}
\hline \multicolumn{4}{|c|}{ $K$ valley AB MoTe$_2$ } \\
\hline
\diagbox{$\textbf{Wavefunction component}$}{$\textbf{layer}$} &A$_2$&B & eigen energy\\
\hline
$\psi_{K,1}$ &$1.000$ & $0.000$  &\textcolor{red}{0} \\
\hline
$\psi_{K,2}$ &$0.000$ & $1.000$ &0\\
\hline
\end{tabular}

\begin{tabular}{|c|c|c|c|c|}
\hline \multicolumn{5}{|c|}{ $K$ valley ABA MoTe$_2$ } \\
\hline
\diagbox{$\textbf{Wavefunction component}$}{$\textbf{layer}$} &A$_2$&B & A$_3$ &eigen energy\\
\hline
$\psi_{K,1}$ &$0.500$ & $0.000$ & $0.500$&\textcolor{red}{21} \\
\hline
$\psi_{K,2}$ &$0.000$ & $1.000$ & $0.000$&0 \\
\hline
$\psi_{K,3}$ &$0.500$ & $0.000$ & $0.500$ &-21\\
\hline
\end{tabular}
\caption{Wave-function components of the HVB obtained from the continuum model for AB- and ABA-stacked MoTe$_2$ at the $K$ valley. The states $\{\psi_{K,1},\psi_{K,2},\psi_{K,3}\}$ are defined in Fig.~\ref{untwisted_fit}.}
\label{table:untwist_k_wave}
\end{table}

Using the models in \cref{eq_AB_K_final} and \cref{eq_ABA_K_final}, we fit the  HVBs and analyze the wave-function components of the HVB in the AB- and ABA-stacked MoTe$_2$ at the $K$ valley. 
The fitting parameters are given in \cref{table:untwist_k_parameter}, and the continuum models at the $K$ valley, fitted to the DFT bands, are shown as red curves in \cref{untwisted_fit}. 
The wave-function components of the HVB obtained from the continuum model are listed in \cref{table:untwist_k_wave}. 
For the AB-stacked bilayer MoTe$_2$, since the $d_{x^2-y^2}\pm i d_{xy}$ orbitals from the A$_2$ and B layers are independent of each other, they remain degenerate at the $K$ valley. 
For the ABA-stacked trilayer MoTe$_2$, the $d_{x^2-y^2}\pm i d_{xy}$ orbitals from the A$_2$ and A$_3$ layers couple with each other and form two hybridized states, with the energy band of orbital from the B layer sandwiched in between. 
These results are consistent with the DFT calculations presented in \cref{untwist_orbital} and discussed in \cref{appendix:untwist_orbital}.

In the A-AB and A-ABA $t$MoTe$_2$, the HVBs at the $K$ valley are mainly derived from $\psi_{K,1}$ of the A$_2$ (A-AB) or A$_2$,A$_3$ (A-ABA) layers in the multilayer sector, together with the $d_{x^2-y^2}\pm i d_{xy}$ orbital of the A$_1$ layer. 
The on-site energy $\epsilon_K$ and the dispersion mass $m_l$ in \cref{k_intralyer_potential} can also be approximately obtained: $\epsilon_K=0$\,meV and $m_l \sim 0.6m_e$ for the A-AB \tmt, and $\epsilon_K=21$\,meV and $m_l \sim 0.6m_e$ for the A-ABA \tmt\ (see \cref{table:untwist_k_parameter} and \cref{table:untwist_k_wave}).

\subsection{Two-orbital continuum model of A-AB and A-ABA $t$MoTe$_2$ at $K/K'$ valley}
\label{appendix:continuum_k}

In the following three subsections, we discuss the continuum models of A-AB and A-ABA $t$MoTe$_2$ at both $\Gamma$ valley and $K/K'$ valley. 
We define the moir\'e lattice vectors as $\textbf{a}_{M,1}=a_M(\tfrac{\sqrt{3}}{2},\tfrac{1}{2})^T$ and $\textbf{a}_{M,2}=a_M(0,1)^T$, where $a_M=3.52$\,\AA. 
The corresponding moir\'e reciprocal lattice vectors are $\textbf{b}_{M,1}=b_M(1,0)^T$ and $\textbf{b}_{M,2}=b_M(\tfrac{1}{2},\tfrac{\sqrt{3}}{2})^T$. 
Here, the moir\'e lattice constant is given by $a_M=\tfrac{a_0}{2\sin(\tfrac{\theta}{2})}$, and the reciprocal lattice constant is $b_M=\tfrac{4\pi}{\sqrt{3}a_M}$.

We first construct a two-orbital continuum model for A-AB and A-ABA $t$MoTe$_2$ at the $K/K'$ valleys. 
spin-valley and orbital–valley locking require that the low-energy band components from the A (B) layers at the $K$ valley are spin-up (spin-down) and $d_{x^2-y^2}+ i d_{xy}$ ($d_{x^2-y^2}- i d_{xy}$), respectively.
Furthermore, the orbitals from the A layers at the $K$ valley couple negligibly with those from the B layer due to spin-valley locking and spin $U(1)$ symmetry. 
Therefore, the continuum model can be constructed separately for the A layers and the B layer. 
The $K/K'$-valley continuum model, based on the $d_{x^2-y^2} \pm i d_{xy}$ orbitals of the monolayer sector and the highest-energy layer-hybridized $d_{x^2-y^2} \pm i d_{xy}$ states of the A layers in the 2H-stacked multilayer sector, can be written as follows
\begin{equation}
\label{eq8}
H^{K}_{\eta}=\int d^2r\left ( \begin{matrix}
  \psi^{\dagger}_{\eta,\textbf{r},b}& \psi^{\dagger}_{\eta,\textbf{r},t}
\end{matrix} \right ) \begin{pmatrix}
 h_{\eta,b}(\textbf{r}) & t_{\eta}(\textbf{r})\\
 t^{*}_{\eta}(\textbf{r}) &  h_{\eta,t}(\textbf{r})
\end{pmatrix}\begin{pmatrix}
 \psi_{\eta,\textbf{r},b}\\
 \psi_{\eta,\textbf{r},t}
\end{pmatrix}\ ,
\end{equation}
where $\eta=\pm 1$ denotes the $K/K'$ valleys, $b$ and $t$ label the $d_{x^2-y^2} \pm i d_{xy}$ orbitals from the A$_1$ and A$_2$ layers in A-AB-$t$MoTe$_2$, 
while in A-ABA-$t$MoTe$_2$ they label the $d_{x^2-y^2} \pm i d_{xy}$ orbital of the A$_1$ layer and the highest-energy layer-hybridized $d_{x^2-y^2} \pm i d_{xy}$ state from the A$_2$ and A$_3$ layers respectively. 
Under the symmetry constraints imposed by $\mathcal{T}$ and $C_3$, the intralayer energy term takes the form

\begin{equation}
\label{eq9}
h_{\eta,l}(\textbf{r}) = (-\frac{\hbar^2 \nabla^2}{2 m_{l}}+(-)^l \frac{\epsilon_K}{2}) + V_{\eta,l}(\textbf{r})
\end{equation}

The absence of $C_{2y}$ symmetry allows $V_{\eta,b}(\textbf{r}) \neq V_{\eta,t}(\textbf{r})$, and introduces an onsite energy $\epsilon_K$ in both A-AB and A-ABA $t$MoTe$_2$ to describe the energy difference induced by interlayer hybridization between the orbitals of the monolayer sector and those of the 2H-stacked multilayer sector. 
From the DFT results, we find $\epsilon_K=0$\,meV for A-AB $t$MoTe$_2$ and $\epsilon_K=21$\,meV for A-ABA $t$MoTe$_2$, as discussed in \cref{appendix:untwisted_kp}. 
To fit the top two valence bands at the $\pm K$ valleys, it is sufficient to consider only the first-harmonic terms of the intralayer and interlayer couplings. 
Under the symmetry constraints imposed by $\mathcal{T}$ and $C_3$ (see also Ref.~\cite{jia2023moir}), these terms can be written as
\begin{align}
\label{eq10}
V_{\eta,l}(\textbf{r}) = \sum_{i=1}^3 V_{l}&(e^{-i(-)^l \psi_{l}}e^{i\textbf{g}_i\textbf{r} }+
e^{i(-)^l \psi_{\eta,l}}e^{-i\textbf{g}_i\textbf{r} }) \\
\label{eq11}
&t_{\eta}(\textbf{r}) = w \sum_{1}^3e^{-i\eta  \textbf{q}_i\cdot \textbf{r}}
\end{align}
Here, $\textbf{q}_1=b_M (0,\tfrac{1}{\sqrt{3}})^T$, $\textbf{q}_2=C_3\textbf{q}_1$, and $\textbf{q}_3=C_3^{-1}\textbf{q}_1$, while time-reversal symmetry $\mathcal{T}$ requires $V_{+,l}=V_{-,l}\equiv V_{l}$. 
Based on the moir\'e translation symmetry, the Hamiltonian in \cref{eq8} can be rewritten in the moir\'e reciprocal space. 
The Fourier transform of the basis is given by
\begin{equation}
\label{eq12}
\psi^{\dagger}_{\eta, \textbf{r},l}=\frac{1}{\sqrt{\mathcal{V}}}\sum_{k}\sum_{\textbf{Q} \in Q_l^\eta}
\psi_{k-\textbf{Q},l}^{\dagger}e^{-i(\textbf{k}-\textbf{Q})\textbf{r} }
\end{equation}
where $Q_l^\eta = \{\textbf{G}_M+\eta(-)^l \textbf{q}_1\}$, and the Hamiltonain in moir\'e space is expressed as 

\begin{equation}
\label{eq13}
H_{\eta}^{K}=\sum_{\textbf{k}}\sum_{\textbf{Q},\textbf{Q'}\in Q_l^\eta}\psi^{ \dagger}_{\eta,\textbf{k}-\textbf{Q},l}h^{K}_{\eta,\textbf{Q}\textbf{Q'}}(\textbf{k})\psi_{\eta, \textbf{k}-\textbf{Q'}}
\end{equation}
and $h^{K}_{\eta,\textbf{Q}\textbf{Q'}}(k)$ is given by

\begin{equation}
\begin{aligned}
\label{eq14}
h^{K}_{\eta,\textbf{Q}\textbf{Q'}}(\textbf{k})&=[-\frac{\hbar^2(\textbf{k}-\textbf{Q})}{2m_l}+(-)^l \frac{\epsilon_K}{2}]\delta_{\textbf{Q},\textbf{Q'}}\\
&+V_{b}\sum_{i=1}^3 [e^{i\eta \psi_{b}}\delta_{\textbf{Q},\textbf{Q'}-\textbf{g}_i}+
e^{-i\eta \psi_{b}}\delta_{\textbf{Q},\textbf{Q'}+\textbf{g}_i}]_{\textbf{Q},\textbf{Q'} \in Q_b^\eta}\\&+V_{t}\sum_{i=1}^3 [e^{-i\eta \psi_{t}}\delta_{\textbf{Q},\textbf{Q'}-\textbf{g}_i}+
e^{i\eta \psi_{t}}\delta_{\textbf{Q},\textbf{Q'}+\textbf{g}_i}]_{\textbf{Q},\textbf{Q'} \in Q_t^\eta}\\
&+w\sum_{1}^3[\delta_{\textbf{Q},\textbf{Q'}-\textbf{q}_i}+\delta_{\textbf{Q},\textbf{Q'}+\textbf{q}_i}]
\end{aligned}
\end{equation}

Due to the breaking of $C_{2y}$ symmetry, there is no symmetry constraint on $V_b=V_t$ and the $K/K'$ valley band pairs split along $\Gamma_M-K_M-M_M$ path. Thus, it is necessary to fit $K$ and $K'$ valleys separately using the same parameters. 
The fitted bands at $K/K'$ valley of A-AB and A-ABA $t$MoTe$_2$ with various twist angles are shown in \cref{2orbital_band-A-AB} and \cref{3orbital_band-A-ABA}. The fitting parameters are shown in \cref{table:K_valley_2orbital}.

Moreover, the layer-resolved orbital analysis in \cref{appendix:moire_band_orbital} reveals a contribution from the $d_{x^2-y^2}\mp i d_{xy}$ orbital of the B layer to the $K/K'$-valley bands, which can be captured within a single-orbital model. 
The continuum model based on the single $d_{x^2-y^2}\mp i d_{xy}$ orbital of the B layer at the $\eta$ valley in reciprocal space is given by

 \begin{equation}
\begin{split}
\label{eq15}
h^{B}_{\eta,\textbf{Q}\textbf{Q'}}(\textbf{k})&=[-\frac{\hbar^2(\textbf{k}-\textbf{Q})}{2m_B}]\delta_{\textbf{Q},\textbf{Q'}}+V_B\sum_{i=1}^3 [e^{i\eta \psi_B}\delta_{\textbf{Q},\textbf{Q'}-\textbf{g}_i}+
e^{-i\eta \psi_B}\delta_{\textbf{Q},\textbf{Q'}+\textbf{g}_i}]_{\textbf{Q},\textbf{Q'} \in Q_t^\eta}
\end{split}
\end{equation}

The fitting results for the B layer are shown as the orange bands in \cref{full_band_3.89} of the main text and in \cref{2orbital_band-A-AB,3orbital_band-A-ABA} of \cref{appendix:result_parameter}, with parameters $m_B=0.52m_e$, $V_B=16$\,meV, and $\psi_B=-105^\circ$ for A-AB $t$MoTe$_2$, and $m_B=0.58m_e$, $V_B=9.5$\,meV, and $\psi_B=-95^\circ$ for A-ABA $t$MoTe$_2$. 
The spin Chern numbers of these bands are found to be zero. 
Moreover, the parameters of the single-orbital model for the B layer at the $K/K'$ valleys remain unchanged with twist angle, which is due to the negligible coupling between the B-layer states at the $K$ valley and those of the A layers, rendering the moir\'e potential ineffective.

\subsection{Three-orbital continuum model of A-ABA $t$MoTe$_2$ at $K/K'$ valley}
\label{appendix:continuum_three_orbital}

Beyond the two-orbital model, we also construct a more complete continuum model for A-ABA $t$MoTe$_2$ that includes three orbitals: the $d_{x^2-y^2}\pm i d_{xy}$ orbitals from the A$_1$, A$_2$, and A$_3$ layers at the $K/K'$ valleys. 
While the two-orbital model only includes the highest-energy layer-hybridized state from the A$_2$ and A$_3$ layers and therefore describes only the two lowest-lying moir\'e valence bands at the $K/K'$ valleys, the three-orbital model incorporates both $d_{x^2-y^2}\pm i d_{xy}$ orbitals from the A$_2$ and A$_3$ layers and can accurately capture up to three moir\'e valence bands originating from the A layers at each valley. 
The corresponding real-space Hamiltonian is given by

\begin{equation}
\label{eq:three_model_real}
H^{K}_{\eta,three}=\int d^2r\left ( \begin{matrix}
  \psi^{\dagger}_{\eta,\textbf{r},A_1}& \psi^{\dagger}_{\eta,\textbf{r},A_2}&\psi^{\dagger}_{\eta,\textbf{r},A_3}
\end{matrix} \right ) \begin{pmatrix}
 h_{\eta,A_1}(\textbf{r}) & t_{\eta,A_1A_2}(\textbf{r}) & t_{\eta,A_1A_3}(\textbf{r})\\
 t^{*}_{\eta,A_1A_2}(\textbf{r}) &  h_{\eta,A_2}(\textbf{r}) &t_{\eta,A_2A_3}(\textbf{r})\\t^{*}_{\eta,A_1A_3}(\textbf{r})&t^{*}_{\eta,A_2A_3}(\textbf{r}) & h_{\eta,A_3}(\textbf{r}) 
\end{pmatrix}\begin{pmatrix}
 \psi_{\eta,\textbf{r},A_1}\\
 \psi_{\eta,\textbf{r},A_2}\\
 \psi_{\eta,\textbf{r},A_3}
\end{pmatrix}
\end{equation}
Here, $\eta=\pm 1$ denotes the $K/K'$ valleys. The A$_1$ site belongs to the bottom monolayer sector, whereas A$_2$ and A$_3$ belong to the top 2H-stacked multilayer sector. 
The high-energy hybridized state formed by the $d_{x^2-y^2}\pm i d_{xy}$ orbitals from A$_2$ and A$_3$ constitutes the single hybridized state of the 2H-stacked multilayer sector in \cref{eq8}. 
Considering only the first harmonic, the intralayer couplings take the same form as in \cref{eq9}, which are written as

\bea
\label{eq:three_model_intra}
h_{\eta,l}(\textbf{r}) &= -\frac{\hbar^2 \nabla^2}{2 m_{l}} + \varepsilon_l+ V_{\eta,l}(\textbf{r})\\
V_{\eta,A_1}(\textbf{r}) &= \sum_{i=1}^3 V_{A_1}(e^{i \psi_{A_1}}e^{i\textbf{g}_i\textbf{r} }+
e^{-i \psi_{\eta,A_1}}e^{-i\textbf{g}_i\textbf{r} })\\
V_{\eta,A_2/A_3}(\textbf{r}) &= \sum_{i=1}^3 V_{A_2/A_3}(e^{-i \psi_{A_2/A_3}}e^{i\textbf{g}_i\textbf{r} }+
e^{i \psi_{\eta,A_2/A_3}}e^{-i\textbf{g}_i\textbf{r} })
\eea
Here, $\varepsilon_l$ ($l=A_1,A_2,A_3$) represents the energy difference between layers induced by the external displacement field, which is set to zero in the following discussion. 
$t_{\eta,A_1A_2}(\textbf{r})$ and $t_{\eta,A_1A_3}(\textbf{r})$ are interlayer couplings, retained up to the first harmonic and having the same form as \cref{eq11}. 
$t_{\eta,A_2A_3}(\textbf{r})$ is an interlayer coupling within the 2H-stacked multilayer sector, kept at the zeroth harmonic and representing the hybridization energy between the A$_2$ and A$_3$ layers. 
Under $\mathcal{T}$ symmetry, this term is real and is found to equal $\varepsilon_K$ (\cref{eq9}) in the fitting results (see \cref{3orbital_band-A-ABA} and \cref{table:K_valley_3orbital}). 
These interlayer couplings can then be written as

\bea
\label{eq:three_model_inter}
t_{\eta,A_1A_2/A_1A_3}(\textbf{r}) &= w_{A_1A_2/A_1A_3} \sum_{1}^3e^{-i\eta  \textbf{q}_i\cdot \textbf{r}} \\ t_{\eta,A_2A_3}(\textbf{r}) &= w_{A_2A_3}=\varepsilon_K
\eea

The fitting results obtained from the three-orbital continuum model for A-ABA $t$MoTe$_2$ at the $K/K'$ valleys with different twist angles, along with the comparison to the two-orbital model, are presented in \cref{3orbital_band-A-ABA}. 
The corresponding fitting parameters of the three-orbital model are listed in Table~\ref{table:K_valley_3orbital}.

\subsection{Continuum model of A-AB and A-ABA $t$MoTe$_2$ at $\Gamma$ valley }
\label{appendix:continuum_gamma}

In this subsection, we construct a continuum model for both A-AB and A-ABA $t$MoTe$_2$ at the $\Gamma$ valley. 
According to the layer-resolved orbital analysis in \cref{appendix:moire_band_orbital}, the low-energy bands at $\Gamma$ mainly originate from the highest-energy layer-hybridized $d_{z^2}$ state, which has negligible interlayer coupling with the $d_{z^2}$ orbital of the A$_1$ layer.

We first give the general form of a two-orbital model for the $\Gamma$ valley and then simplify it to a single-orbital model in the following discussion.
The two-orbital continuum model incorporates both the $d_{z^2}$ orbital of the A$_1$ layer and the highest-energy layer-hybridized $d_{z^2}$ state of the 2H-stacked multilayer sector. 
The general form of this two-orbital model for the $\Gamma$ valley in real space can be written as

\begin{equation}
\label{eq1}
H_{\Gamma}=\int d^2r\left ( \begin{matrix}
  \psi^{\dagger}_{\textbf{r},b}& \psi^{\dagger}_{\textbf{r},t}
\end{matrix} \right ) \begin{pmatrix}
 h_{\Gamma,b}(\textbf{r}) & t_{\Gamma}(\textbf{r})\\
 t^{\dagger}_{\Gamma}(\textbf{r}) &  h_{\Gamma,t}(\textbf{r})
\end{pmatrix}\begin{pmatrix}
 \psi_{\textbf{r},b}\\
 \psi_{\textbf{r},t}
\end{pmatrix}
\end{equation}
where $\psi^{\dagger}_{\textbf{r},l} = \{\psi^{\dagger}_{\textbf{r},l,\uparrow}, \psi^{\dagger}_{\textbf{r},l,\downarrow}\}$, and both $h_{\Gamma,l}(\textbf{r})$ and $t_{\Gamma}(\textbf{r})$ are $2\times2$ matrices acting on the spin degree of freedom. 
Similar to the symmetry analysis of AB- and ABA-stacked MoTe$_2$ at the $\Gamma$ valley in \cref{appendix:continuum_gamma}, the linear Rashba term is forbidden by $C_3$, $\mathcal{T}$, and $M_z$ symmetries in the monolayer (see also Ref.~\cite{jia2023moir}). 
The general form of $h_{\Gamma,l}(\textbf{r})$ is then given by

\begin{equation}
\label{eq2}
h_{\Gamma,l}(\textbf{r}) = (\frac{\hbar^2 \nabla^2}{2 m_{\Gamma,l}} +(-)^l \frac{\epsilon_\Gamma}{2})s_0 + V_{\Gamma,l}(\textbf{r})
\end{equation}
Here, the model breaks the $C_{2y}$ symmetry relative to that of AA-stacked $t$MoTe$_2$. 
As a result, there is no constraint requiring $V_{\Gamma,b}(\textbf{r}) = V_{\Gamma,t}(\textbf{r})$, and the onsite energy difference $\epsilon_\Gamma$ is large due to strong interlayer hybridization at the $\Gamma$ valley. 
From the DFT results, we obtain $\epsilon_\Gamma = 350$\,meV for A-AB $t$MoTe$_2$ and $\epsilon_\Gamma = 311$\,meV for A-ABA $t$MoTe$_2$, as discussed in \cref{appendix:untwisted_kp}.

To describe the low-energy bands at the $\Gamma$ valley, it is sufficient to include the first-harmonic term of $V_{\Gamma,l}(\textbf{r})$ and the zeroth-harmonic term of $t_{\Gamma}(\textbf{r})$. 
Under the relevant symmetry constraints, these terms take the following form
\begin{align}
\label{eq3}
V_{\Gamma,l}(\textbf{r}) = \sum_{i=1}^3 V_{\Gamma,l}&(e^{i(-)^l \psi_{\Gamma,l}}e^{i\textbf{g}_i\textbf{r} }+
e^{-i(-)^l \psi_{\Gamma,l}}e^{-i\textbf{g}_i\textbf{r} })s_0 \\
\label{eq4}
&t_{\Gamma}(\textbf{r}) = w_{\Gamma}s_0
\end{align}
Here, $\textbf{g}_1=b_{M,1}$, $\textbf{g}_2=C_3 b_{M,1}=-b_{M,1}+b_{M,2}$, and $\textbf{g}_3=C_3^{-1} b_{M,1}=-b_{M,2}$. 
We then perform the Fourier transformation into reciprocal space as follows
\begin{equation}
\label{eq5}
\psi^{\dagger}_{\textbf{r},l}=\frac{1}{\sqrt{\mathcal{V}}}\sum_{k}\sum_{\textbf{G}_M}
\psi_{k-\textbf{G}_M,l}^{\dagger}e^{-i(\textbf{k}-\textbf{G}_M)\textbf{r} }
\end{equation}

\begin{equation}
\label{eq6}
H_{\Gamma}=\sum_{\textbf{k}}\sum_{\textbf{G}_M,\textbf{G'}_M}\sum_{l,l'}\psi^{\dagger}_{\textbf{k}-\textbf{G}_M,l}h^{\Gamma}_{ll',\textbf{G}_M\textbf{G'}_M}(\textbf{k})\psi_{\textbf{k}-\textbf{G'}_M,l'}
\end{equation}
By substituting the explicit forms of $h_{\Gamma,l}(\textbf{r})$ (\cref{eq2,eq3}) and $t_{\Gamma}(\textbf{r})$ (\cref{eq4}) into \cref{eq1}, and using the relation in \cref{eq5}, we obtain $h^{\Gamma}_{ll',\textbf{G}_M\textbf{G'}_M}(\textbf{k})$ in the moir\'e reciprocal space as 
\begin{equation}
\begin{split}
\label{eq7}
h^{\Gamma}_{ll',\textbf{G}_M\textbf{G'}_M}(\textbf{k})=&[\frac{\hbar^2(\textbf{k}-\textbf{G}_M)^2}{2m_{\Gamma,l}}+(-)^l\frac{\epsilon_\Gamma}{2}]\delta_{\textbf{G}_M,\textbf{G'}_M}\delta_{l,l'}\\
+&V_{\Gamma,l}\sum_{i=1}^3 (e^{i(-)^l \psi_{\Gamma,l}}\delta_{\textbf{G}_M,\textbf{G'}_M-\textbf{g}_i}+
e^{-i(-)^l \psi_{\Gamma,l}}\delta_{\textbf{G}_M,\textbf{G'}_M+\textbf{g}_i})\delta_{l,l'}\\
+&w_{\Gamma}\delta_{\textbf{G}_M,\textbf{G'}_M}\delta_{\overline{l},l'}
\end{split}
\end{equation}
Here, $\textbf{G}_M = n_1\textbf{b}_{M,1}+n_2\textbf{b}_{M,2}$. 
Nevertheless, the low-energy bands at the $\Gamma$ valley originate mainly from the 2H-stacked multilayer sector in both A-AB and A-ABA $t$MoTe$_2$, with only negligible coupling to the A$_1$ layer, as shown in \cref{moire-orbital}. 
Thus, the two-orbital model discussed above can be simplified to a single-orbital model—representing the hybridized $d_{z^2}$ states of the 2H-stacked multilayer sector—to describe the low-energy bands at the $\Gamma$ valley, neglecting the interlayer coupling. 
The simplified model is then written as
\begin{equation}
\begin{split}
\label{eq_single_gamma}
h^{\Gamma}_{\textbf{G}_M\textbf{G'}_M}(\textbf{k})&=[\frac{\hbar^2(\textbf{k}-\textbf{G}_M)^2}{2m_{\Gamma,t}}]\delta_{\textbf{G}_M,\textbf{G'}_M}\\
&+V_{\Gamma,t}\sum_{i=1}^3 (e^{i \psi_{\Gamma,t}}\delta_{\textbf{G}_M,\textbf{G'}_M-\textbf{g}_i}+
e^{-i\psi_{\Gamma,t}}\delta_{\textbf{G}_M,\textbf{G'}_M+\textbf{g}_i})
\end{split}
\end{equation}
We use \cref{eq_single_gamma} to describe the $\Gamma$-valley bands. 
The results for both A-AB and A-ABA $t$MoTe$_2$ at different twist angles are shown in \cref{A-AB_ABA_gamma}, and the parameters of the single-orbital continuum model are summarized in \cref{table:G_valley_2orbital}.

\subsection{Charge density fitting to the DFT results}
\label{appendix:charge_density_model}

In this subsection, we present a systematic and quantitative scheme to evaluate how well the charge density obtained from the continuum model matches that from the DFT calculation. 
We use two complementary metrics: (i) a normalized correlation, $\rho_{ratio}$, which captures similarity in spatial patterns, and (ii) an absolute difference, $\Delta\rho$, which measures discrepancies. Precise definitions and computational details are given below.  

In the continuum model, the real-space charge density corresponding to a state at $\mathbf{k}_M$ in the $K/K'$ valley is expressed as
\bea
\label{charge_density_K}
\rho_{\eta,n,\bsl{k}_M}(\bsl{r}) = \frac{1}{\Omega}\sum_{l}\left| \sum_{\bsl{Q}\in \Q_l} e^{\ii (\bsl{k}_M-\bsl{Q})\cdot \bsl{r} }\psi_{\eta, n, l , \bsl{k}_M-\bsl{Q}} \right|^2
\eea
where $\eta$ denotes the $K/K'$ valley, $n$ is the eigenstate index, and $l$ is the layer index. 
$\Q_l$ denotes the set of moir\'e reciprocal vectors, defined as $\mathcal{Q}_l=\{n_1 \textbf{G}_1+n_2 \textbf{G}_2+(-)^l\textbf{q}_1\}$, which form a hexagonal reciprocal lattice structure. Similarly, the real-space charge density of a state at the $\Gamma$ valley and momentum $\bsl{k}_M$ is given by
\bea
\label{charge_density_G}
\rho_{\Gamma,n,\bsl{k}_M}(\bsl{r}) = \frac{1}{\Omega}\sum_{l}\left| \sum_{\bsl{G}\in \G} e^{\ii (\bsl{k}_M-\bsl{G})\cdot \bsl{r} }\psi_{\Gamma, n, l , \bsl{k}_M-\bsl{G}} \right|^2
\eea
where the notations $n$ and $l$ follow the same convention as in \cref{charge_density_K}, and $\G = \{n_1 \textbf{G}_1+n_2 \textbf{G}_2\}$ form a triangular reciprocal lattice.

We denote the charge density obtained from the DFT calculation as $\rho_{DFT}$ and that from the continuum model as $\rho_{model}$. 
To ensure a fair comparison, we first smooth $\rho_{DFT}$ by performing a Fourier transformation into the moir\'e reciprocal space using the same $\G$ set as in the continuum model, and then apply an inverse Fourier transformation back to real space on the same discrete $\{\textbf{r}\}$ grid as used in the continuum model. After this smoothing procedure, both $\rho_{DFT}(\textbf{r})$ and $\rho_{model}(\textbf{r})$ can be evaluated at arbitrary positions in real space. 
We then define two standard quantities to quantify the degree of agreement between $\rho_{DFT}$ and $\rho_{model}$, both of which can be computed numerically. 
The first quantity is the correlation coefficient $\rho_{ratio}$ between $\rho_{DFT}$ and $\rho_{model}$, which is defined as
\bea
\label{ratio1}
\rho_{\text{ratio}} = \frac{ \int_{\Omega} \rho_{\text{model}}(\bsl{r}) \rho_{\text{DFT}}(\bsl{r}) d^2r}{\sqrt{ \int_{\Omega} \rho_{\text{model}}(\bsl{r}) \rho_{\text{model}}(\bsl{r})  d^2r \int_{\Omega} \rho_{\text{DFT}}(\bsl{r}) \rho_{\text{DFT}}(\bsl{r})  d^2r } }
\eea
The closer $\rho_{ratio}$ is to 1, the better the agreement between $\rho_{DFT}$ and $\rho_{model}$. 
The second quantity is the integral of the difference $\Delta \rho$ between $\rho_{DFT}$ and $\rho_{model}$, defined as follows:

\bea
\label{ratio2}
\Delta\rho = \int_{\Omega} |\rho_{\text{model}}(\bsl{r})-\rho_{\text{DFT}}(\bsl{r})| d^2r
\eea
The closer $\Delta \rho$ is to 0, the better the agreement between $\rho_{DFT}$ and $\rho_{model}$.

The calculated $\rho_{model}(\textbf{r})$, $\rho_{ratio}$, and $\Delta \rho$ are presented in several cases. 
For A-ABA $t$MoTe$_2$, the results from the top $\Gamma$-valley bands at $\Gamma_M$ and $K_M$ with twist angles of $4.41^\circ$ and $5.09^\circ$ are labeled in \cref{A_ABA_Gamma_fit}. 
For A-AB $t$MoTe$_2$, the results from the top $K/K'$-valley band at $\Gamma_M$ and $K_M$ with a twist angle of $3.89^\circ$ are labeled in \cref{A_AB_K_3.89}. 
For A-ABA $t$MoTe$_2$, the results from the top $K/K'$-valley band at $\Gamma_M$ and $K_M$ with a twist angle of $4.41^\circ$ are labeled in \cref{A_ABA_K_4.41}.

\subsection{Fitting results and parameters}
\label{appendix:result_parameter}

\begin{figure}[H]
    \centering 
    \includegraphics[width=0.90\textwidth]{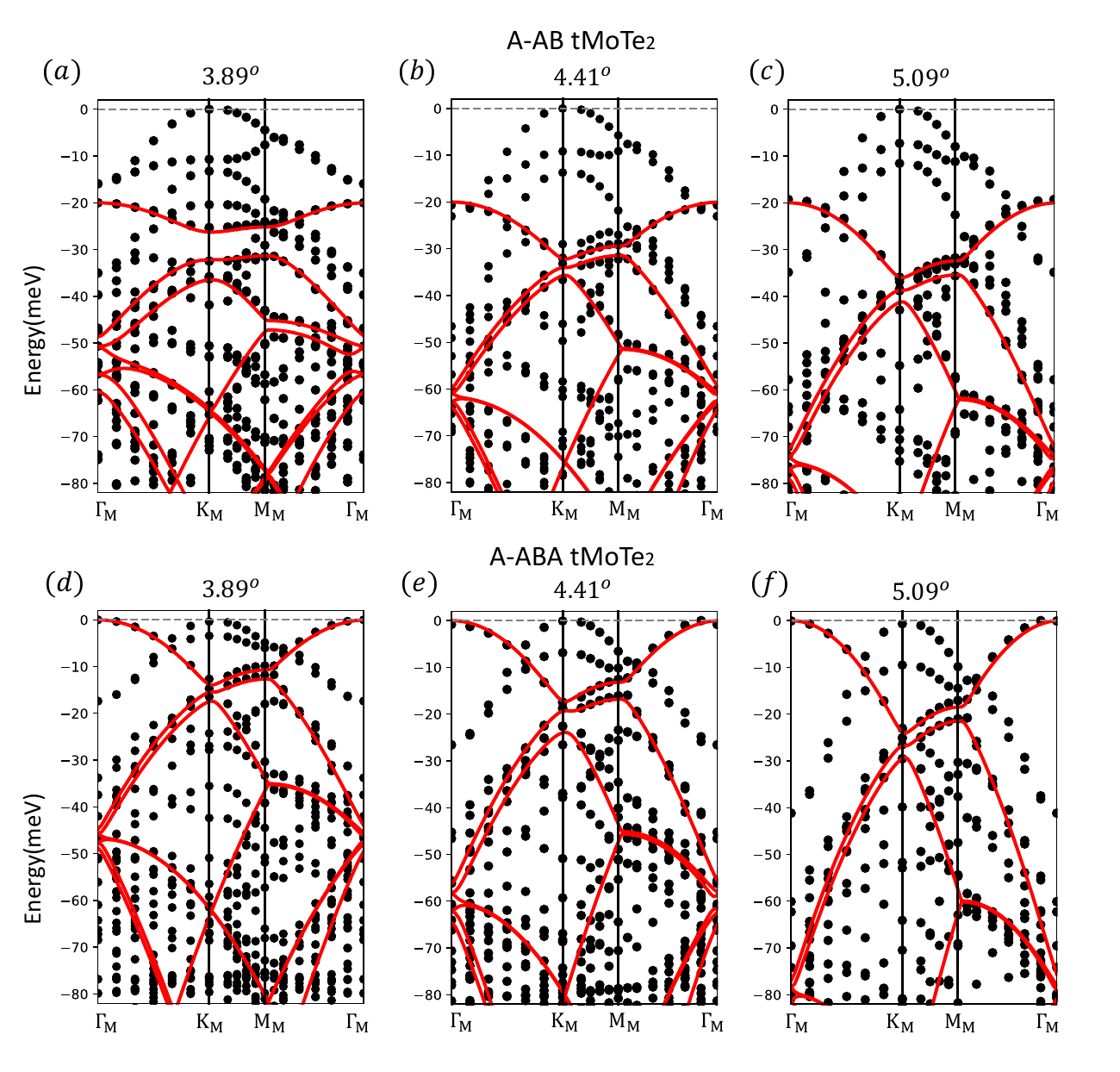}
    \caption{    Fitting results of the $\Gamma$-valley bands for $3.89^\circ$, $4.41^\circ$, and $5.09^\circ$ A-AB (a-c) and A-ABA (d-f) $t$MoTe$_2$ using the single-orbital continuum model. 
The red lines represent the bands calculated by the continuum model, while the black dots represent the DFT bands calculated by VASP.
    \label{A-AB_ABA_gamma}}
\end{figure}

\begin{figure}[H]
    \centering 
    \includegraphics[width=0.90\textwidth]{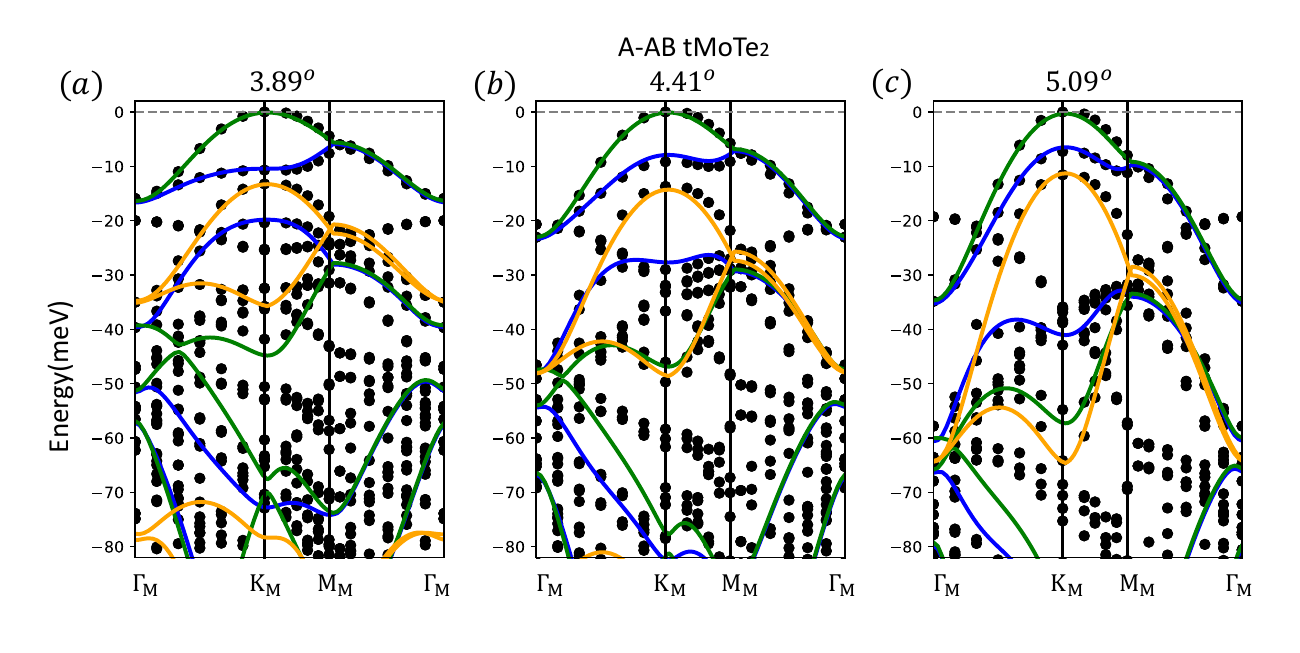}
    \caption{ 
    Fitting results of the $K/K'$-valley bands for $3.89^\circ$ (a), $4.41^\circ$ (b), and $5.09^\circ$ (c) A-AB $t$MoTe$_2$ using the two-orbital continuum model for the $A$ layers and the single-orbital model for the B  layer. 
The blue and green lines represent the $K/K'$-valley bands of the A layers, while the orange lines represent those of the B layer. 
The black dots denote the DFT bands obtained from VASP.
    \label{2orbital_band-A-AB}}
\end{figure}

\begin{figure}[H]
    \centering 
    \includegraphics[width=0.90\textwidth]{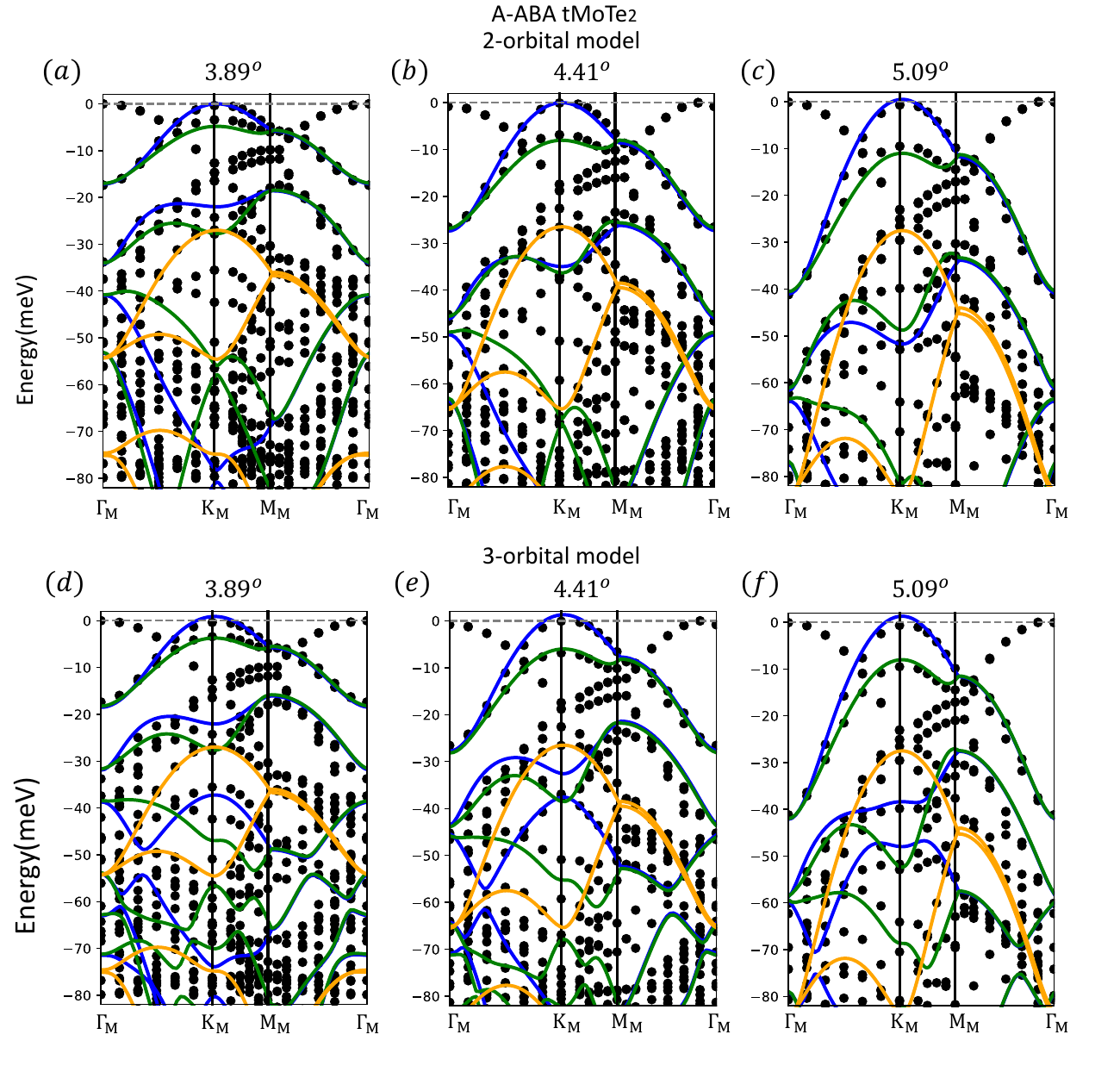}
    \caption{    Fitting results of the $K/K'$-valley bands for $3.89^\circ$, $4.41^\circ$, and $5.09^\circ$ A-ABA $t$MoTe$_2$ using a two-orbital model (a–c) or a three-orbital model (d–f) for the $A$ layers, together with a single-orbital model for the B layer. 
The blue and green lines represent the $K/K'$-valley bands of the $A$ layers, while the orange lines represent those of the B layer. 
The black dots denote the DFT bands obtained from VASP.
\label{3orbital_band-A-ABA}}
\end{figure}

In this part, we present the detailed fitting results and fitting parameters of the continuum models constructed in \cref{appendix:continuum_gamma,appendix:continuum_k,appendix:continuum_three_orbital}. 
The $\Gamma$-valley bands of both A-AB and A-ABA $t$MoTe$_2$ with twist angles of $3.89^\circ$, $4.41^\circ$, and $5.09^\circ$ are fitted using the single-orbital continuum model, as shown in \cref{A-AB_ABA_gamma}. 
The spin Chern numbers of the top isolated bands with nearly spin degeneracy are all zero. 
The top two valence bands derived from the A layers of A-AB $t$MoTe$_2$ at the $K/K'$ valley are fitted using the two-orbital continuum model, as shown in \cref{2orbital_band-A-AB}. 
The spin Chern numbers of the first and second top valence bands at the $K/K'$ valley are both $C_{\uparrow/\downarrow}=\pm 1$.
We fit the top two or three valence bands originating from the A layers of A-ABA $t$MoTe$_2$ at the $K/K'$ valley using the corresponding two- or three-orbital continuum models, as shown in \cref{3orbital_band-A-ABA}. 
The spin Chern numbers of the first and second top valence bands at the $K/K'$ valley are $C_{\uparrow/\downarrow}=\pm 1$ and $C_{\uparrow/\downarrow}=\mp 1$, respectively. 
We also fit the top valence band derived from the B layer at the $K/K'$ valley using a single-orbital continuum model. 
These results are plotted as orange lines in \cref{2orbital_band-A-AB} and \cref{3orbital_band-A-ABA}, and the corresponding fitting parameters are provided in \cref{appendix:continuum_k} and \cref{table:B_valley_orbital}.

\begin{figure}[H]
    \centering 
    \includegraphics[width=1.05\textwidth]{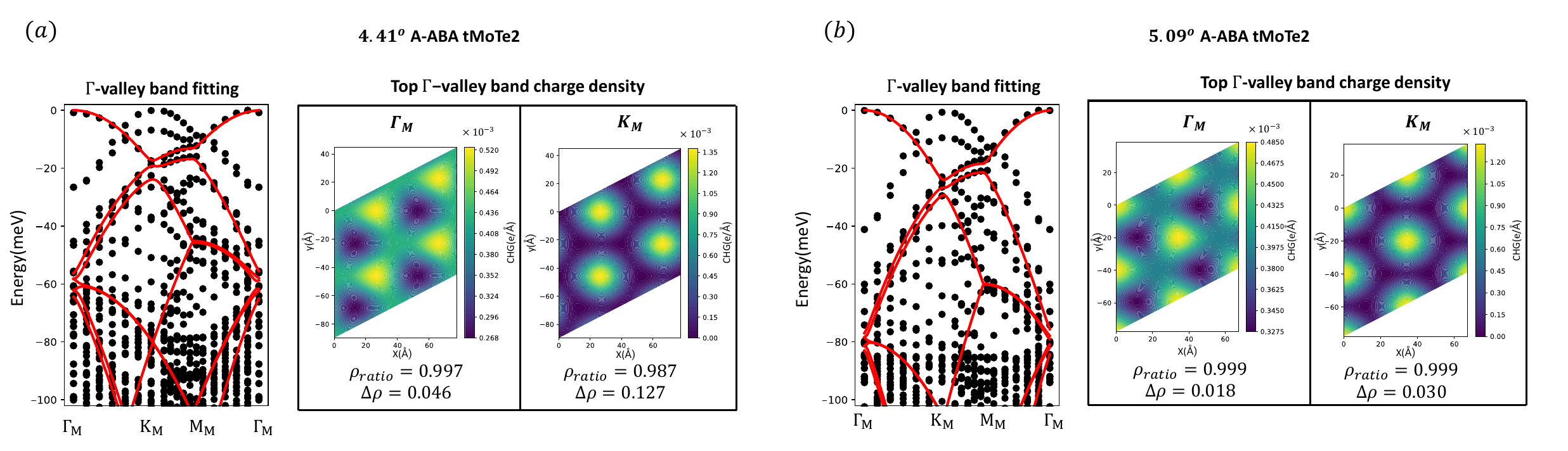}
    \caption{    Fitting results of the bands and charge density at the $\Gamma$ valley for A-ABA $t$MoTe$_2$ with twist angles of $4.41^\circ$ (a) and $5.09^\circ$ (b). 
The black dots represent the DFT results, and the red lines represent the bands calculated from the $\Gamma$-valley continuum model. 
The charge density for the top valence band at $\Gamma_M$ and $K_M$ in the $\Gamma$ valley is calculated using the single-orbital continuum model.
The quantities $\rho_{ratio}$ and $\Delta \rho$ are calculated using Eq.~\ref{ratio1} and Eq.~\ref{ratio2}, respectively.\label{A_ABA_Gamma_fit}}
\end{figure}

\begin{figure}[H]
    \centering 
    \includegraphics[width=1.05\textwidth]{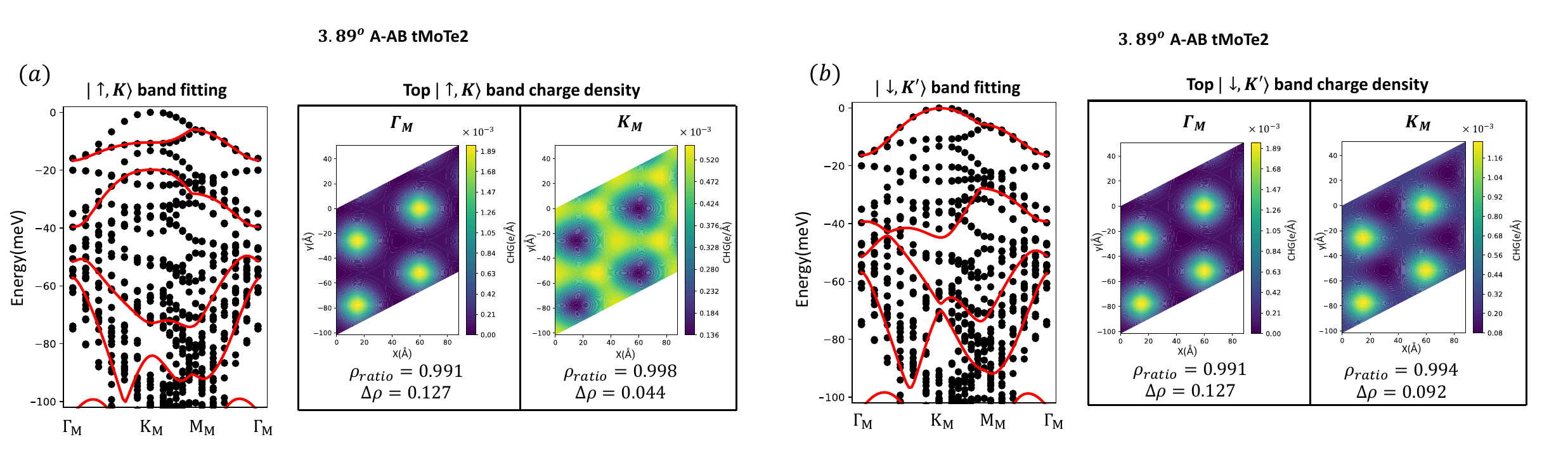}
    \caption{     Fitting results of the bands and charge density at the $K$ (a) and $K'$ (b) valleys of A-AB $t$MoTe$_2$ with a twist angle of $3.89^\circ$. 
The black dots represent the DFT results, while the red lines represent the $K/K'$-valley continuum model results. 
The charge density for the top valence band at $\Gamma_M$ and $K_M$ is calculated using the two-orbital continuum model at the $K/K'$ valley. 
The quantities $\rho_{ratio}$ and $\Delta \rho$ are calculated using \cref{ratio1} and \cref{ratio2}, respectively.
    \label{A_AB_K_3.89}
    }
\end{figure}

\begin{figure}[H]
    \centering 
    \includegraphics[width=1.05\textwidth]{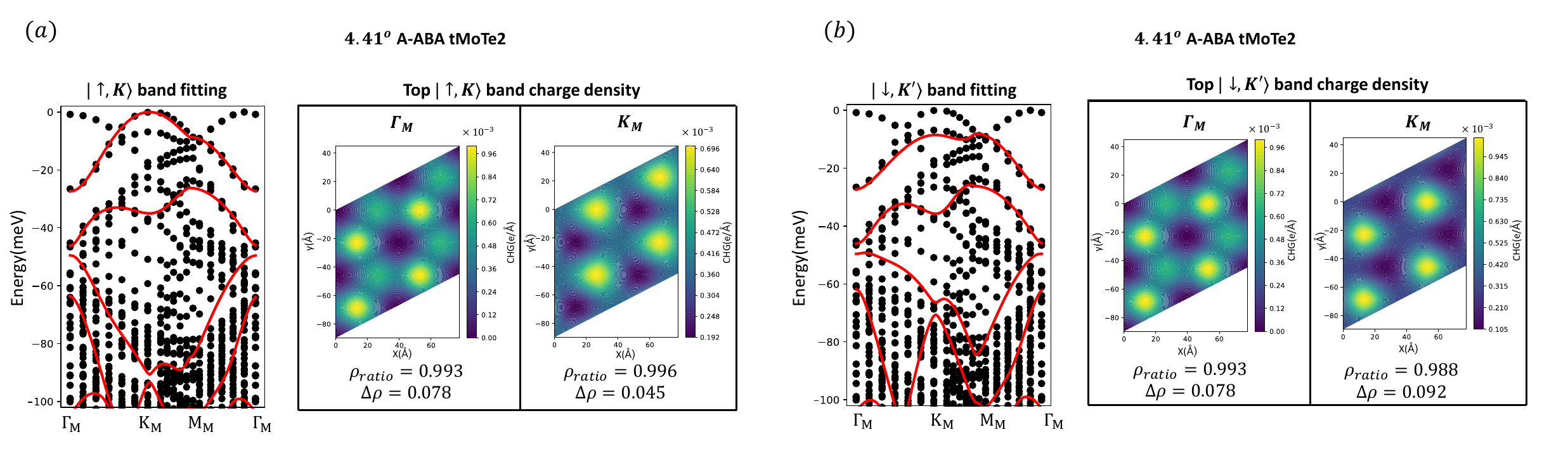}
    \caption{ Fitting results of the bands and charge density at the $K$ (a) and $K'$ (b) valleys of A-ABA $t$MoTe$_2$ with a twist angle of $4.41^\circ$. 
The black dots represent the DFT results, while the red lines represent the $K/K'$-valley continuum model results. 
The charge density for the top valence band at $\Gamma_M$ and $K_M$ is calculated using the two-orbital continuum model at the $K/K'$ valley. 
The quantities $\rho_{ratio}$ and $\Delta \rho$ are calculated using Eq.~\ref{ratio1} and Eq.~\ref{ratio2}, respectively.\label{A_ABA_K_4.41}}
\end{figure}

We also fit the charge density distributions at $\Gamma_M$ and $K_M$ using the two standard quantities defined in \cref{ratio1} and \cref{ratio2}, to quantify the agreement between the charge densities obtained from the continuum model and the DFT results. 
These comparisons are shown in \cref{A_ABA_Gamma_fit,A_AB_K_3.89,A_ABA_K_4.41} for the top valence band at the $\Gamma$ and $K/K'$ valleys in A-AB and A-ABA \tmt, where the results from the continuum models show excellent agreement with those from DFT calculation, with $\rho_{ratio}\!\to\! 1$ and $\rho_\Delta\!\to\! 0$.

As for the fitting parameters, those of the single-orbital continuum model at the $\Gamma$ valley are listed in \cref{table:G_valley_2orbital}. 
The parameters of the two-orbital continuum model at the $K/K'$ valley are given in \cref{table:K_valley_2orbital}. 
The hybridized energy $\epsilon_K$ and the effective dispersion mass $m_l$ can also be approximately extracted from untwisted multilayer MoTe$_2$ (see \cref{appendix:untwisted_kp}), with slight adjustments to account for the effect of the moir\'e potential. 
Moreover, the differences between $V_b,\psi_b$ and $V_t,\psi_t$ reflect the degree of $C_{2y}$-symmetry breaking. 
The parameters of the three-orbital continuum model at the $K/K'$ valley for A-ABA $t$MoTe$_2$ are summarized in \cref{table:K_valley_3orbital}, which share the same effective dispersion mass and hybridized energy as the two-orbital model at each twist angle. 
The values of the first-harmonic interlayer and intralayer coupling terms also show good agreement between the two-orbital and three-orbital continuum models for A-ABA $t$MoTe$_2$.

\begin{table*}[h]
\renewcommand{\arraystretch}{1.5}
\setlength{\tabcolsep}{9pt}
\centering
\begin{tabular}{|c|c|c|c|}

\hline

\diagbox{$\textbf{Twist angle}$}{$\textbf{Parameter}$} &$m_{\Gamma,t}(m_e)$ &  $V_{\Gamma,.t}(meV)$ &  $\psi_{\Gamma,t}(^\circ)$ \\
\hline
$3.89^\circ$ A+AB & ${2.3}$ &  $3.0$ & $-30$  \\
\hline
$4.41^\circ$ A+AB & ${2.3}$ &  $1.0$ & $-30$ \\
\hline
$5.09^\circ$ A+AB & ${2.3}$ &  $1.5$ & $-30$  \\
\hline
$3.89^\circ$ A+ABA & ${1.6}$ &  $1.0$ & $-85$   \\
\hline
$4.41^\circ$ A+ABA & ${1.6}$ &  $2.0$ & $-75$  \\
\hline
$5.09^\circ$ A+ABA & ${1.6}$ &  $1.5$ & $-30$  \\
\hline
\end{tabular}
\caption{Values of the parameters at different twist angles for the $\Gamma$-valley single-orbital continuum model of A-AB and A-ABA $t$MoTe$_2$.  }
\label{table:G_valley_2orbital}
\end{table*}

\begin{table*}[h]
\renewcommand{\arraystretch}{1.0}
\setlength{\tabcolsep}{5pt}
\centering
\begin{tabular}{|c|c|c|c|}
\hline
\diagbox{$\textbf{Twist angle}$}{$\textbf{Parameter}$} &$m_{B}(m_e)$ & $V_{B}(meV)$& $\psi_{B}(^\circ)$  \\
\hline
$3.89^\circ$,$4.41^\circ$,$5.09^\circ$ A+AB &$0.52$ & $16$ & $-105$  \\
\hline
$3.89^\circ$,$4.41^\circ$,$5.09^\circ$ A+ABA &$0.58$ & $9.5$ & $-95$ \\
\hline
\end{tabular}
\caption{Values of the parameters in the single-orbital continuum model for the B layer in A-AB and A-ABA $t$MoTe$_2$.  }
\label{table:B_valley_orbital}
\end{table*}

\begin{table*}[h]
\renewcommand{\arraystretch}{1.5}
\setlength{\tabcolsep}{9pt}
\centering

\resizebox{\linewidth}{!}{
\begin{tabular}{|c|c|c|c|c|c|c|c|c|c|}

\hline
\diagbox{$\textbf{Twist angle}$}{$\textbf{Parameter}$} &$m_{b}(m_e)$&$m_{t}(m_e)$ & $V_{b}(meV)$& $V_{t}(meV)$ & $\psi_{b}(^\circ)$& $\psi_{t}(^\circ)$ &$w(meV)$  & $\epsilon_K(meV)$ \\
\hline
$3.89^\circ$ A+AB &$0.60$ & $0.60$ & $18.0$ & $9.2$ & $-116$ &$-101$ &$-19$ & $0$ \\
\hline
$4.41^\circ$ A+AB &$0.62$ & $0.62$ & $18.0$ & $11.3$ & $-116$ &$-101$ &$-18$ & $0$ \\
\hline
$5.09^\circ$ A+AB &$0.64$ & $0.64$ & $18.0$ & $11.8$ & $-116$ &$-101$ &$-18$ & $0$ \\
\hline
$3.89^\circ$ A+ABA &$0.60$ & $0.60$ & $18.5$ & $9.5$ & $-116$ &$-106$ &$-13.0$ & $21$ \\
\hline
$4.41^\circ$ A+ABA &$0.62$ & $0.62$ & $18.0$ & $10.0$ & $-116$ &$-106$ &$-13.8$ & $21$ \\
\hline
$5.09^\circ$ A+ABA &$0.62$ & $0.62$ & $18.0$ & $10.5$ & $-116$ &$-106$ &$-13.8$ & $21$ \\
\hline
\end{tabular}}
\caption{Values of the parameters at different twist angles for the two-orbital continuum model of the A layers at the $K/K'$ valley in A-AB and A-ABA $t$MoTe$_2$.}
\label{table:K_valley_2orbital}
\end{table*}

\begin{table*}[h]
\renewcommand{\arraystretch}{1.5}
\setlength{\tabcolsep}{9pt}
\centering

\resizebox{\linewidth}{!}{
\begin{tabular}{|c|c|c|c|c|c|c|c|c|c|c|}

\hline
\diagbox{$\textbf{Twist angle}$}{$\textbf{Parameter}$} &\makecell[c]{$m_{l}(m_e)$\\($l=A_1,A_2,A_3$)}&  $V_{A_1}(meV)$& $V_{A_2}(meV)$ &$V_{A_3}(meV)$ & $\psi_{A_1}(^\circ)$& $\psi_{A_2}(^\circ)$ &$\psi_{A_3}(^\circ)$ &$w_{A_1A_2}(meV)$  & $w_{A_1A_3}(meV)$  &$w_{A_2A_3}(meV)$\\
\hline
$3.89^\circ$ A+ABA &$0.60$ & $18.5$ & $9.5$ & $9.5$ & $-98$ &$-116$ & $-71$ &$-13.0$ & $1$ & $21$ \\
\hline
$4.41^\circ$ A+ABA &$0.62$ & $18.5$ & $10.0$ & $9.0$ & $-99$ &$-111$ & $-61$ &$-14.0$ & $1$ & $21$ \\
\hline
$5.09^\circ$ A+ABA &$0.64$ & $18.5$ & $10.0$ & $9.0$ & $-100$ &$-111$ & $-61$ &$-14.0$ & $1$ & $21$ \\
\hline
\end{tabular}}
\caption{Values of the parameters at different twist angles for the three-orbital continuum model at the $K/K'$ valley of the $A$ layers in A-ABA $t$MoTe$_2$
}
\label{table:K_valley_3orbital}
\end{table*}

\subsection{Berry curvature and quantum metric}
\label{appendix:curvature and metric}
In this subsection, we present the results of the Berry curvature and quantum metric for the top two bands from the $K$ valley of the A layers and the nearly spin-degenerate top bands from the $\Gamma$ valley for both A-AB and A-ABA $t$MoTe$_2$.

\begin{figure}[H]
    \centering 
    \includegraphics[width=1.00\textwidth]{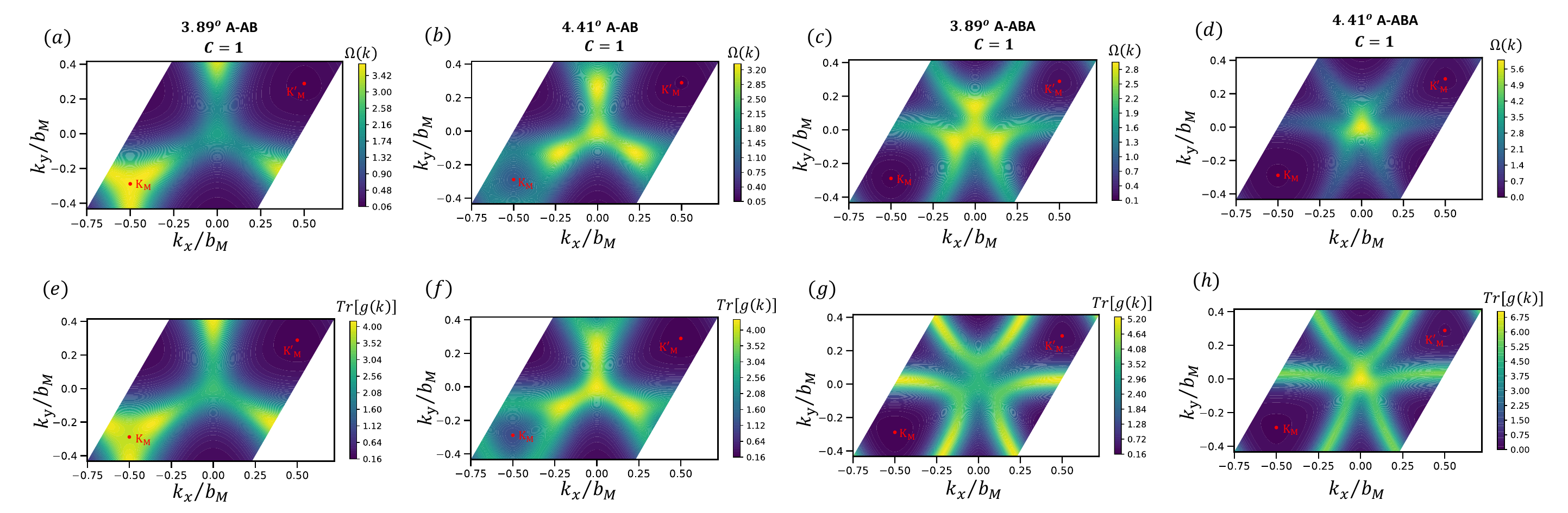}
    \caption{     The distribution of the Berry curvature $\Omega_n(\textbf{k})$ (a-d) and the quantum metric $\text{Tr}[g_{n}(\textbf{k})]$ (e-h) for the top band from the $K$ valley of A layers in A-AB and A-ABA $t$MoTe$_2$ at twist angles of $3.89^\circ$ and $4.41^\circ$. Here, $b_{M}=\frac{4\pi}{\sqrt{3}a_M}$, where $a_M=\frac{a_0}{2\sin(\frac{\theta}{2})}$.
\label{fig:curvature_k_first} }
\end{figure}

\begin{figure}[H]
    \centering 
    \subfloat{\label{curvature_gamma_top:a}}
    \subfloat{\label{curvature_gamma_top:b}}
    \subfloat{\label{curvature_gamma_top:c}}
    \subfloat{\label{curvature_gamma_top:d}}
    \subfloat{\label{curvature_gamma_top:e}}
    \subfloat{\label{curvature_gamma_top:f}}
    \subfloat{\label{curvature_gamma_top:g}}
    \subfloat{\label{curvature_gamma_top:h}}
    \includegraphics[width=1.00\textwidth]{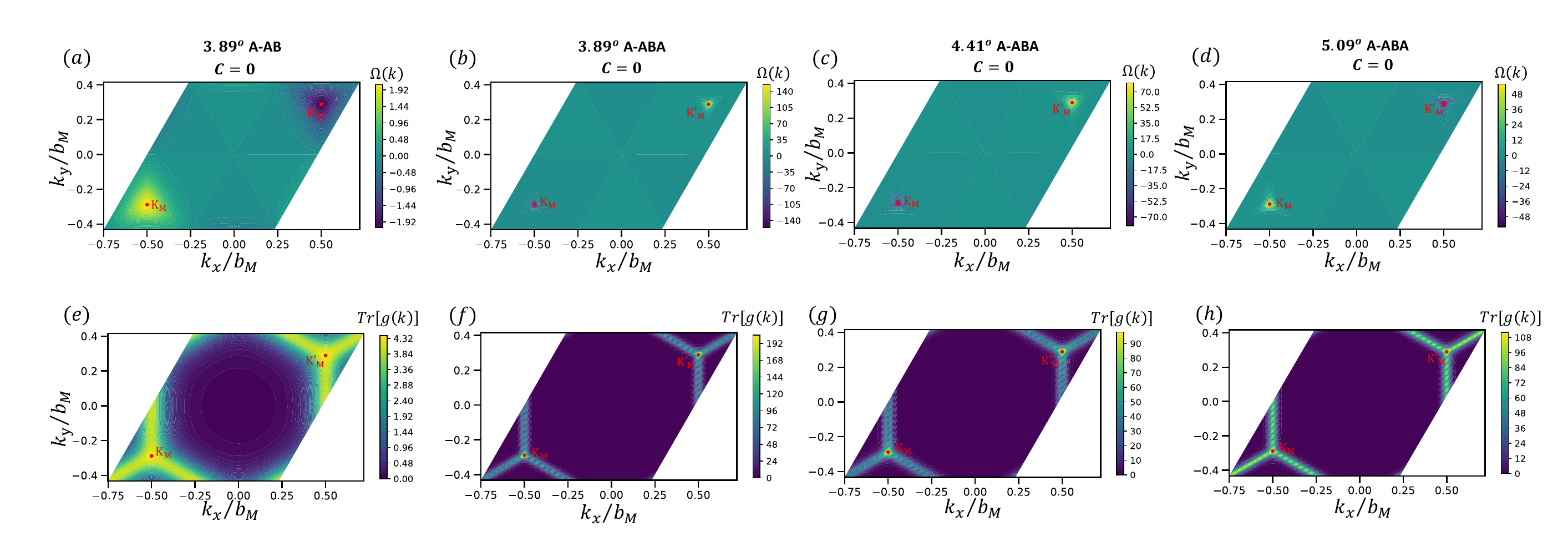}
    \caption{ The distribution of the Berry curvature $\Omega_n(\textbf{k})$ (a-d) and quantum metric $\text{Tr}[g_{n}(\textbf{k})]$ (e-h) for the nearly spin-degenerate top two bands from the $\Gamma$ valley, for $3.89^\circ$ A-AB and $3.89^\circ$, $4.41^\circ$, $5.09^\circ$ A-ABA $t$MoTe$_2$.
\label{fig:curvature_gamma_top} }
\end{figure}

The Berry curvature $\Omega_n(\textbf{k})$ and quantum metric tensor $g_n(\textbf{k})$ of the targeted $n$-th band can be written in terms of the projection operator, $P_n(\textbf{k})=\ket{u_n(\textbf{k})}\bra{u_n(\textbf{k})}$, as follows
\bea
&\Omega_n(\textbf{k})=\ii \sum_{ij}\epsilon_{ij}Tr[P_n(\textbf{k})\partial_{k_i}P_n(\textbf{k})\partial_{k_j}P_n(\textbf{k})] \\
&g_{n,ij}(\textbf{k})=\frac{1}{2}Tr[\partial_{k_i}P_n(\textbf{k})\partial_{k_j}P_n(\textbf{k})]
\eea
where $i,j=x,y$. Here, we show the distribution of $\Omega_n(\textbf{k})$ and $\text{Tr}[g_{n}(\textbf{k})]$ for the top bands from the $K$ valley of A-layers in A-AB and A-ABA $t$MoTe$_2$ at $3.89^\circ$ and $4.41^\circ$ (\cref{fig:curvature_k_first}), as well as those for the top bands from the $\Gamma$ valley in A-AB at $3.89^\circ$ and in A-ABA at $3.89^\circ$, $4.41^\circ$, and $5.09^\circ$ (\cref{fig:curvature_gamma_top}).

Several unusual properties emerge from these results. For the top band from the $K$ valley (\cref{fig:curvature_k_first}), the spin Chern numbers for both A-AB and A-ABA $t$MoTe$_2$ are the same ($C_\uparrow=1$), but the distributions of $\Omega_n(\textbf{k})$ and $\text{Tr}[g_{n}(\textbf{k})]$ are significantly different. Larger values of $\Omega_n(\textbf{k})$ and $\text{Tr}[g_{n}(\textbf{k})]$ are observed near $\Gamma_M$ in A-ABA $t$MoTe$_2$ (similar to the case in the AA-stacked bilayer $t$MoTe$_2$, see Ref.~\cite{jia2023moir}), whereas they are observed near $K_M$ in A-AB $t$MoTe$_2$ with a twist angle of $3.89^\circ$. Moreover, as the twist angle increases, the regions of large values converge towards the $\Gamma_M$ point for both A-AB and A-ABA $t$MoTe$_2$.

For the top bands from the $\Gamma$ valley (\cref{fig:curvature_gamma_top}), the spin Chern numbers for both A-AB and A-ABA $t$MoTe$_2$ are 0, with two opposite peaks of $\Omega_n(\textbf{k})$ at $K_M$ and $K_M'$. Moreover, a band inversion in A-ABA $t$MoTe$_2$, occurring between $5.09^\circ$ and $4.41^\circ$, is reflected in the distribution of $\Omega_n(\textbf{k})$, with opposite peaks at $K_M$ for these two twist angles (\cref{curvature_gamma_top:b} and \cref{curvature_gamma_top:c}). This is consistent with the change in the $C_3$ eigenvalues (\cref{twist_band}). The Wilson loops, shown in \cref{fig:WilsonLoop}, further confirm that the $Z_2$ indices remain zero despite the band inversion between $5.09^\circ$ and $4.41^\circ$.

\section{Moiré-Model-Building Without Fitting}
\label{appendix:universal_model}

In this section, we construct accurate continuum models using our universal Moiré-Model-Building method~\cite{zhang2024universal}. We first calculate the Truncated Atomic Plane Wave (TAPW)~\cite{PhysRevB.107.125112,Chen_2024,shi2024moireopticalphononsdancing,zhang2024universal} Hamiltonian with the OpenMX software package~\cite{ozaki_Variationally_2003,ozaki_Numerical_2004,ozaki_Efficient_2005} to obtain an accurate representation of the electronic properties of moiré systems, and then extract the continuum models directly from the DFT results without fitting for both the $\Gamma$ and $K/K'$ valleys in A-AB and A-ABA $t$MoTe$_2$.

\subsection{Truncated Atomic Plane Wave calculation}
\label{appendix:tapw}

\begin{figure}[H]
    \centering 
    \includegraphics[width=1.0\textwidth]{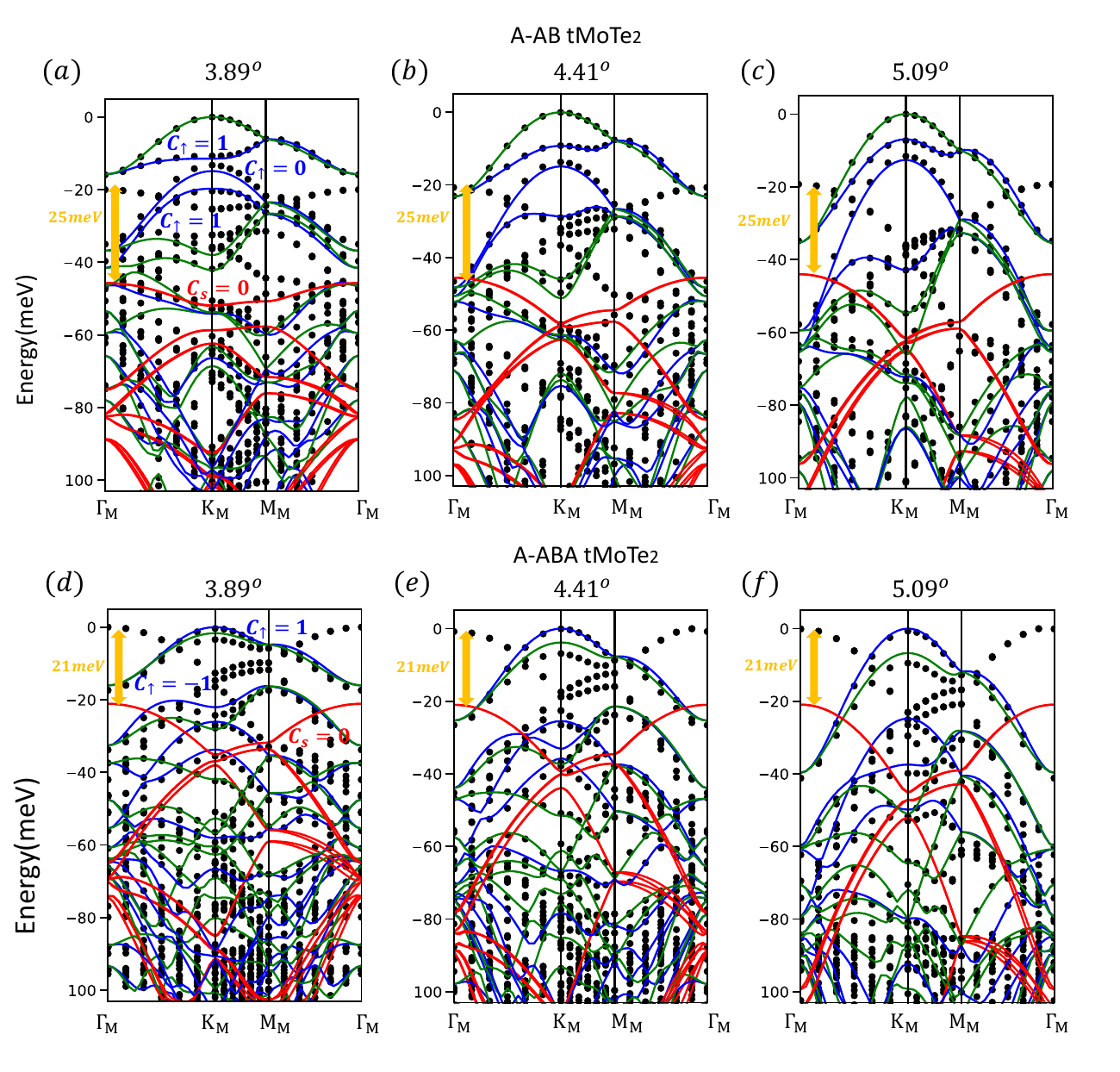}
    \caption{ The valley-projected electronic structures obtained from the TAPW method for $3.89^\circ$ A-AB (a) and A-ABA (b) $t$MoTe$_2$. Black dot lines represent the moir\'e bands calculated using VASP, while the solid lines correspond to the valley-projected bands calculated from the TAPW Hamiltonian. Red lines indicate $\Gamma$ valley bands, blue lines indicate $K$ valley bands, and green lines indicate $K'$ valley bands. The spin Chern numbers of the top several bands from the $\Gamma$ and $K/K'$ valleys are labeled in the figures.
\label{tapw_band}}
\end{figure}

First, we perform DFT calculations in OpenMX using the same relaxed crystal structures as in VASP. The pseudo-atomic orbital (PAO) basis functions for the OpenMX calculations are chosen as Mo7.0-$s3p2d2$ and Te7.0-$s3p2d2f1$, where "7.0" denotes the cutoff radius of the PAO functions in Bohr units. Here, $s3p2d2$ represents 3 $s$-orbitals, 2 sets of $p$-orbitals, and 2 sets of $d$-orbitals for Mo atoms $(3 + 2 \times 3 + 2 \times 5 = 19)$, while $s3p2d2f1$ for Te atoms represents 3 $s$-orbitals, 2 sets of $p$-orbitals, 2 sets of $d$-orbitals, and 1 set of $f$-orbitals $(3 + 2 \times 3 + 2 \times 5 + 7 = 26)$.
Then, we construct a TAPW-basis Hamiltonian from the full Hamiltonian obtained from OpenMX. We project the \textit{ab initio} tight-binding Hamiltonian onto the atomic plane waves localized around the $K$, $K'$, and $\Gamma$ valleys separately, and present the valley-projected electronic structures of $3.89^\circ$ A-AB and A-ABA $t$MoTe$_2$ in \cref{tapw_band}. The electronic structures obtained from the TAPW Hamiltonian are in good agreement with those from VASP, except that the $\Gamma$ and $K$ valleys now differ by approximately 20 meV, which does not affect the model within each valley.
Spin Chern numbers are also calculated from our TAPW-basis Hamiltonian (labeled in \cref{tapw_band}), which are consistent with those obtained from the fitted continuum models.

\subsection{Procedure for Constructing the Continuum Model}
\label{appendix:universal_model}
Below, we briefly summarize the procedure used to construct the continuum model from the DFT Hamiltonian, $H_{\text{DFT}}$, on both the full and reduced $\bsl{Q}$ lattices. In our application to the $K$ and $\Gamma$ valleys of $t$MoTe$_2$, we first extract an effective Hamiltonian in the low-energy subspace by projecting $H_{\text{DFT}}$ onto the relevant orbital basis. We then diagonalize the $\bsl{Q}$-blocks of $H_{\text{DFT}}$ and retain only the branch closest to the Fermi level. Subsequently, using a second-order perturbation procedure, we integrate out the high-energy states, thereby obtaining a low-energy effective Hamiltonian $H_{\text{DFT}}^{\text{low-energy}}$ (or $H_{\text{DFT}}^{\text{reduced}}$ when a truncated $\bsl{Q}$ lattice is employed). A detailed derivation can be found in ~\cite{zhang2024universal}.

\subsubsection{Application to the $K$ Valley in A--AB and A--ABA Systems}
\label{appendix:universal_K_valley}

At the $K$ valley of $t$MoTe$_2$, SOC locks each valley to a single spin polarization at low energies: the A-layers contribute only the spin–up combination ($\ket{d_{x^2-y^2}+i\,d_{xy},\uparrow}$), while the B layer contributes only the spin–down partner ($\ket{d_{x^2-y^2}-i\,d_{xy},\downarrow}$). Accordingly, the continuum Hamiltonian at the $K$ valley separates into two independent sectors: an \emph{A-sector} built from the spin–up $d+id$ orbitals, and a \emph{B-sector} built from the spin–down $d-id$ orbitals. Although A–AB and A–ABA differ in the number of A-layer orbitals, they share the same block-diagonal structure, which corresponds to two-orbital (\cref{appendix:continuum_k}) or three-orbital (\cref{appendix:continuum_three_orbital}) continuum models for the A-layers block of A-AB and A-ABA $t$MoTe$_2$, and a single-orbital (\cref{appendix:continuum_k}) continuum model for the B-layer block of both A-AB and A-ABA $t$MoTe$_2$. We introduce
\begin{equation}
\Psi^\dagger_{A,\bsl{k}} =
\begin{cases}
\bigl(\psi^\dagger_{A_1,\bsl{k}},\,\psi^\dagger_{A_2,\bsl{k}}\bigr), 
& \text{A--AB},\\[4pt]
\bigl(\psi^\dagger_{A_1,\bsl{k}},\,\psi^\dagger_{A_2,\bsl{k}},\,\psi^\dagger_{A_3,\bsl{k}}\bigr),
& \text{A--ABA},
\end{cases}
\quad
\Psi^\dagger_{B,\bsl{k}} = \bigl(\psi^\dagger_{B_1,\bsl{k}}\bigr),
\end{equation}
so that
\begin{equation}
H^{K}= H^{K}_A + H^{K}_B
=\sum_{\bsl{k}}
\begin{pmatrix}\Psi_{A,\bsl{k}}^\dagger & \Psi_{B,\bsl{k}}^\dagger\end{pmatrix}
\begin{pmatrix}
H^K_A(\bsl{k}) & 0\\
0 & H^K_B(\bsl{k})
\end{pmatrix}
\begin{pmatrix}\Psi_{A,\bsl{k}}\\\Psi_{B,\bsl{k}}\end{pmatrix},
\end{equation}
with $H^K_A$ acting on the spin–up $d+id$ manifold of the A layers, and $H^K_B$ acting on the spin–down $d-id$ manifold of the B layer.

In the general continuum model, each block $h^{K}_{ll'}(\bsl{k})$ is expanded as:
\begin{equation}
h^{K}_{ll'}(\bsl{k})
=
\sum_{M_z,M_{z^*}\ge0}
\sum_{\bsl{G}_M}
t^{\,M_zM_{z^*}}_{\,l,l';\bsl{G}_M}\,
Y^{\,M_zM_{z^*}}_{\,l,l',\bsl{p}}(\bsl{k})
+\text{symm.\ terms},
\quad
\bsl{p} = \bsl{q}_l - \bsl{q}_{l'} + \bsl{G}_M,
\end{equation}
where \textit{symm.\ terms} denote the additional terms required by the the three-fold rotation symmetry $C_{3z}$ and Hermitian conjugation, ensuring that the full Hamiltonian satisfies all the symmetry constraints discussed below.
 The basis matrices are defined by:
\begin{equation}
\bigl[Y^{\,M_zM_{z^*}}_{\,l,l',\bsl{p}}(\bsl{k})\bigr]_{\bsl{Q}_1,\bsl{Q}_2}
=
\delta_{l_{\bsl{Q}_1},l}\,\delta_{l_{\bsl{Q}_2},l'}\,
(\bsl{k}-\bsl{Q}_1)_z^{M_z}\,(\bsl{k}-\bsl{Q}_1)_{z^*}^{M_{z^*}}\,
\delta_{\bsl{Q}_1,\bsl{Q}_2+\bsl{p}},
\end{equation}
with $(\bsl{k}-\bsl{Q})_z=(k_x-Q_x)+i(k_y-Q_y)$ and $(\bsl{k}-\bsl{Q})_{z^*}=(k_x-Q_x)-i(k_y-Q_y)$.  By construction, no spin indices appear in these blocks because each layer contributes only one spin-polarized orbital.

\paragraph{Physical meaning of the coefficients.}
The coefficients $t^{\,M_zM_{z^*}}_{\,l,l';\bsl{G}_M}$ are extracted by projecting the \emph{ab-initio} Hamiltonian onto the basis functions.
\begin{itemize}
\item $l=l'$, $\bsl{G}_M=0$, $M_z=M_{z^*}=0$ $\Longrightarrow$ on–site term;
\item $l=l'$, $\bsl{G}_M=0$, $M_z+M_{z^*}\neq0$ $\Longrightarrow$ kinetic term;
\item $l=l'$, $\bsl{G}_M\neq0$ $\Longrightarrow$ intralayer moiré potential;
\item $l\neq l'$ $\Longrightarrow$ interlayer coupling.
\end{itemize}

\paragraph{Symmetry constraints.}
Within a single $K$ valley the only spatial symmetry is the threefold rotation $C_{3z}$.  We denote by $\lambda_d(l)$ the phase acquired by the Mo $d$–orbital on layer $l$ under a $120^\circ$ rotation:
\begin{equation}
\lambda_d(l)=
\begin{cases}
+\dfrac{\pi}{3}, & l\in\{A_1,A_2,\dots\}\ (\text{spin--up }d+id),\\[6pt]
-\dfrac{\pi}{3}, & l=B_1\ (\text{spin--down }d-id).
\end{cases}
\end{equation}
In the combined plane‐wave ($\bsl{Q}$) and layer ($l$) basis, $C_{3z}$ is represented by the matrix:
\begin{equation}
[D(C_{3z})]_{\bsl{Q},l;\,\bsl{Q}',l'}
=
\delta_{l,l'}\,
\delta_{\bsl{Q},\,C_{3z}\bsl{Q}'}\,
e^{\,i\lambda_d(l)}.
\end{equation}

\begin{figure}[!hbtp]
    \centering
    \includegraphics[width=1\linewidth]{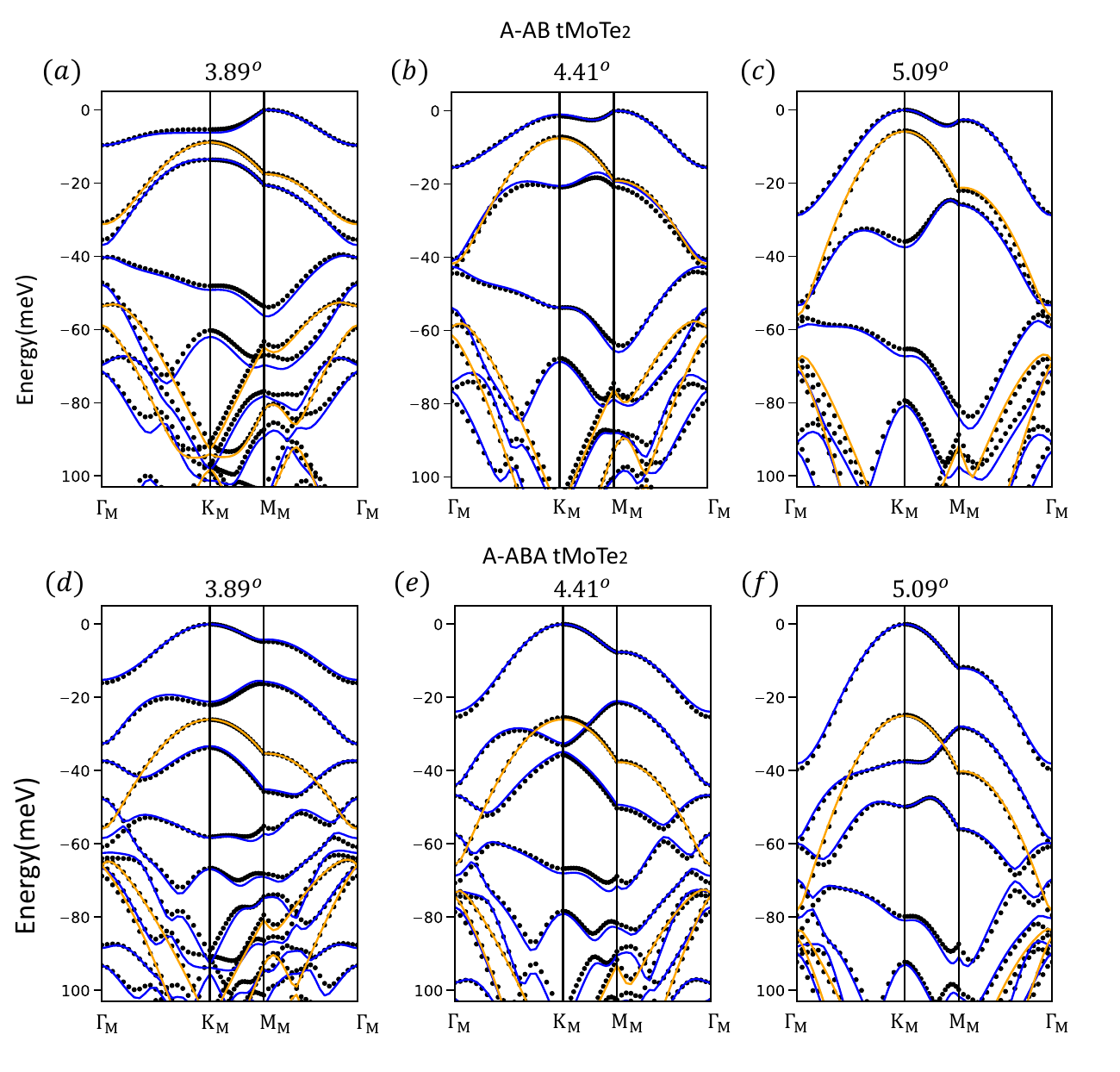}
    \caption{    (a)–(c) Reduced-model results for the $K$ valley of A-AB $t$MoTe$_2$. 
(d)–(f) Reduced-model results for the $K$ valley of A-ABA $t$MoTe$_2$. 
Black dots represent the DFT band structures at the $K$ valley, while the blue and orange lines denote the continuum-model calculations. 
Blue lines correspond to contributions from the A-layer(s), and orange lines correspond to contributions from the B layer.}
    \label{fig:ruduced_K_model}
\end{figure}

\begin{figure}[!hbtp]
    \centering 
    \includegraphics[width=1.00\textwidth]{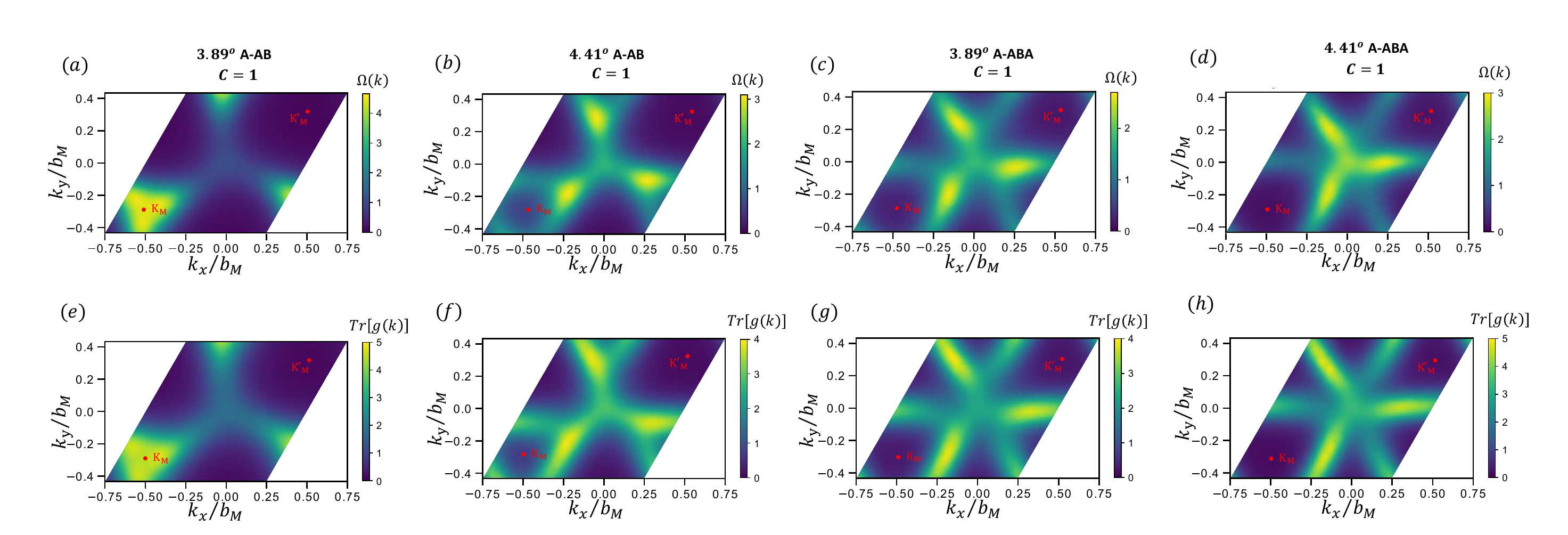}
    \caption{ The distributions of Berry curvature $\Omega_n(\textbf{k})$ (a–d) and quantum metric $\text{Tr}[g_{n}(\textbf{k})]$ (e–h) for the first top band from the $K$ valley are shown for $3.89^\circ$ and $4.41^\circ$ twist angles of A-AB and A-ABA $t$MoTe$_2$. 
$b_{M} = \frac{4\pi}{\sqrt{3}a_M}$, where $a_M = \frac{a_0}{2 \sin(\theta/2)}$. 
These results are obtained from the accurate continuum model at the $K$ valley. \label{fig:universal_curvature_k_first} }
\end{figure}

When this rotation acts on the continuum basis‐function matrices $Y^{\,M_zM_{z^*}}_{l,l',\bsl{p}}(\bsl{k})$, the orbital phases $\lambda_d(l)$ cancel within each A‐ or B‐sector (since $\lambda_d(l)=\lambda_d(l')$), and only the factors $(k_x\pm i k_y)^{M_z}$ pick up a phase:
\begin{equation}
D(C_{3z})\,Y^{\,M_zM_{z^*}}_{l,l',\bsl{p}}(\bsl{k})\,D(C_{3z})^{-1}
=
e^{\,-i\frac{2\pi}{3}(M_z-M_{z^*})}\;
Y^{\,M_zM_{z^*}}_{\,l,l',\,C_{3z}\bsl{p}}\bigl(C_{3z}\bsl{k}\bigr).
\end{equation}
Requiring the Hamiltonian to commute with $C_{3z}$ then enforces:
\begin{equation}
t^{\,M_zM_{z^*}}_{\,l,l';\bsl{G}_M}
=
e^{\,-i\frac{2\pi}{3}(M_z-M_{z^*})}\;
t^{\,M_zM_{z^*}}_{\,l,l';\,C_{3z}\bsl{G}_M}.
\end{equation}
\cref{fig:ruduced_K_model} exhibits the results of the reduced accurate continuum model for the $K$ valley of A-AB and A-ABA $t$MoTe$_2$ with various twist angles. The model parameters are listed in ~\cite{IOP_eln2}. The Berry curvature and quantum metric distributions of the top valence bands from the $K$ valley, calculated from the reduced accurate continuum model, are shown in \cref{fig:universal_curvature_k_first}. These distributions are consistent with those obtained from the fitted continuum models (see \cref{fig:curvature_k_first}), except that the mirror symmetry $M_y$ is broken due to the higher harmonic terms in the accurate continuum models.

% -------------------------------------------------------------------------
\subsubsection{Application to the $\Gamma$ Valley in A--AB and A--ABA Systems}
\label{appendix:universal_gamma_valley}

At the $\Gamma$ valley, the only low–energy degrees of freedom are the Mo $d_{z^{2}}$ orbitals on the specified layers, each carrying a spin $\sigma=\uparrow,\downarrow$.  We introduce the corresponding basis operators with spin degree of freedom as
\begin{equation}
\psi^\dagger_{l,\sigma,\bsl{k}},
\qquad
l\in
\begin{cases}
\{A_2,B_1\}, & \text{A--AB},\\
\{A_2,B_1,A_3\}, & \text{A--ABA},
\end{cases}
\quad
\sigma\in\{\uparrow,\downarrow\}.
\end{equation}
The continuum Hamiltonian then takes the form:
\begin{equation}
H^\Gamma
=
\sum_{\bsl{k}}\!
\sum_{l,l'}\!\sum_{\sigma,\sigma'}
\psi^\dagger_{l,\sigma,\bsl{k}}\;
h^\Gamma_{\,l\sigma,\;l'\sigma'}(\bsl{k})\;
\psi_{l',\sigma',\bsl{k}},
\end{equation}
where
\begin{equation}
h^\Gamma_{\,l\sigma,\;l'\sigma'}(\bsl{k})
=
\sum_{M_z,M_{z^*}\ge0}
\sum_{\bsl{G}_M}
t^{\,M_zM_{z^*}}_{\,l\sigma,\;l'\sigma';\bsl{G}_M}\,
Y^{\,M_zM_{z^*}}_{\,l\sigma,\;l'\sigma',\bsl{p}}(\bsl{k})
+\text{symm.\ terms},
\qquad
\bsl{p}=\bsl{q}_l-\bsl{q}_{l'}+\bsl{G}_M.
\end{equation}
This model corresponds to the single-orbital model with approximate spin degeneracy, as described in \cref{appendix:continuum_gamma}. The single orbital in \cref{appendix:continuum_gamma} is the highest-energy layer-hybridized $d_{z^2}$ state of $\{A_2,B_1\}$ ($\{A_2,B_1,A_3\}$) layers in A-AB (A-ABA) $t$MoTe$_2$.
\textit{symm.\ terms} denote the additional terms required by three‐fold rotation symmetry $C_{3z}$, time–reversal symmetry $\hat T$, and Hermitian conjugation.  The basis matrices are defined by:
\begin{equation}
\bigl[
Y^{\,M_zM_{z^*}}_{\,l\sigma,\;l'\sigma',\bsl{p}}(\bsl{k})
\bigr]_{\bsl{Q}_1\sigma_1,\bsl{Q}_2\sigma_2}
=
\delta_{l_{\bsl{Q}_1},l}\,\delta_{\sigma_1,\sigma}\,
\delta_{l_{\bsl{Q}_2},l'}\,\delta_{\sigma_2,\sigma'}\,
(\bsl{k}-\bsl{Q}_1)_z^{M_z}\,(\bsl{k}-\bsl{Q}_1)_{z^*}^{M_{z^*}}\,
\delta_{\bsl{Q}_1,\bsl{Q}_2+\bsl{p}},
\end{equation}
with $(\bsl{k}-\bsl{Q})_z=(k_x-Q_x)+i(k_y-Q_y)$ and
$(\bsl{k}-\bsl{Q})_{z^*}=(k_x-Q_x)-i(k_y-Q_y)$.

\paragraph{Physical meaning of the coefficients.}
The coefficients $t^{\,M_zM_{z^*}}_{\,l\sigma,\;l'\sigma';\bsl{G}_M}$ are obtained by projection and have the following interpretation:
\begin{itemize}
  \item $l=l'$, $\sigma=\sigma'$, $\bsl{G}_M=0$, $M_z=M_{z^*}=0$ $\Longrightarrow$ on‐site energy.
  \item $l=l'$, $\sigma=\sigma'$, $\bsl{G}_M=0$, $M_z+M_{z^*}\neq0$ $\Longrightarrow$ kinetic term.
  \item $l=l'$, $\sigma=\sigma'$, $\bsl{G}_M\neq0$ $\Longrightarrow$ intralayer moiré potential.
  \item $l\neq l'$, $\sigma=\sigma'$ $\Longrightarrow$ interlayer coupling.
  \item $\sigma\neq\sigma'$ (any $l,l',\bsl{G}_M$) $\Longrightarrow$ spin‐flip coupling.
\end{itemize}

\paragraph{Symmetry constraints.}
At the $\Gamma$ valley the continuum Hamiltonian is invariant under both three‐fold rotation $C_{3z}$ and time–reversal $\mathcal{T}$.

\textit{Three-fold rotation $C_{3z}$.}  
In the composite index $(\bsl{Q},l,\sigma)$ the representation is:
\begin{equation}
[D(C_{3z})]_{\bsl{Q},l,\sigma;\,\bsl{Q}',l',\sigma'}
=
\delta_{l,l'}\,
\delta_{\bsl{Q},\,C_{3z}\bsl{Q}'}\,
\bigl[e^{-i\pi\sigma_z/3}\bigr]_{\sigma'\sigma}.
\end{equation}
Because the spin matrix carries the entire phase, rotation of a:
basis matrix yields
\begin{equation}
D(C_{3z})\;
Y^{\,M_zM_{z^*}}_{l\sigma,l'\sigma',\bsl{p}}(\bsl{k})\;
D(C_{3z})^{-1}
=
e^{\,-i\frac{2\pi}{3}(M_z-M_{z^*})}\;
Y^{\,M_zM_{z^*}}_{\,l\sigma,l'\sigma',\,C_{3z}\bsl{p}}
\bigl(C_{3z}\bsl{k}\bigr),
\end{equation}
which imposes:
\begin{equation}
t^{\,M_zM_{z^*}}_{\,l\sigma,\;l'\sigma';\bsl{G}_M}
=
e^{\,-i\frac{2\pi}{3}(M_z-M_{z^*})}\;
t^{\,M_zM_{z^*}}_{\,l\sigma,\;l'\sigma';\,C_{3z}\bsl{G}_M}.
\end{equation}

\begin{figure}[!hbtp]
    \centering
    \includegraphics[width=1\linewidth]{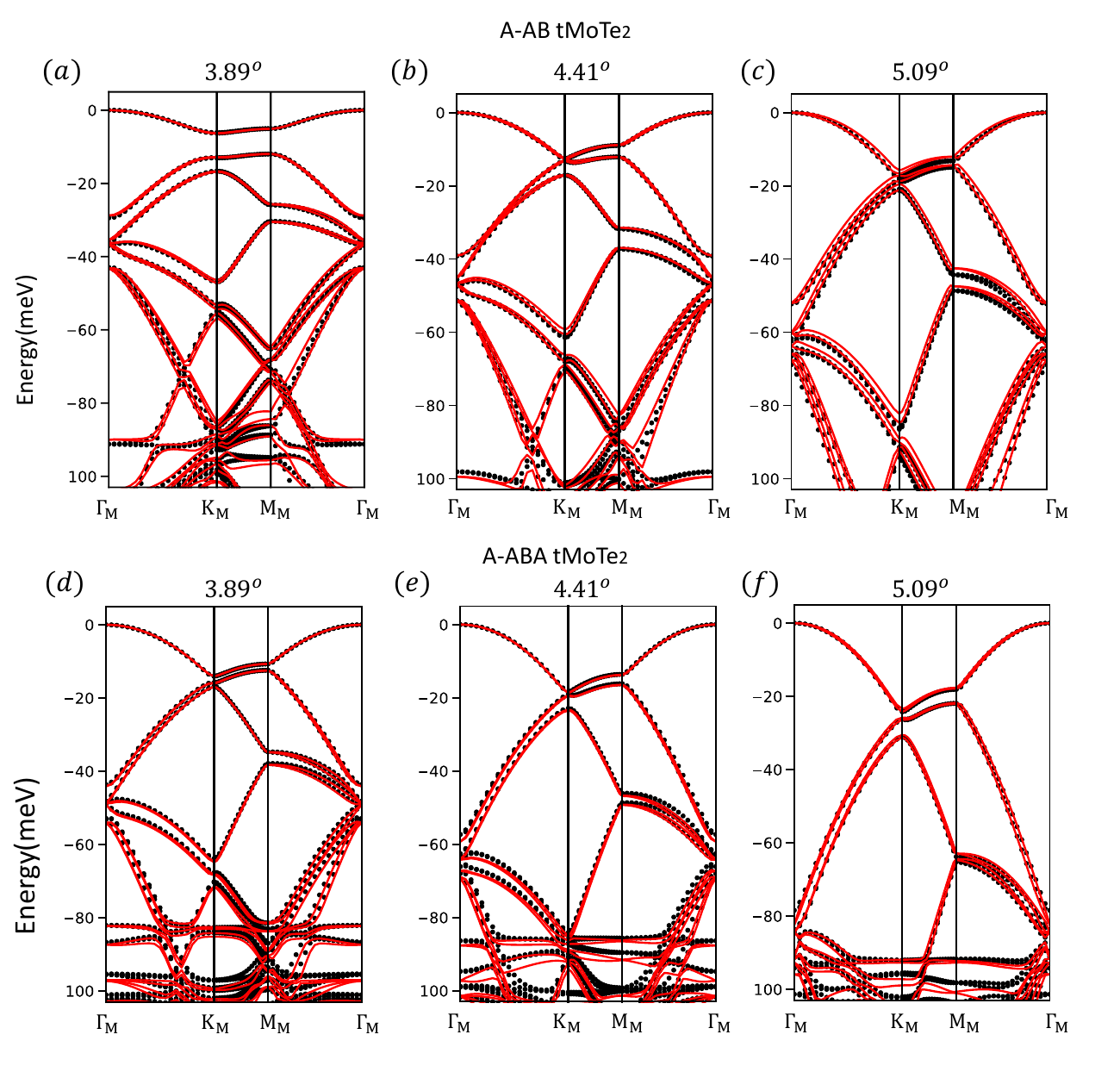}
    \caption{(a)–(c) Reduced model results for the $\Gamma$ valley of A-AB $t$MoTe$_2$. 
(d)–(f) Reduced model results for the $\Gamma$ valley of A-ABA $t$MoTe$_2$. 
Black dots indicate the DFT band structures at the $\Gamma$ valley, while the red lines show the corresponding continuum-model calculations.}
    \label{fig:ruduced_gamma_model}
\end{figure}

\textit{Time reversal $\hat T$.}  
With
$D(T)_{\bsl{Q},l,\sigma;\,\bsl{Q}',l',\sigma'}=
\delta_{l,l'}\,
\delta_{\bsl{Q}',-\bsl{Q}}\,
[i\sigma_y]_{\sigma'\sigma}\,\mathcal K$,
one finds
\begin{equation}
D(T)\;
Y^{\,M_zM_{z^*}}_{\,l\sigma,l'\sigma',\bsl{p}}(\bsl{k})\;
D(T)^{-1}
=
(-1)^{M_z+M_{z^*}}\;
(i\sigma_y)\;
\bigl[
Y^{\,M_{z^*}M_z}_{\,l\bar\sigma,\,l'\bar\sigma',\,-\bsl{p}}(-\bsl{k})
\bigr]^{*}
\;(i\sigma_y),
\end{equation}
leading to the coefficient constraint:
\begin{equation}
t^{\,M_zM_{z^*}}_{\,l\sigma,\;l'\sigma';\bsl{G}_M}
=
(-1)^{M_z+M_{z^*}}\;
\bigl[
t^{\,M_{z^*}M_z}_{\,l'\bar\sigma',\,l\bar\sigma;\,-\bsl{G}_M}
\bigr]^{*},
\qquad
\bar\uparrow=\downarrow,\;
\bar\downarrow=\uparrow.
\end{equation}
These symmetry relations are imposed when the continuum parameters are extracted from first-principles calculations. \cref{fig:ruduced_gamma_model} shows the results of the reduced models for the $\Gamma$ valley of A-AB and A-ABA $t$MoTe$_2$ with various twist angles, and the model parameters are listed in ~\cite{IOP_eln2}. Using this reduced accurate continuum model, we evaluate the spin polarization of the low-energy states to verify the robustness of the spin $U(1)$ symmetry at the $\Gamma$ valley. Because of the weak SOC at the $\Gamma$ valley, the spin-polarization direction of each pair of nearly spin-degenerate states varies across the moiré Brillouin zone. To correct for this, we employ the following procedure to rotate each pair of nearly degenerate states to maximize their spin polarization.

\begin{itemize}
\item We denote the two nearly degenerate eigenstates at the $\Gamma$ valley as $\ket{\psi_1(k)}$ and $\ket{\psi_2(k)}$. The goal is to determine the transformation coefficients that yield two orthogonal states, $\ket{\phi_1(k)}$ and $\ket{\phi_2(k)}$, with maximized spin polarization

\bea
\label{eq:polarization_state}
\ket{\phi_1(k)} = c_1 \ket{\psi_1(k)} + c_2 \ket{\psi_2(k)}\\
\ket{\phi_2(k)} = c_2^* \ket{\psi_1(k)} - c_1^* \ket{\psi_2(k)}
\eea
The spin operator is denoted as $M_{sp} = s_z \otimes I$. The spin polarization of the original two states is given by $P_i = \bra{\psi_i(k)} M_{sp} \ket{\psi_i(k)}$ ($i=1,2$), and we define $Q = |Q| e^{i\gamma} = \bra{\psi_1} M_{sp} \ket{\psi_2}$. The spin polarization $\tilde{P}_i$ of the two target states, $\ket{\phi_i}$, is then expressed as

\bea
\tilde{P_1} = |c_1|^2 P_1 + |c_2|^2 P_2 + 2Re [c_1^*c_2 Q]\\
\tilde{P_2} = |c_2|^2 P_1 + |c_1|^2 P_2 -2Re [c_1^*c_2 Q]
\eea
We extract $c_1$ and $c_2$ by maximizing $\tilde{P}_1$ (and minimizing $\tilde{P}_2$). Note that they satisfy the normalization condition $|c_1|^2 + |c_2|^2 = 1$, so we can write $|c_1| = \cos(\theta)$ and $|c_2| = \sin(\theta)$. The analytical expressions for $c_1$ and $c_2$ are then given by

\bea
\label{eq:spin_polarization}
cos(2\theta)=&\frac{P_1-P_2}{\sqrt{(P_1-P_2)^2+4|Q|^2}}\\
c_1=\sqrt{1+\frac{cos(2\theta)}{2}}&\;,\; c_2=\sqrt{1-\frac{cos(2\theta)}{2}}e^{-i\gamma}
\eea

\end{itemize}

\begin{figure}[!hbtp]
    \centering 
    \includegraphics[width=1.00\textwidth]{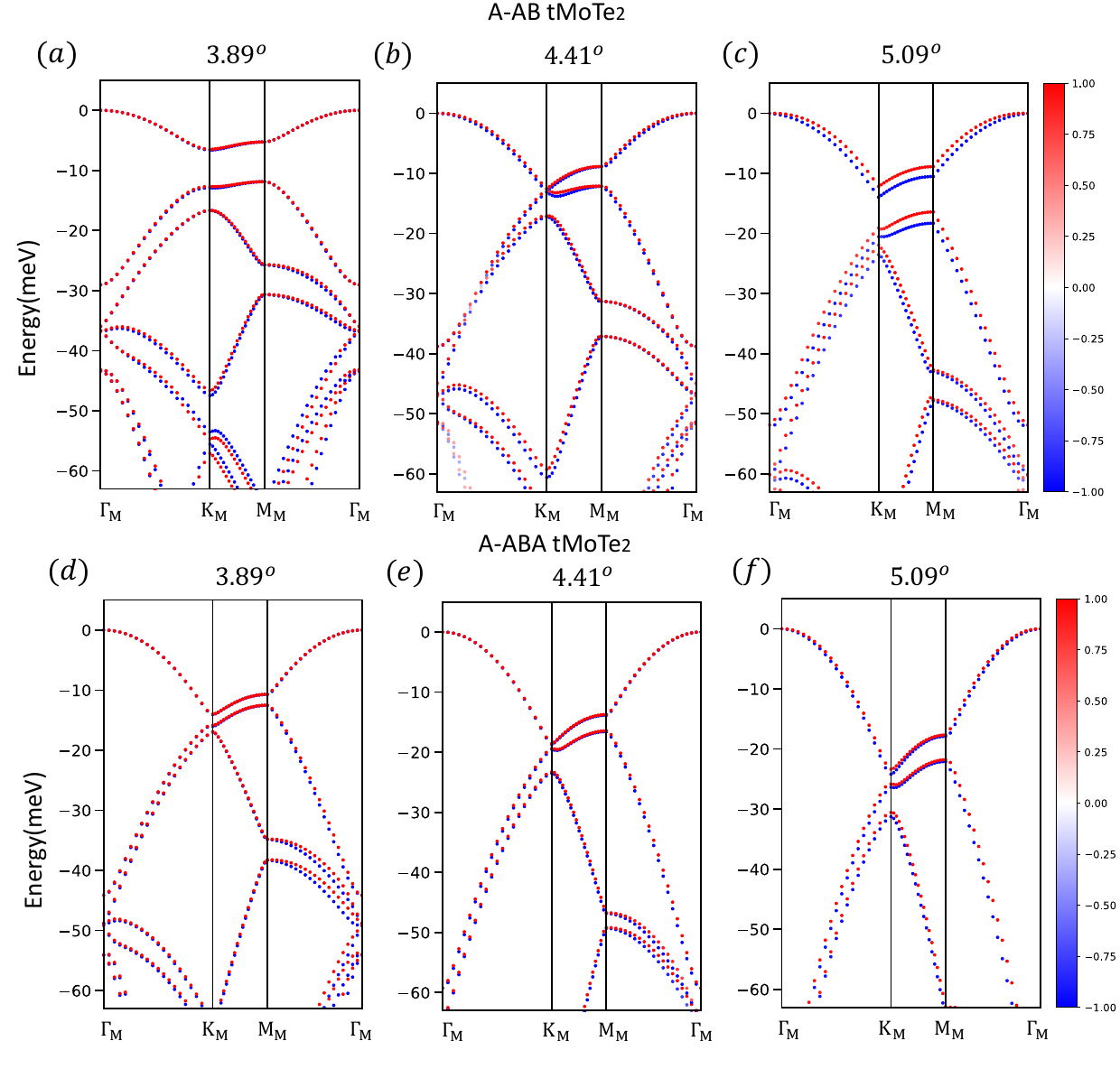}
    \caption{ Spin polarization $\tilde{P_i}$ of the rotated low-energy states $\ket{\phi_i}$, weighted on the energy bands at the $\Gamma$ valley in A-AB and A-ABA $t$MoTe$_2$.
    These results are obtained from the accurate continuum models.
  \label{fig:spin_polar_reduce_gamma} }
\end{figure}

Using the parameters in \cref{eq:spin_polarization}, we obtain $\ket{\phi_i}$ from \cref{eq:polarization_state} and calculate $\tilde{P}_i$ of the low-energy states at different $k$-points, shown as the color weight in \cref{fig:spin_polar_reduce_gamma}. The results exhibit strong spin polarization and, consequently, robust spin $U(1)$ symmetry in the low-energy states at the $\Gamma$ valley for both A-AB and A-ABA $t$MoTe$_2$.

\begin{figure}[!hbtp]
    \centering 
    \subfloat{\label{WilsonLoop:a}}
    \subfloat{\label{WilsonLoop:b}}
    \subfloat{\label{WilsonLoop:c}}
    \subfloat{\label{WilsonLoop:d}}
    \subfloat{\label{WilsonLoop:e}}
    \subfloat{\label{WilsonLoop:f}}
    \includegraphics[width=1.00\textwidth]{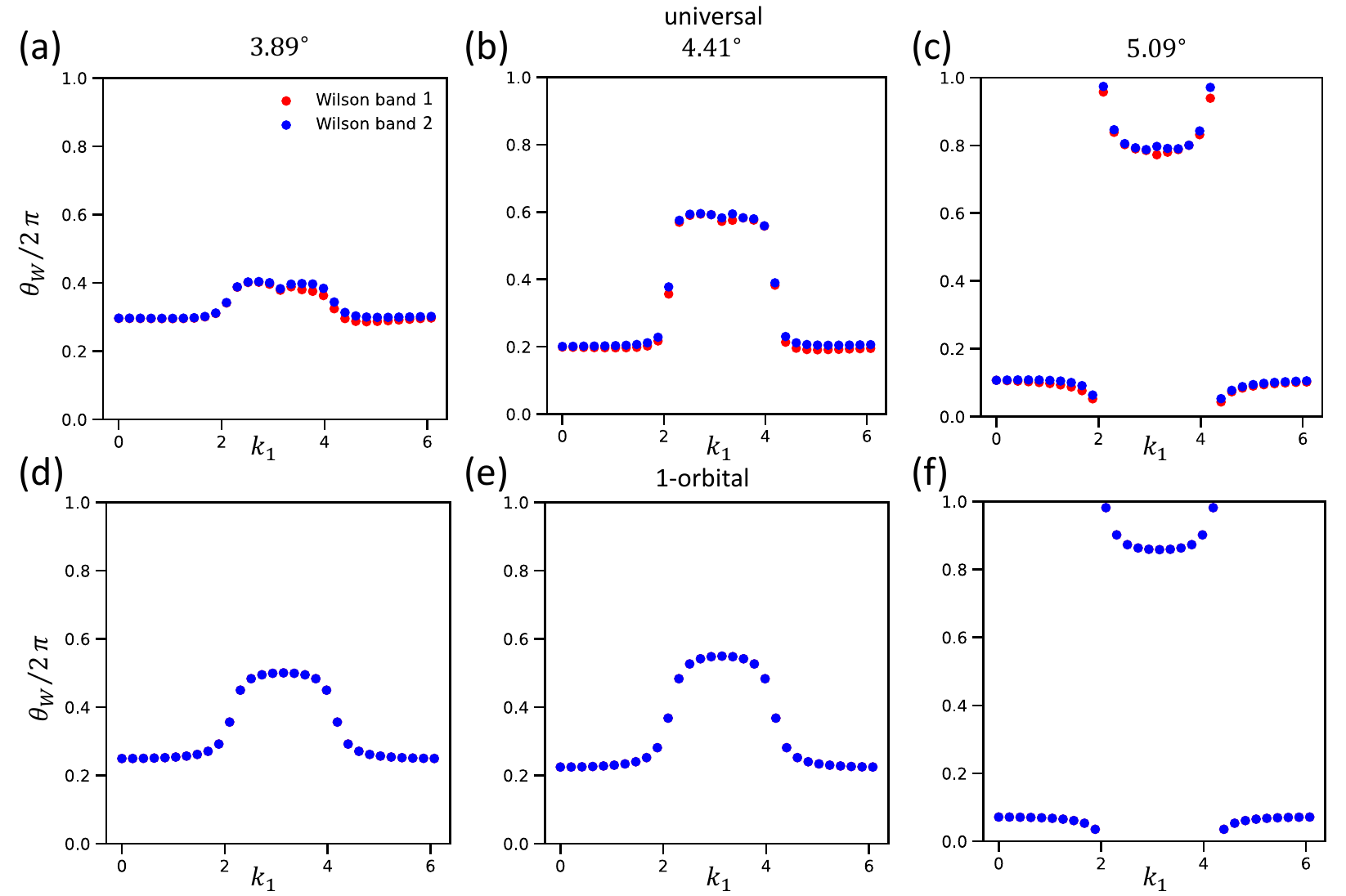}
    \caption{Wilson loops of the nearly spin-degenerate top $\Gamma$-valley bands in A-ABA \tmt\ along $b_{M_1}$. Panels (a)–(c) show the results calculated from the accurate continuum model for twist angles between $3.89^\circ$, $4.41^\circ$ and $5.09^\circ$, while (d)–(f) display those obtained from the fitted single-orbital continuum model.
  \label{fig:WilsonLoop} }
\end{figure}

In A–ABA \tmt, a band inversion occurs between the low-energy $\Gamma$-valley bands when the twist angle increases from $4.41^\circ$ to $5.09^\circ$. To further probe the associated topological properties, we computed the Wilson loops \cite{Wilsonloop_PhysRevB.84.075119} of the nearly spin-degenerate top $\Gamma$-valley bands along $b_{M_1}$ using the accurate continuum model, as shown in \cref{WilsonLoop:a}-\cref{WilsonLoop:c}. The Wilson loops exhibit a clear distinction across the band inversion, yet both the $Z_2$ index and the Chern number remain zero throughout. To benchmark these results, we also calculated the corresponding Wilson loops using the fitted single-orbital continuum model in \cref{WilsonLoop:d}-\cref{WilsonLoop:f}, which yield consistent results. Notably, because the fitted model includes only the first-harmonic terms, the top $\Gamma$-valley bands from this model are exactly spin-degenerate, leading to two coincident Wilson loop lines. Nevertheless, the inclusion of higher-harmonic terms in the accurate continuum model slightly lifts this degeneracy, resulting in a small splitting in both the top $\Gamma$-valley energy bands and their corresponding Wilson loops.

\section{Localized Wannier functions and tight-binding models of A-ABA \tmt}
\label{appendix:TB_model}

In this section, we present the details of constructing maximally localized Wannier functions (MLWFs) and tight-binding (TB) models for the top two bands from the $K$ valley ($C_\uparrow = \pm 1$) and the top bands from the $\Gamma$ valley ($C_s = 0$) with nearly spin degeneracy in A-ABA $t$MoTe$_2$.

\subsection{Procedure of Wannierization for continuum models}
\label{appendix:wann_method}

In this work, we construct MLWFs and extract TB models directly from continuum models using the WANNIER90 package~\cite{MOSTOFI2008685}.
Previous studies have successfully constructed minimal TB models of targeted moiré bands in twisted graphene~\cite{wann_PhysRevResearch.1.033072,song_1,song_2,Yu_2023_PhysRevB,herzogarbeitman2024} using this method. Here, we generalize this method to construct TB models from continuum models at multiple valleys in twisted TMDs, which can also be applied to other moir\'e systems.

In the WANNIER90 package, the MLWFs and TB models are constructed by minimizing the total Wannier function spread in real space~\cite{Wannier_PhysRevB.56.12847}, which reads

\bea
\Omega=\sum_n [ \bra{W_n^\eta(\textbf{0})} r^2 \ket{W_n^\eta(\textbf{0})} - (\bra{W_n^\eta(\textbf{0})} \textbf{r} \ket{W_n^\eta(\textbf{0})})^2 ]
\eea
Here, $\ket{W_n^\eta(\textbf{R})}$ denotes the n-th Wannier function localized at the unit cell $\textbf{R}$ for the $\eta$ valley in the moir\'e system. All that is required is to generate four input files (".win", ".eig", ".mmn", ".amn") from the continuum models. The procedure for generating these files from continuum models differs in several aspects from the standard procedure based on conventional first-principles calculations. Below, we present the detailed steps and key considerations for generating these four files.

\begin{itemize}
  \item The procedure for preparing the ".win" and ".eig" files is straightforward. The ".win" file requires information about the moir\'e unit cell, the number of Wannier functions, window settings, the coordinates of $k$-points, and other relevant parameters. The ".eig" file requires the eigenvalues of the targeted moiré bands at the $k$-points specified in the ".win" file.

    \item The ".mmn" file requires the overlap matrix elements $M_{mn}^{\textbf{k},\textbf{b}}$ of the periodic part $[u_{m,\textbf{k}}]$ of the moir\'e Bloch wave functions $[\psi_{m,\textbf{k}}]$, which are given by
    \bea
    \label{eq:mmn}
    M_{mn}^{\textbf{k},\textbf{b}}=\braket{u_{m,\textbf{k}}}{u_{n,\textbf{k}+\textbf{b}}}
    \eea
  The key point is that the basis of the Bloch states in continuum models is the plane wave basis set $\ket{\alpha, \textbf{k}-\textbf{Q}^\eta}$, where $\{\textbf{Q}^\eta\}$ are the plane wave vectors at the $\eta$ valley, and $\alpha$ denotes the layer, spin, and orbital degrees of freedom. Once $\textbf{k}+\textbf{b}$ crosses a MBZ boundary, a unitary matrix $S(\textbf{G}_M)$ is required to eliminate the random phase problem by transforming $[\psi_{m,\textbf{k}+\textbf{b}-\textbf{G}_M}]$ to $[\psi_{m,\textbf{k}+\textbf{b}}]$, which satisfies

  \bea
  \ket{\psi_{m,\textbf{k}+\textbf{b}}} = S(\textbf{G}_M) \ket{\psi_{m,\textbf{k}+\textbf{b}-\textbf{G}_M}}
  \eea
  Where $\textbf{G}_M$ denotes the moir\'e reciprocal vector and $(\textbf{k}+\textbf{b}-\textbf{G}_M)$ is in the first MBZ.

  \item The ".amn" file requires the initial trial wave functions, which we construct using the Fourier transform of Gaussian functions localized at the expected Wannier centers:
  \bea
  &[\psi_{init}^{\textbf{r}_0}]_{\alpha,\textbf{k} -\textbf{Q}^\eta }=\frac{1}{N_r}\sum_{\textbf{r}} e^{i(\textbf{k}-\textbf{Q}^\eta)\cdot \textbf{r}} \;\psi^{gauss}(\textbf{r};\sigma_0,\textbf{r}_0) \\
  &\psi^{gauss}(\textbf{r};\sigma_0,\textbf{r}_0) = \frac{1}{2\pi\sigma_0^2}\; exp[-\frac{(\textbf{r}-\textbf{r}_0)}{2\sigma_0^2}]
  \eea
  Here, $[\psi_{init}^{\textbf{r}_0}]_{\alpha,\textbf{k} -\textbf{Q}^\eta }$ is the $(\alpha,\textbf{k} -\textbf{Q}^\eta)$ component of the initial trial function, and $\psi^{\text{gauss}}(\textbf{r};\sigma_0,\textbf{r}_0)$ is a standard Gaussian function localized at $\textbf{r}_0$. In A-ABA $t$MoTe$_2$ systems, for the top two moiré bands from the $K$ valley with $C_\uparrow=\pm 1$, we use two trial functions localized at the M-XMX and X-MXM regions, which correspond to the same localized regions as in AA bilayer $t$MoTe$_2$~\cite{devakul_magic_2021}. For the top moiré band from the $\Gamma$ valley, with nearly spin-degenerate states and $C_s=0$, we use trial functions localized at the M-XMX region for $3.89^\circ$ and $4.41^\circ$, and at the A-ABA region for $5.09^\circ$ A-ABA $t$MoTe$_2$, according to the charge density distributions (see \cref{fig7}).

\end{itemize}

After running the WANNIER90 code, we obtain the Hamiltonian information in the file "\_hr.dat" and the $\textbf{k}$-dependent unitary matrix $U^\eta_{mn}(\textbf{k})$ in the file "\_u.mat". The MLWFs, $\ket{W^\eta_n(\textbf{R})}$, are then expressed as follows:

\bea
\label{eq:wannier}
&\ket{\tilde{\psi}^\eta_{n\textbf{k}}}=\sum_{m=1,2}U^\eta_{nm}(\textbf{k})\ket{\psi^\eta_{m\textbf{k}}} \\
&\ket{W^\eta_n(\textbf{R})}=\frac{1}{\sqrt{N_k}}\sum_{\textbf{k}}e^{-i\textbf{k}\cdot\textbf{R}}\ket{\tilde{\psi}^\eta_{n\textbf{k}}}
\eea

\subsection{Results of Wannierization at $K/K'$ and $\Gamma$ valley in A-ABA \tmt}
\label{appendix:wann_result}
In this part, we present our results of Wannierization based on the continuum models at the $K/K'$ and $\Gamma$ valleys.

\begin{figure}[H]
    \centering 
    \includegraphics[width=1.00\textwidth]{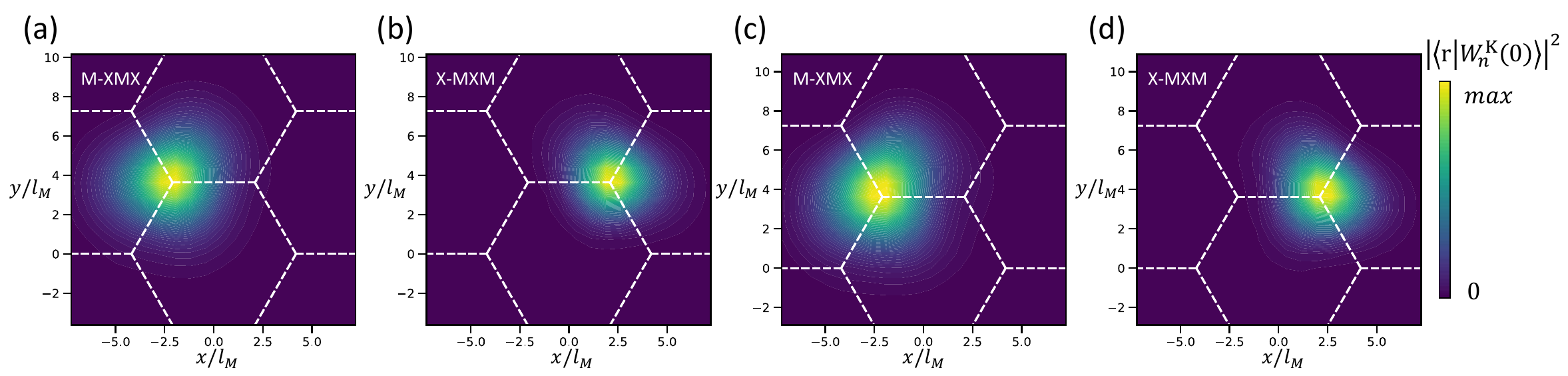}
    \caption{    Wannier function distributions $|\braket{\textbf{r}}{W^K_n(\textbf{0})}|^2$ of the two orbitals at the $K$ valley, extracted from the three-orbital continuum model (a,b) and the accurate continuum model (c,d) for A-ABA $t$MoTe$_2$. The two Wannier functions are localized at the M-XMX and X-MXM regions, respectively. The moiré lattice length is defined as $l_M = a_M / (\frac{4\pi}{\sqrt{3}})$.\label{fig:wann_K}}
\end{figure}

\begin{figure}[H]
    \centering 
    \includegraphics[width=1.00\textwidth]{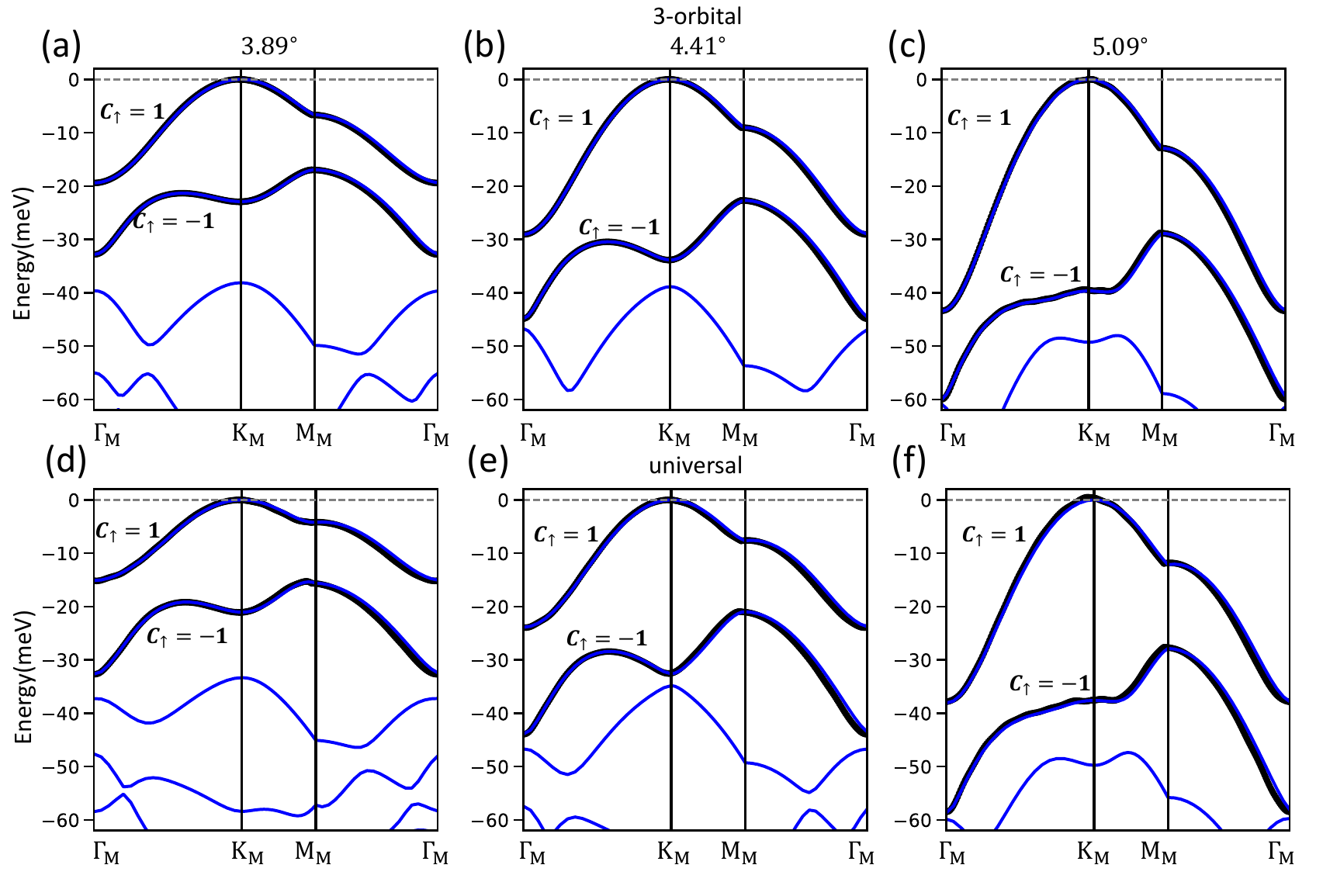}
    \caption{ Comparisons of the energy bands calculated from the two-band TB model (black lines) with those from the three-orbital (a–c) and accurate (d–f) continuum models (blue lines) at the $K$ valley for different twist angles. \label{fig:TB_angle_K} }
\end{figure}

We employ Eq.~\ref{eq:wannier} to calculate the real-space distributions of MLWFs obtained from WANNIER90. \cref{fig:wann_K} shows the Wannier function distributions, $|\braket{\textbf{r}}{W^K_n(\textbf{0})}|^2$, of the top two valence bands at the $K$ valley for A-ABA $t$MoTe$_2$ with a $3.89^\circ$ twist angle, calculated using both the three-orbital continuum model (\cref{appendix:continuum_three_orbital}) and the accurate continuum model (\cref{appendix:universal_K_valley}). These Wannier functions are localized at the M-XMX and X-MXM regions, giving consistent results between the two models. Different from AA bilayer $t$MoTe$_2$, the spreads of the two Wannier functions are not identical, with a larger spread at the M-XMX region due to the $C_{2y}$ symmetry breaking. The comparisons of the energy bands calculated from the two-band TB models and the continuum models at the $K$ valley for different twist angles are shown in \cref{fig:TB_angle_K}, exhibiting excellent agreement.

\begin{figure}[H]
    \centering 
    \includegraphics[width=1.00\textwidth]{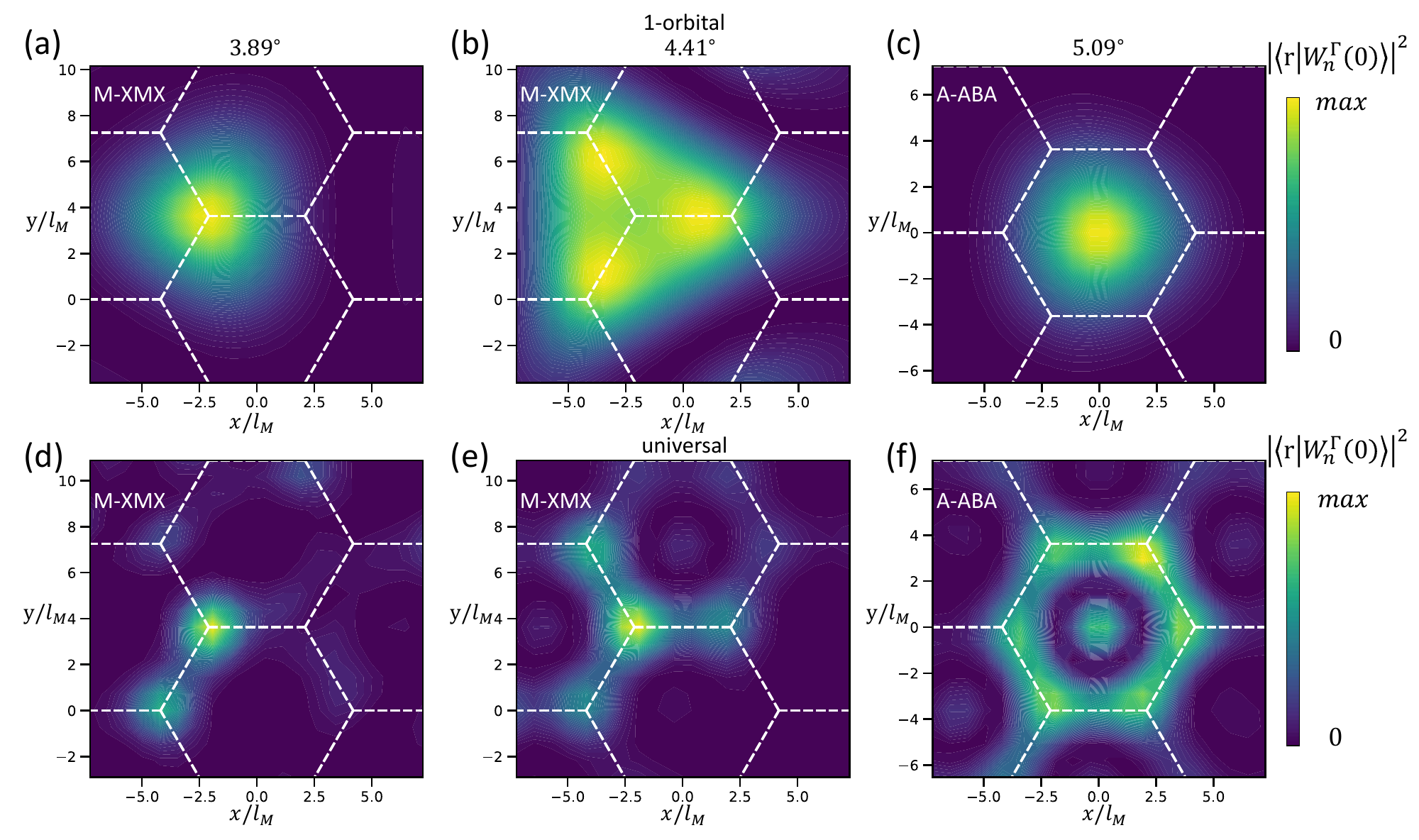}
    \caption{ Wannier function distributions $|\braket{\textbf{r}}{W^\Gamma_n(\textbf{0})}|^2$, extracted from the single-orbital (a–c) and accurate (d–f) continuum models at the $\Gamma$ valley in A-ABA $t$MoTe$_2$ for various twist angles.
\label{fig:wann_G} }
\end{figure}

\begin{figure}[H]
    \centering 
    \includegraphics[width=1.00\textwidth]{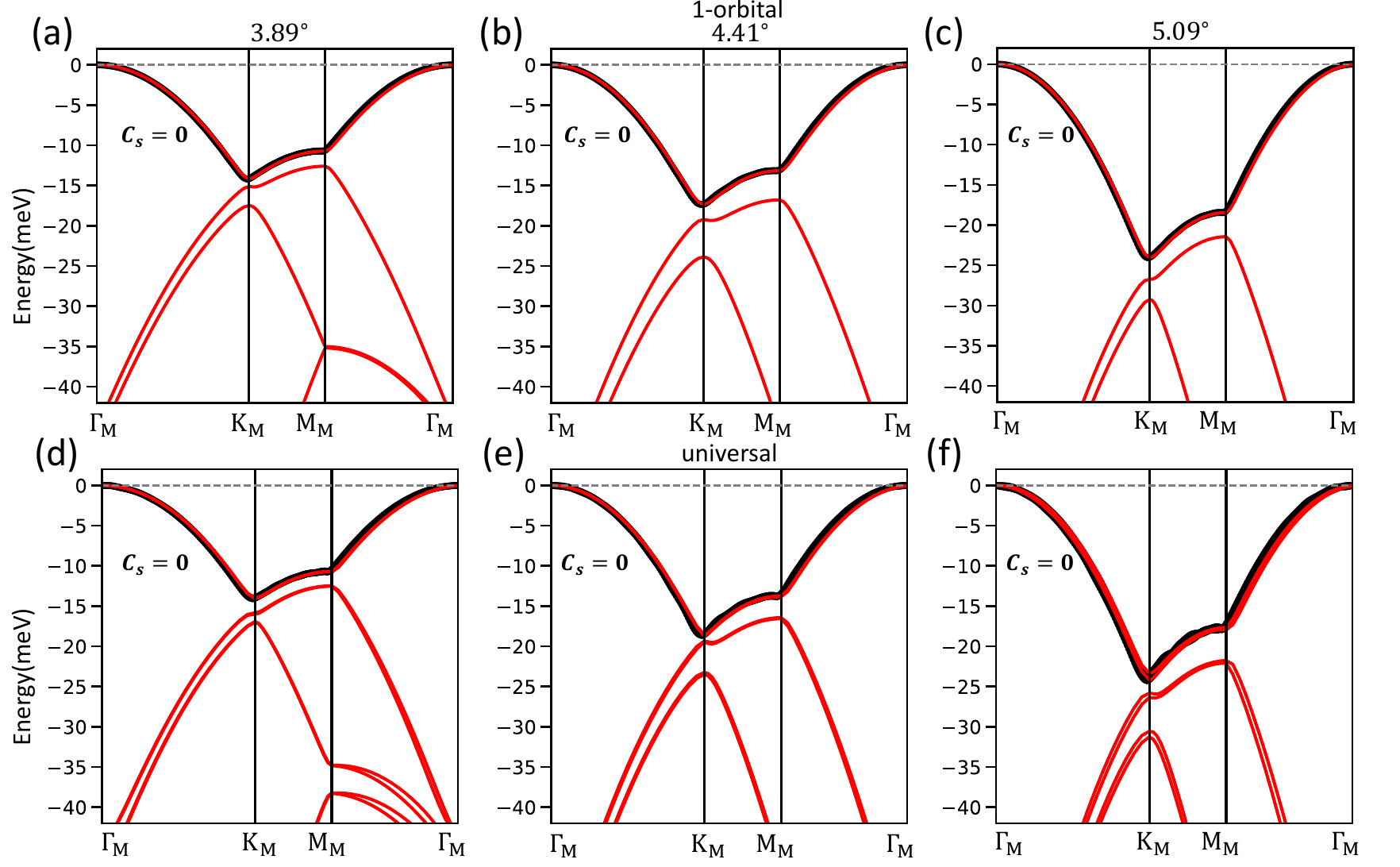}
    \caption{ Comparisons of the energy bands calculated from the single-band TB model (black lines) with those from the single-orbital (a–c) and accurate (d–f) continuum models (red lines) at the $\Gamma$ valley for different twist angles.
\label{fig:TB_angle_G} }
\end{figure}

\cref{fig:wann_G} shows the Wannier function distributions, $|\braket{\textbf{r}}{W^\Gamma_n(\textbf{0})}|^2$, of the top valence bands with nearly spin degeneracy at the $\Gamma$ valley of A-ABA $t$MoTe$_2$. The distributions are calculated using both the one-orbital continuum model (\cref{appendix:continuum_gamma}) and the accurate continuum model (\cref{appendix:universal_gamma_valley}) for different twist angles.
These two Wannier functions are localized at the M-XMX region for $3.89^\circ$ and $4.41^\circ$, and at the A-ABA region for $5.09^\circ$, yielding consistent results with only minor shape differences. The shift of the Wannier function center with varying twist angle is caused by the band inversion between $4.41^\circ$ and $5.09^\circ$, consistent with the trend observed in the charge density distributions (see \cref{A_ABA_Gamma_fit}). The comparisons of energy bands calculated from the TB models and the continuum models at the $\Gamma$ valley for different twist angles are shown in \cref{fig:TB_angle_G}, also exhibiting excellent agreement.

\end{document}